\documentclass[reprint,amsmath,amssymb, aps, rmp]{revtex4-2}

\usepackage{graphicx}
\usepackage{dcolumn}
\usepackage{bm}
\usepackage{hyperref}

\usepackage{comment}
\usepackage[compat=1.1.0]{tikz-feynman} 

\usepackage[utf8x]{inputenc}
\usepackage{relsize}
\usepackage{tikz}
\usepackage{soul}
\setstcolor{red}

\DeclareMathAlphabet{\mathdutchcal}{U}{dutchcal}{m}{n}
\SetMathAlphabet{\mathdutchcal}{bold}{U}{dutchcal}{b}{n}
\DeclareMathAlphabet{\mathdutchbcal}{U}{dutchcal}{b}{n}

\usepackage{float}

\usepackage{subfigure}          
\usepackage{bm}
\usepackage{enumerate}
\usepackage[capitalise]{cleveref}
\usepackage{bbold}
\usepackage{dsfont}

\hypersetup{
	colorlinks   = true, 
	urlcolor     = blue, 
	linkcolor    = blue, 
	citecolor   = blue 
}

\newcommand{\ket}[1]{|#1\rangle}
\newcommand{\bra}[1]{\langle#1|}
\newcommand{\bracket}[2]{\langle#1|#2\rangle}

\newcommand{\vev}[1]{\langle#1\rangle}

\newcommand{\ai}{a_i}
\newcommand{\aidag}{a_i^\dagger}
\newcommand{\mb}[1]{\mathbf{#1}}
\newcommand{\bbeta}{\boldsymbol\beta}
\newcommand{\eeta}{\boldsymbol\eta}
\newcommand{\aalpha}{\boldsymbol\alpha}
\newcommand{\pphi}{\boldsymbol\phi}
\newcommand{\ggamma}{\boldsymbol\gamma}

\newcommand{\del}{\partial}

\begin{document}

\title{Field theories and quantum methods for stochastic reaction-diffusion systems}

\author{Mauricio J. del Razo}
\email{m.delrazo@fu-berlin.de}
 \affiliation{Department of Mathematics and Computer Science, Freie Universit{\"a}t Berlin}
\affiliation{and Zuse Institute Berlin}

\author{Tommaso Lamma}
\email{t.lamma@math.leidenuniv.nl}
\affiliation{Mathematical Institute, Leiden University}

\author{Wout Merbis}
\email{w.merbis@uva.nl}
\affiliation{Dutch Institute for Emergent Phenomena, Institute for Theoretical Physics, University of Amsterdam
}

\date{\today}

\begin{abstract}
Complex systems are composed of many particles or agents that move and interact with one another. In most real-world applications, these systems involve a varying number of particles/agents that change due to interactions with the environment or their internal dynamics. The underlying mathematical framework to model these systems must incorporate the spatial transport of particles/agents and their interactions, as well as changes to their copy numbers, all of which can be formulated in terms of stochastic reaction-diffusion processes. However, the standard probabilistic representation of these processes can be overly complex because of the combinatorial aspects arising due to the non-linear interactions and varying particle numbers. In this manuscript, we review the main field theory representations of stochastic reaction-diffusion systems, which handle these issues ``under--the--hood''. First, we focus on bringing techniques familiar to theoretical physicists ---such as second quantization, Fock space, path integrals and quantum field theory--- back into the classical domain of reaction-diffusion systems. We demonstrate how various field theory representations, which have evolved historically, can all be unified under a single basis-independent representation. We then extend existing quantum-based methods and notation to work directly on the level of the unifying representation, and we illustrate how they can be used to consistently obtain previous known results in a more straightforward manner, such as numerical discretizations and relations between model parameters at multiple scales. Throughout the work, we contextualize how these representations mirror well-known models of chemical physics depending on their spatial resolution, as well as the corresponding macroscopic (large copy number) limits. The framework presented here may find applications in a diverse set of scientific fields, including physical chemistry, theoretical ecology, epidemiology, game theory and socio-economical models of complex systems, specifically in the modeling and multi-scale simulation of complex systems with varying numbers of particles/agents. The presentation is done in a self-contained educational and unifying manner such that it can be followed by researchers across several fields.
\end{abstract}

\keywords{Chemical reaction networks, path integral, field theory, stochastic systems, Markov process}

\maketitle

\tableofcontents

\section{Introduction}

\subsection{Motivation and scope}
Complex systems are generally composed out of many particles (or agents) that move and interact with one another; they are used to model systems ranging from biochemical processes in living organisms \cite{anderson2015stochastic,winkelmann2020stochastic} to social and economic organizations such as opinion dynamics \cite{castellano2009statistical,helfmann2023modelling} or power, transportation and communication systems \cite{newman2018networks,helbing2001traffic}. In most real-world applications, these systems are open, which means they allow the exchange of energy and material with their surroundings. For instance, living cells exchange molecules and energy with their environment, driving a nonequilibrium cycle of energy consumption, waste production and heat dissipation that is fundamental for life \cite{qian2007phosphorylation}. The underlying mathematical framework to model a large number of these systems is stochastic reaction-diffusion, where the diffusion part models the movement of particles/agents in the space of interest, and the reaction part models interactions that change the number of particles in the system, such as chemical reactions, material exchange with the environment, or similar processes \cite{bressloff2014stochastic,del2024open,del2024dynamics}. Thus, there is an enormous interest in developing theory, techniques and numerical schemes to investigate stochastic reaction-diffusion systems in open settings.

One of the main difficulties of handling open systems, i.e. with varying numbers of particles, is the ``on--the--fly'' modification of the dimension of the system. For any classical or stochastic dynamical system, one can write the differential equations for the dynamics of a fixed number of particles, but as soon as a particle is added or removed, one would need to add or remove a dimension to the system of equations. This can be resolved by lifting the dynamics to the level of probability densities living in Fock space \cite{fock1932,dirac1929quantum}, as we will introduce later. This works whenever individual particles or agents of the same species are assumed to be indistinguishable, as is often the case in biochemical systems and other relevant applications. Inspired by this approach to many-body quantum mechanics, often called `second quantization', Fock space methods for stochastic reaction-diffusion systems has been developed by several authors, spread over several decades \cite{schonberg1952application,schonberg1953application,kadanoff1968transport,felderhof1971spin,doi1976second,doi1976stochastic,zel1978mass,grassberger1980fock,mikhailov1981path,mikhailov1981path2,peliti1985path}.
Some of the advantages are that this approach is designed to work directly at the probabilistic level; one can perform calculations, approximations and derive emergent models at different scales without having to handle the combinatorics explicitly.

\begin{figure*}
	\centering
	\includegraphics[width=0.95\textwidth]{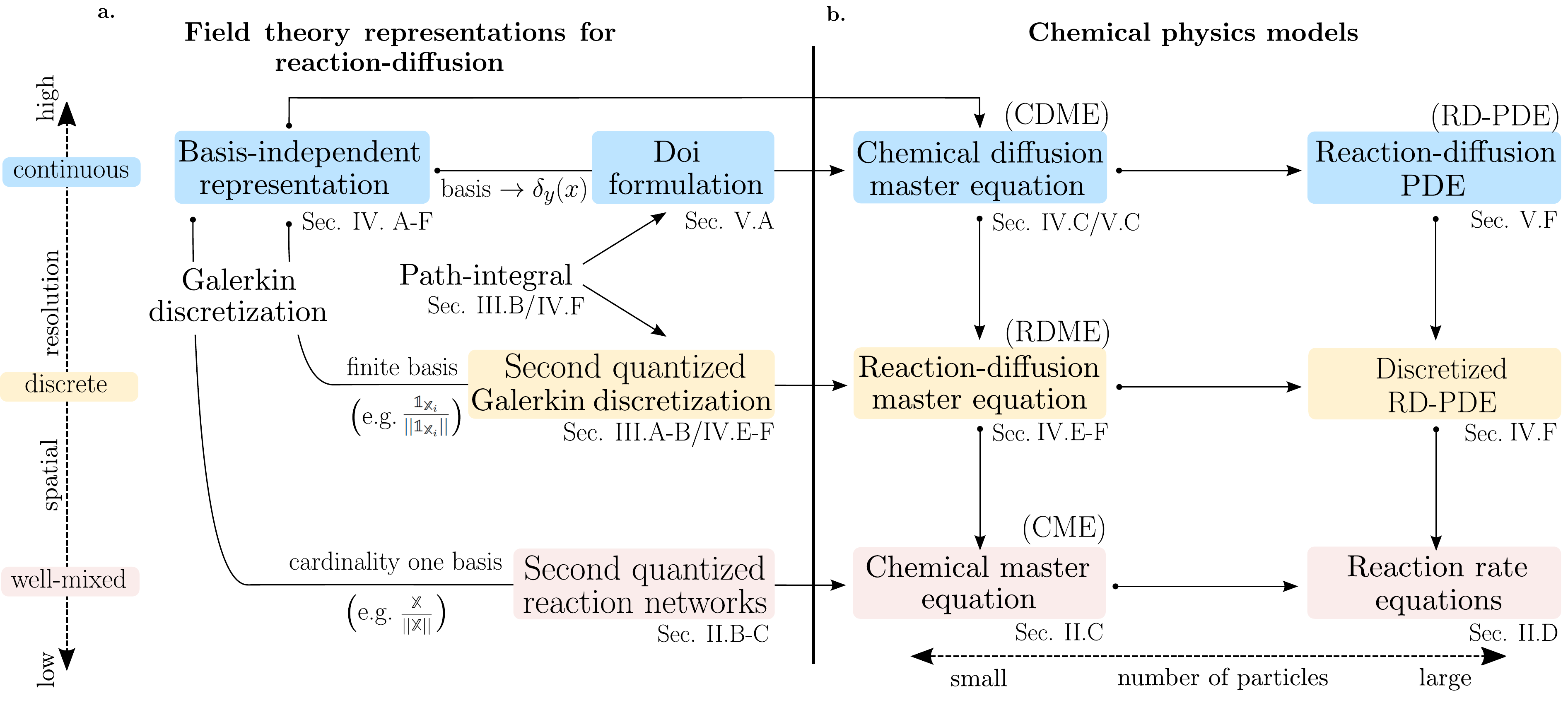}
\caption{Diagram showing the relations between the main theories and models discussed in this review, along with the section(s) where they are discussed. \textbf{a.} All the field theory representations emerge as special cases of the basis-independent representation of \cref{sec:stochMecRD}. These are categorized according to their spatial resolution determined by the basis of the space/subspace used. \textbf{b.} The stochastic chemical physics models corresponding to the field theory representations are shown depending on their spatial resolution. As the number of particles becomes larger, deterministic macroscopic models emerge.
The basis-independent formulation can be used to formulate the CDME, and thus other chemical physics models like the discretized reaction-diffusion partial differential equation (RD-PDE) for arbitrary choices of basis functions (\cref{sec:pelitipathint}). 
\label{fig:maindiag} }	
\end{figure*}

In this work, we review Fock space and field theoretic methods for stochastic reaction-diffusion systems. We show how the different field theory representations developed historically can be framed in one unifying representation.
We further contextualize how the different field theory representations mirror well-known models of chemical physics depending on their spatial resolution, as well as the corresponding macroscopic (large copy number) limits, which are often studied by applied mathematicians, e.g. \cite{erban2009stochastic,flegg2012two,kang2019multiscale,kim2017stochastic,kostre2021coupling,kurtz1971limit,del2024open,isaacson2022mean,isaacson2021reaction}. In specific, the wonderful books and reviews \cite{anderson2015stochastic,erban2020stochastic,schnoerr2017approximation, smith2019spatial, smith2018spatially,winkelmann2020stochastic} summarize many of these results. The relations between the field theory representations and chemical physics models at different spatial resolution and particle number are all condensed in \cref{fig:maindiag}. 

In particular, we present here a basis-independent representation following \cite{del2021probabilistic,delRazo2}) in the bra-ket notation from quantum mechanics \cite{dirac_braket}. We review well-known approaches, like the Doi formulation and the Doi-Peliti path integral, and we show how these are recovered as special cases of the basis-independent representation. We also illustrate how to perform Galerkin discretizations and how to obtain emergent macroscopic reaction-diffusion partial differential equations (PDEs), as well as how to derive a discretized macroscopic PDE.
Finally, we formulate alternative representations, such as the flux formulation and the stochastic concentrations, which are essential in establishing connections with tangential fields. It is our aim to present the basis-independent results using a uniform (bra-ket) notation, in a general setting and to highlight connections with existing models. 

In what is left of the introduction, we give a more detailed overview and references of previous relevant work, as well as the scope of the theory, various applications and complementary approaches that are not treated in detail in this review. Readers impatient for the formalism can jump directly to \cref{sec:quantmethodsCRN}.

\subsection{Overview of field theories in chemical physics}
The simplest model of chemical reactions systems is given by the reaction rate equations formulated by the law of mass action \cite{horn1972general,beard2008chemical,keener2009mathematical,murray2003book}. This consist of a system of deterministic, ordinary differential equations for the evolution of concentrations in a well-mixed setting. This means that diffusion is assumed to occur at a much faster timescale than reactions; the chemical components are distributed uniformly in space, and their spatial distribution does not play a role in the dynamics. It is well-known in the chemical physics community that these differential equations are only valid for large concentrations and that they emerge as large copy-number limit of the underlying stochastic model for well-mixed reaction networks \cite{anderson2015stochastic}. The stochastic model is called the chemical master equation (CME) \cite{gillespie1992rigorous,qian2010chemical,schnoerr2017approximation}. Unlike the rate equations, the state of the system is given by the number of particles and not their concentrations. The CME takes into account the inherent stochastic nature of reactions and describes the time evolution of the probability mass function, which yields the probability of having any given number of particles of each species at any given time. The CME is usually formulated for any chemical reaction network using standard methods; however, one can reformulate it using field-theoretic inspired methods \cite{munsky2006finite,Baez_2017,baez2018quantum}. This yields a second quantized formulation of reaction networks based on classical Fock space representations, illustrated in the bottom right row of \cref{fig:maindiag}. We begin the presentation of these methods in \cref{sec:quantmethodsCRN}, since it serves as a smooth introduction into the topic. 

The concepts of Fock space and second quantization were first introduced in 1932 by the physicist Vladimir Fock \cite{fock1932} within the scope of many-body quantum mechanics. 
In contrast to `first quantization', where each quantum particle needs to be characterized by its state, second quantization tracks for each possible state \textit{how many particles} are present. This allows one to easily deal with indistinguishable entities whose permutations leave the state of the system unchanged. These techniques prepared the ground for quantum field theory \cite{peskin2018introduction} and its applications in statistical field theory \cite{abrikosov} with the introduction of creation and annihilation operators, previously introduced in \cite{dirac1929quantum}. The second-quantized description hinges on the formulation of a Hamiltonian operator in terms of these creation and annihilation operators, allowing for a systematic way to discuss symmetries via Noether's theorem, seamlessly handling of varying particle numbers, and easily facilitating generalizations to multiple species of particles.
A first adaptation of Fock space methods in classical Hamiltonian systems can be found in two consecutive papers \cite{schonberg1952application,schonberg1953application} by Sch{\"o}nberg, where the second quantization procedure is carried out on the Liouville equation of classical statistical mechanics. These techniques have been applied and extended to formulate reaction-diffusion systems in \cite{doi1976second}. For the case of well-mixed reaction networks, where spatial dynamics no longer play a role, the framework is considerably simpler and one can use it to formulate and perform calculations with the CME, often with more ease than with its conventional formulation \cite{Baez_2017}. 

In the case of reaction-diffusion processes with full spatial resolution, one needs to keep track not only of the probability of having a certain number of particles of each species, but also their locations. Due to its complexity, the master equation for these processes is not simple to formulate. 
The master equation for stochastic reaction-diffusion processes was first constructed in two seminal papers by Masao Doi \cite{doi1976second,doi1976stochastic}. He used a field-theoretical representation---here referred to as the Doi formulation (\cref{fig:maindiag})---to ease the burden of writing the equation by hand and aid in analytical calculations. The microscopic description is probabilistic, but stochastic in the sense of classical probability theory, not quantum mechanical in the sense of amplitudes. However, we can still employ a Fock space representation to describe the probability distributions governing the systems dynamics \cite{doi1976second, doi1976stochastic,grassberger1980fock}. About a decade later, Peliti expanded on this work by discretizing the formulation of Doi into a lattice and constructing a path integral---the Doi-Peliti path integral \cite{peliti1985path}--- with the purpose of calculating observables, estimating perturbations and exploring the renormalization group in this context \cite{peliti1986renormalisation}. To facilitate referral to the master equation for reaction-diffusion processes (in analogy to the CME), it has recently been termed the chemical diffusion master equation (CDME), partly due to the term ``reaction-diffusion master equation'' already being used for the spatially discretized case. The underlying particle-based process described by the CDME---and the CDME itself---have also been referred to as the Doi model \cite{isaacson2013convergent}, and in special cases, the volume reactivity or $\lambda-\rho$ model \cite{erban2009stochastic,smith2019spatial}. In this manuscript, we refer to it is as the CDME to emphasize we specifically mean the general master equation for particle-based reaction-diffusion process.

\subsection{Application scope}
The applications of stochastic reaction-diffusion systems are remarkably broad, as they form the underlying model for a diverse range of complex systems. Consequently, field theory methods for reaction-diffusion systems have permeated several different fields, ranging from physics to biochemistry, population dynamics and neuroscience. Early work in physics has been mainly focused on the renormalization group analysis \cite{cardy1999field,tauber2005applications}, computation of critical exponents \cite{janssen1981nonequilibrium} and the characterization of dynamical phase-transitions \cite{hinrichsen2000non} in diffusion-controlled chemical reaction networks. Broadly speaking, two types of reactions are considered; the relaxational models \cite{peliti1986renormalisation,lee1994renormalization,lee1994scaling,lee1995renormalization}, where the system eventually reaches an absorbing state, or models with an absorbing state phase transition \cite{janssen1981nonequilibrium,grassberger1981phase,cardy1985epidemic,cardy1998field,janssen2005field,janssen2005survival}, where there is competition between the absorbing state and particle creation. In both cases, it is possible to use the Doi-Peliti field theory formulation to compute critical exponents in an expansion close to the critical dimension using the renormalization group, as we will review in \cref{sec:perturbandrenormalize}. Other applications in physics concern diffusion-limited reactions \cite{doi1976stochastic} and aggregation processes \cite{peliti1985field,sandow1993aggregation,rudavets1993phase}, nonequilibrium systems \cite{rose1979renormalized}, kinetic models such as the (totally) asymmetric exclusion process \cite{sasamoto1998exact,antal2000asymmetric} or kinetically contrained models of glassy dynamics \cite{ritort2003glassy,schulz1997analytical,garrahan2007dynamical,garrahan2009first,garrahan2018aspects,whitelam2004dynamic}, multiple species pair annihilation \cite{deloubriere2002multispecies,hilhorst2004segregation}, the contact process \cite{deroulers2004field}, sandpile models \cite{wiese2016coherent,wiese2022theory} and applications to exemplary stochastic processes \cite{nekovar2016field,bothe2021doi}.

Doi-Peliti field theories have found further applications in different research fields, marking their use as a general tool for analysing stochastic systems. In cellular and molecular biology, field theories exist for stochastic gene expression \cite{sasai2003stochastic}, stem cell differentiation \cite{zhang2014stem}, stochastic gene switching \cite{walczak2005self}, non-equilibrium bacterial dynamics \cite{thompson2011lattice}, Michaelis–Menten enzyme kinetics \cite{santos2015fock}, RNA dynamics \cite{vastola2021analytic}, and tumor cell migration models \cite{deroulers2009modeling}. On larger length scales, field theories have been developed to study ecological pattern formation \cite{hernandez2004clustering,butler2009robust,butler2011fluctuation}, predator-prey dynamics \cite{butler2009predator} and the spatio-temporal fluctuations in stochastic Lotka-Volterra models \cite{mobilia2007phase,tauber2012population}, swarming and flocking behavior in animals \cite{scandolo2023active,cavagna2023natural} and epidemic spreading models \cite{merbis2021exact,merbis2022logistic,de2022fock}. Another interesting domain of application in recent years is neuroscience, where field theory methods are used to study fluctuations in neural activity \cite{buice2007field,buice2010systematic,bressloff2010stochastic} and the universal properties of neuronal avalanches \cite{garcia2018field,pausch2020time}. 
This broad (yet inexhaustive) application domain provides great potential for the basis-independent representation of reaction-diffusion systems to aid in developing multi-scale simulations and analysis across many complex systems, constructing bridges between traditionally disparate communities.

An overarching application is employing the field theory framework to obtain emergent models at coarser scales, yielding mathematical bridges and relations between parameters of models at different scales. which serves the development of multi-scale numerical schemes. In this work, we illustrate this for chemical reaction networks, where one often deals with large numbers of reactants and reaction products. Applying the methodology to a CDME for bimolecular (non-linear) reaction systems (see \cref{sec:emerRD}), we obtain not only an emergent reaction-diffusion PDE, but also a precise connection between the parameters at the particle level and those of the PDE, enabling consistent multi-scale simulation. Such connections have also been obtained formally \cite{smith2019spatial,kostre2021coupling,del2024open} and rigorously \cite{isaacson2021reaction,isaacson2022mean} using standard methods; we show here how field theory methods can facilitate these calculations.

From a more general computational perspective, it is often convenient to focus on the dynamics in coarser and/or specific regions of phase space. We are then interested in a formalism that allows us to project the process on a subspace of the whole function space, by means of a Galerkin discretization \cite{klus2018data,prinz2011markov,schutte2015critical}. This organically leads to the \textit{reaction-diffusion master equation} (RDME) (see \cref{fig:maindiag}), where the discretization can be on a lattice \cite{baras1996reaction}, on metastable regions \cite{winkelmann2016spatiotemporal}, unstructured meshes \cite{engblom2009simulation,isaacson2018unstructured} or others. Simulation schemes and software for such equations have been explored in detail in several works \cite{drawert2012urdme,erban2024multi,engblom2009simulation,fange2010stochastic,hellander2020hierarchical,hellander2012reaction,isaacson2009reaction2}. One known issue with RDMEs was their lack of convergence to the underlying particle dynamics in the microscopic limit \cite{fange2010stochastic,hellander2012reaction,isaacson2008relationship,isaacson2009reaction2}, which was eventually solved in  \cite{isaacson2013convergent}. One advantage of the Galerkin discretization presented here is that the convergence of the resulting RDME is guaranteed. 

\subsection{Other relevant and complementary works}
Although we cover a large range of topics in this work, there is a large body of work which we do not delve into. Much of this has been presented in detail in the literature and specifically in wonderful reviews. We choose to limit ourselves here to Fock space and field theory methods for master equations describing reaction-diffusion systems (see \cref{fig:maindiag}). Field theory methods have been applied to stochastic differential equations since pioneering work of Feynman-Kac \cite{kac1949distributions}, Onsager-Machlup \cite{onsager1953fluctuations} and its formulation due to Martin, Siggia, Rose \cite{martin1973statistical}, de Dominicis \cite{dominicis1976techniques}, Janssen \cite{janssen1976lagrangean} and Bausch, Janssen, and Wagner \cite{bausch1976renormalized}. The connection between all these path integral formulations, together with their connection to forward and backward equations, can be found in the excellent review by Weber and Frey \cite{weber2017master}.
Other reviews of the Doi-Peliti formalism are by Mattis and Glasser \cite{mattis1998uses}, and by T\"auber, Howard and Vollmayr-Lee \cite{tauber2005applications}, which focuses on the application of renormalization techniques to reaction-diffusion system (see also \cite{tauber2014critical}). Exactly solvable models are discussed in \cite{schutz2001exactly} and a good field theory review catered to neuroscientists is \cite{chow2015path}. 
Other notable mathematical works have investigated similar topics using rigorous measure theory \cite{kolokoltsov2010nonlinear}.

While there is undoubtedly some overlap with previous works in presenting the Doi-Peliti path integral formalism, this work is meant as a self-contained pedagogical introduction, leading into the less-known basis-independent formulation.  We focus mainly on the practical use of the formalism and less on its mathematical rigor. Our main objective is to show how this formulation is related to previous work pertaining to second quantized stochastic systems and reaction-diffusion models, and to highlight its unifying characteristics and the potential for future applications. This is achieved by developing the quantum-inspired methods and notation at the level of the basis-independent representation, enabling organic connections with previous models. We will further exemplify the robustness of the framework by deriving well-posed numerical discretizations and emerging macroscopic models in a straightforward manner.

\subsection{Organization of this paper}
The paper is organized as follows: \Cref{sec:quantmethodsCRN} introduces chemical reaction networks in the well-mixed setting, the reaction rate equations and how to formulate the CME using Fock space methods, i.e. the second quantized reaction networks formalism. It further shows how to recover the rate equations from the CME, as well as how to write generating functions and some remarks related to large deviation theory.

\Cref{sec:continuum} presents the Doi-Peliti path integral formalism for reaction-diffusion systems. We start by introducing discretized spatial dynamics, followed by the derivation of the path integral by taking the continuum limit. We then comment on the perturbative expansion and the renormalization group analysis for reaction-diffusion systems. We close the section with an overview of various application and generalizations of the Doi-Peliti path integral.  

\Cref{sec:stochMecRD} introduces the basis-independent representation, which constitutes a unifying field theory representation for stochastic reaction-diffusion processes. We introduce all Fock space machinery for the generalized case, and we use it formulate the chemical diffusion master equation (CDME) for generic single species chemical reactions, as well as for multiple species non-linear reactions. We show how to apply a Galerkin discretization to obtain the second quantized Galerkin representation. More specifically, we impose indicator functions as an approximate basis, recovering the starting point from \cref{sec:continuum}, as well as the reaction-diffusion master equation (RDME) used by chemical physicists. Finally, starting from a generic Galerkin discretization, we present how to write the path integral formalism for the general case. As the discretization is not specified, this results in a basis-independent discretized path integral formulation capable of handling spatially-dependent reaction rates. 

\Cref{sec:specialapplications} discusses limiting cases and alternative representations, specifically in the context of the Doi formulation. First, the original Doi formulation from \cite{doi1976second} is recovered as a special limiting case of the basis-independent representation, when Dirac delta functions are used instead of basis functions, previously shown in \cite{delRazo2}. Historically, the spatial discretization of this formulation is the starting point of the Doi-Peliti path integral presented in \cref{sec:continuum}. Using the Doi formulation, we further write the CDME for a general one species reaction, and we show how to recover the standard representation of the CDME used by physical chemists in terms of scalar probability densities (\cref{fig:maindiag}). We also present how to formulate the master equation in terms of probability fluxes, which is helpful for a thermodynamic analysis. To finalize, we show how to obtain an equation for the dynamics of stochastic concentrations as well as the emergence of macroscopic reaction-diffusion master equations for a non-linear reaction.

\section{Quantum methods for chemical reaction networks}
\label{sec:quantmethodsCRN}

In this section, we will focus on well-mixed chemical reaction networks, where diffusion is assumed to occur at a much faster timescale than reactions. Hence the chemical components are distributed uniformly in space, and their spatial distribution does not play a role in the dynamics.
We will discuss models at two different levels. At the top level, the dynamics of the population (or: densities/concentrations of chemical compounds) is described by the \textit{rate equations}; a set of ordinary differential equations describing the change in densities of different species over time. These macroscopic equations are deterministic, continuous and often non-linear. This description emerges as the large copy number limit of a microscopic theory at the bottom level, given by the \textit{chemical master equation} (CME)\cite{anderson2015stochastic, gardiner1977poisson,gillespie1992rigorous,qian2010chemical,schnoerr2017approximation}. This equation models the inherently stochastic reaction processes by quantifying the probabilistic change in discrete particle numbers of each species, due to individual reaction events in a bounded domain. This allows one to not only track average quantities, but also study fluctuations. 

We first introduce briefly the rate equations for well-mixed chemical reaction networks, and we formulate the CME in terms of creation and annihilation operators acting on a Fock space for the number of particles. Then, we show the emergence of the rate equations, followed by a discussion on generating functions, large deviation theory and the application scope. In later sections, we will incorporate space and turn this into a field theory for reaction-diffusion.

\subsection{Chemical reaction networks}\label{sec:rateeqn}

A chemical reaction network (CRN) consists out of a set of \textit{species} $\mathcal{S}$, from which one can form linear combinations (complexes), divided into reactant and product complexes. Additionally, the CRN contains a set of reaction rates labeled by $r  \in \mathbb{R}_{\geq 0}$, which can be understood as maps from a reactant complex to a product complex. A more mathematically rigorous definition of a CRN is discussed in \cite{feinberg2019}, however, this is not needed for the current presentation. 

When the expected numbers of particles in the system is very large, random fluctuations in the concentration will be small due to the law of large numbers. The concentrations of each species can then be obtained from the \textit{rate equations} for the CRN, which follows from the \textit{law of mass action}.
The rate equations for each species $i \in \mathcal{S}$, determine the rate of change $\frac{dx_i(t)}{dt}$ for the concentrations $x_i(t)$. Each reaction in the network involving this species will contribute to the rate of change; positively if the species is in the product complex and negatively if it is in the reactant complex. The argument is the same for each reaction, so we can focus on a single reaction first. Let us denote this generic reaction as:
\begin{equation}\label{eq:generic_reaction}
    \sum_{i \in \mathcal{S}} m_i \mathrm{X}_i \to \sum_{i \in \mathcal{S}} n_i \mathrm{X}_i.
\end{equation}
such that $m_i$ and $n_i$ are the \textit{stochiometric coefficients} of the species $i$ in the reactant and product complexes, respectively. Let us suppose that this reaction has rate $r$ and involves $s$ species ($|\mathcal{S}|=s$).
The law of mass action states that the rate of change of $x_i$ due to this reaction is proportional to the concentrations of all reactants, since the reaction only occurs once all of its inputs are present: 
\begin{equation}\label{singletransrateeqn}
	\frac{dx_i}{dt} = r (n_i - m_i) x_1^{m_1} \ldots x_s^{m_s}.
\end{equation}
Here the reaction vector $(n_i - m_i) $ accounts for the effective number of elements of species $i$ created (or destroyed) in the reaction. 

A recurrent example throughout this text will be the general single species reaction:
\begin{equation}\label{singlespeciesreaction}
    k A \to  \ell A
\end{equation}
Straightforward application of \cref{singletransrateeqn} gives the rate equation for the concentration of $A$ particles $x_A$ as
\begin{equation}\label{singlespeciesrateeqn}
    \frac{d x_A}{dt} = r(\ell - k)x_A^k \,.
\end{equation}

The generalization of \cref{singletransrateeqn} to multiple reactions is straightforward. To reduce the notational complexity, we will first introduce the shorthand notation:
\begin{equation}\label{notation}
	\mb{x}^{\mathbf{m}} \equiv x_1^{m_1} \ldots x_s^{m_s}.
\end{equation}
Now each reaction (labelled by $\tau$) has its own rate $r(\tau) \in \mathbb{R}_{\geq 0}$, and consumes $m_i(\tau)$ and produces $n_i(\tau)$ copies of species $i$. To obtain the rate of change in the expected number of species $i$, one simply sums \eqref{singletransrateeqn} over all reactions $\tau$
\begin{equation}\label{general_rateeqn}
		\frac{dx_i}{dt} = \sum_{\tau} r(\tau) (n_i(\tau) - m_i(\tau)) \mb{x}^{\mathbf{m}(\tau)}\,.
\end{equation}
So, the law of mass action allows one to write down a set of ordinary differential equations (ODEs) corresponding to the reaction network. There are many results relating the dynamics of these ODEs to the geometry of the reaction network, such as the famous deficiency theorems \cite{horn1972general,feinberg1972complex}, the Anderson-Craciun-Kurtz theorem \cite{anderson2010product} for the product form of certain equilibrium solutions, or the more recent geometric decomposition into cycles and cocycles proposed in \cite{dal2023geometry}. More information on classical methods to handle the chemical rate equations can be found in the textbooks \cite{beard2008chemical,keener2009mathematical,murray2003book,winkelmann2020stochastic}, among others. Here, we will be more interested in connecting the rate equations to a microscopic theory which describes the number of particles of a given species as fluctuating randomly (i.e. stochastically). 

\subsection{A Fock space for chemical reaction networks}
\label{sec:Fockspace}

The microscopic description of a CRN involving $k$ species is given in terms of the probability distribution $p_\mathbf{n}(t) $ of having $\mathbf{n}=(n_1,\dots,n_s)$ elements of each species, where $n_i$ is the \textit{copy number} for the \mbox{$i$--th} species, at any given time $t$. The dynamics of this probability distribution is given by the chemical master equation (CME) \cite{gardiner1977poisson,gillespie1992rigorous,qian2010chemical,schnoerr2017approximation}, also known in physics as the Kolmogorov-forward equation \cite{kolmogoroff1931analytischen,feller1957boundaries} for the CRN. Although one can formulate the CME using conventional approaches, in this work we will focus on formulating the equation using quantum-inspired methods, which we refer to as the \textit{second quantized formulation} of reaction networks, following \cite{doi1976second}.

Before deriving the equation for $p_\mathbf{n}(t)$, we first establish some concepts and notation. We begin by constructing a vector space---the Fock space---for the probability distribution in the copy number representation, using creation and annihilation operators \cite{doi1976second,grassberger1980fock,peliti1985path}. This essentially exploits the fact that the elements are assumed to be indistinguishable and their copy number can only change by an integer. For some applications, this indistinguishability is an assumption and not a characteristic of the system. In population dynamical or social applications, the individual species or agents are in reality distinguishable, however, we limit our discussion here to systems where the effective model treats individuals as exchangeable. This assumption implies that characteristics which allows individual particles to be identified do not play a significant role in the dynamics of the system. 

We start with defining the empty state $\ket{0}$, the creation operators $ a_i^\dagger$ and annihilation operators $ a_i$ for the species $i$, such that:
\begin{equation}\label{vacuum}
	a_i \ket{0} = 0 \,, \qquad \ket{\mathbf{n}} = \prod_{i \in \mathcal{S}} ( a_i^\dagger)^{n_i} \ket{0} \equiv ({\mathbf{a}}^\dagger)^{\mathbf{n}} \ket{0} \,.
\end{equation}
Here ${\mathbf{a}}^\dagger$ is a vector of creation operators, one for each of the $s$ species in $S$, and $\mathbf{n} \in \mathbb{Z}_{\geq 0}^s$ is a vector containing all of their copy numbers. Raising a vector to a vector power is defined as in \eqref{notation}. 

The state $\ket{\mathbf{n}}$ represents the system containing $\mathbf{n} = (n_1, n_2, \ldots, n_s)$ copies of the species under consideration. As each one of them is treated as indistinguishable, when we remove (annihilate) a particle of species $i$, there are $n_i$ equivalent ways to do so and hence
\begin{equation}\label{removing_ni}
	a_i \ket{\mathbf{n}} = n_i \ket{n_1, \ldots, n_i -1, \ldots, n_s} \,.
\end{equation}
On the other hand, there is only a single way to add a particle of species $i$, and hence
\begin{equation}
	\label{adding_ni}
	 a_i^\dagger \ket{\mathbf{n}} = \ket{n_1, \ldots, n_i+1, \ldots, n_s} \,.
\end{equation}
The above relations imply that the creation and annihilation operators satisfy bosonic commutation relations:
\begin{equation}
	\label{commutators}
	[ a_i,  a_j^\dagger] = \delta_{ij} \,, \qquad [ a_i, a_j] = [  a_i^\dagger,  a_j^\dagger] = 0\,,
\end{equation}
where $[a,b]=ab -ba$ is the commutator.
Note that there is nothing quantum mechanical about non-commuting operators; the non-commutativity just means that there is one more way to first add a particle and then remove one, than to first remove a particle and then add one. 

An important composite operator is the number operator $N_i = \aidag\ai$, as it counts the number of particles of species $i$ in the state $\ket{\mathbf{n}}$
\begin{equation}
	N_i \ket{\mathbf{n}} = n_i \ket{\mathbf{n}} \,.
    \label{eq:partnumOp}
\end{equation}
Hence, the state $\ket{\mathbf{n}}$ is an eigenvector of the number operators $N_i$ with eigenvalues $n_i$.
We can now define the probability vector $\ket{\rho(t)}$ as:
\begin{equation}\label{probvec}
	\ket{\rho(t)} = \sum_{\mathbf{n}} p_\mathbf{n}(t) \ket{\mathbf{n}} \,,
\end{equation}
where $p_\mathbf{n}(t)$ is the probability of having $\mathbf{n}$ copies of each species at time $t$. Technically, $\ket{\rho(t)}$ is a vector in the Fock space, defined as the direct sum of symmetric tensor powers of the single particle vector spaces; the states $\ket{\mathbf{n}}$ form a basis for this space.

To compute expectation values (or any other scalar quantity of interest) we also need dual vectors, such that we can take inner products. The full story is a bit more complicated\footnote{This is because, unlike quantum mechanics, the vector \eqref{probvec} is in general an element of the $L^1$ space of normalizable functions, whose dual is isomorphic to $L^{\infty}$. This is also discussed in \cref{sec:stochMecRD}.}, but for our purpose we can simply define the dual vacuum to be annihilated by the creation operators, and construct a Fock space with annihilation operators:
\begin{equation}\label{vacuum2}
	 \bra{0} \aidag = 0 \,, \qquad  \bra{\mathbf{n}} = \bra{0} \prod_{i \in \mathcal{S}} \frac{1}{n_i!} \ai^{n_i}  \equiv \frac{1}{\mathbf{n}!} \bra{0} {\mathbf{a}}^{\mathbf{n}}  \,.
\end{equation}
Additionally, we require $\bracket{0}{0} = 1$. Due to the normalizing factor of $1/\mathbf{n}! \equiv \prod_i 1/n_i!$ in the definition of the dual state, the states $\bra{\mathbf{m}} $ and $\ket{\mathbf{n}}$ are orthonormal
\begin{equation}\label{innerprod}
	\bracket{\mathbf{m}}{\mathbf{n}} =\delta_{\mathbf{nm}}\, .
\end{equation}
Thus, $p_\mathbf{n}(t) =\bracket{\mathbf{n}}{\rho(t)}$, and the resolution of the identity is
\begin{equation}\label{resolutionofID}
	\mathds{1} =  \sum_{\mathbf{n}} \ket{\mathbf{n}}\bra{\mathbf{n}}.
\end{equation}
A special dual state, whose contraction with any vector signifies summing over all components of $\ket{\rho(t)}$, is the \textit{flat state} denoted as $\bra{\mathbf{1}}$. As the probability vector is normalized, we have that:
\begin{equation} \label{eq:normwellmix}
	\sum_{\mathbf{n}} p_\mathbf{n}(t)  = \bracket{\mathbf{1}}{\rho(t)} = 1\,.
\end{equation}
The flat state signifies summing over all configurations $\ket{\mathbf{n}}$ and hence is given as
\begin{equation}\label{flat_def}
	\bra{\mathbf{1}} = \sum_{\mathbf{n}} \bra{\mathbf{n}} =  \bra{0} e^{\sum_{i\in S} \ai}\,.
\end{equation}
Here the last equality follows from the definition of the dual state (including the normalization $1/\mathbf{n}!$). It is not hard to verify that the flat state is the left-eigenvector of the creation operators, with eigenvalue $1$
\begin{equation}\label{eigeneqFS}
	\bra{\mathbf{1}} \aidag = \bra{\mathbf{1}} \,, \qquad \forall i \in \mathcal{S} \, .
\end{equation} 
We are now able to compute expectation values of observables. 
Any observable $A(\mathbf{n})$ can be cast in terms of creation and annihilation operators $A(\ai, \aidag)$. As observables should not modify the system, they do not change the number of particles and are thus expressed with an equal amount of creation and annihilation operators for each species, i.e. they are \textit{diagonal} operators.  
The expectation value of $A$ at time $t$ is then computed as
\begin{equation}
	\vev{A}_t = \sum_{\mathbf{n}} A(\mathbf{n}) p_\mathbf{n}(t)  = \bra{\mathbf{1}} {A}(\ai,\aidag) \ket{\rho(t)} \,.
\end{equation}
For example, if we wish to compute the expected value of the number of elements of species $i$, then the observable of interest is the number operator $N_i = \aidag \ai$ and the expectation value becomes
\begin{equation}
	\vev{N_i} = \bra{\mathbf{1}} N_i \ket{\rho(t)} = \bra{\mathbf{1}} \ai \ket{\rho(t)}\,.
\end{equation}
The last equation here follows from the definition of $N_i$ along with equation \eqref{eigeneqFS}.

The final definitions we need before proceeding to the dynamical equations for $\ket{\rho(t)}$ are the \textit{Poisson states}. These are defined analogously to coherent states in quantum mechanics, although its interpretation is subtly different. A Poisson state $\ket{\mb{z}}$ is defined in terms of a vector $\mb{z} \in \mathbb{R}^s$ as
\begin{equation}\label{coherentdef}
	\ket{\mb{z}} = e^{\sum_{i \in \mathcal{S}} z_i \aidag} \ket{0}\,.
\end{equation}
Here, the exponential of an operator is defined in terms of its infinite power series: $e^{\aidag} = \sum_n \frac{1}{n!} (\aidag)^n$. The Poisson state defines (when properly normalized) a Poisson distribution over the Fock space, parameterized by the `species vector' $\mathbf{z}$, representing the mean of the distribution. In quantum mechanics, coherent states are defined similarly, however there, the vectors $\mathbf{z}$ may be complex and they lead to Poisson distributions with means $|z_i|^2$.

Similar to the construction above, we can also define dual Poisson states as
\begin{equation}\label{dualcoherentdef}
	\bra{\mathbf{z}} = \bra{0} e^{\sum_{i \in \mathcal{S}} z_i \ai} \,.
\end{equation}
From this definition, we immediately observe that the flat state \eqref{flat_def} is a special case of a dual Poisson state with all $z_i = 1$. A useful property of the (dual) Poisson states is that they are eigenvectors of the creation and annihilation operators, respectively, with eigenvalue $z_i$:
\begin{equation}\label{coherentstateEV}
    \bra{\mathbf{z}}a_i^\dagger = z_i \bra{\mathbf{z}} \,, \qquad a_i \ket{\mathbf{z}} = z_i \ket{\mathbf{z}}
\end{equation}

As observed in \cite{grassberger1980fock}, it can be convenient to define an `inclusive scalar product'. \cref{sec:inclprod} explores this product and demonstrates how it streamlines factorial-moment calculations. Nonetheless, we adopt the inner product of \cref{innerprod} for the bulk of our exposition, as it provides a straightforward and practical framework for the derivations that follow.

\subsection{The chemical master equation} 
\label{sec:mastereqn}

Now we are ready to formulate the CME, which describes how the probability of having a certain number of elements of any given species changes in time. Generically, it will have the form
\begin{equation}\label{mastereqn}
	\frac{d}{dt} \ket{\rho(t)} =  H(\aidag, \ai) \ket{\rho(t)}
\end{equation}
This equation can be written in standard form \cite{qian2010chemical,gillespie1992rigorous,schnoerr2017approximation}, so the second quantized bra-ket notation is simply one possible representation. This representation shows great similarity with the (imaginary time) Schr\"odinger equation, except that here, the `Hamiltonian' ${H}$ is hardly ever Hermitian. In fact, calling $H$ a Hamiltonian may be misleading, since it is in general not Hermitian (as Hamiltonians in quantum mechanics are), nor is it a function specifying the systems energy in equilibrium (as Hamiltonians in equilibrium statistical mechanics are). In the literature $H$ is referred to by many different names, which may depend on the discipline the papers are directed to. We have seen: quasi-Hamiltonian, the infinitesimal generator, the Markov generator, the Liouvillean, the transition rate matrix, or the time-evolution operator, among others. Here, we will mostly stick to calling $H$ \textit{the generator}.

The generator should satisfy a number of properties, such that Eq.~\eqref{mastereqn} can describe the time evolution of a probability distribution. First of all, the total amount of probability should be conserved under time evolution. This means that after evolving the system for some time $t$, starting from an arbitrary normalized state $\ket{\rho(0)}$, the state should remain normalized. Since the formal solution of \eqref{mastereqn} is $\ket{\rho(t)}  = \exp(Ht) \ket{\rho(0)}$, this implies that $\bra{\mathbf{1}} e^{Ht} \ket{\rho(0)} = 1 $. But $\ket{\rho(0)}$ is arbitrary, so that
\begin{equation}\label{infstoch}
	\bra{\mathbf{1}} e^{Ht} = \bra{\mathbf{1}}\,, \quad \forall t \quad \Rightarrow \quad \bra{\mathbf{1}} H = 0 \,.
\end{equation}
To put in words, the sum over all states of $H$ should vanish, or: the flat state is a left eigenvector of any generator, with eigenvalue $0$. 

Secondly, the generator $H$ should not generate distributions which contain negative probabilities. This implies that any operation implemented by $H$ which \textit{changes} the number of particles of a certain species should occur with positive sign, otherwise it might be possible to generate negative probabilities. Together with \eqref{infstoch}, this implies that diagonal terms in $H$ (which do not change the number of particles) should occur with a negative sign, to balance the positive contributions from the reactions which do change the particle numbers.

Lets make this concrete by considering the single species example \cref{singlespeciesreaction} with reaction rate $\lambda$ and then give the general recipe to turn any CRN into a generator $H$. 
The generator contains a term reflecting the \textit{gain} in probability of creating $\ell$ particles due to the reaction. A necessary condition for this to occur is that the reaction products are present, so the positive contribution reflecting the change in particle numbers, or \textit{the gain}, is $\lambda (a^\dagger)^\ell (a)^k$, or in words: first annihilate $k$ particles and then create $\ell$ particles.

\begin{figure}
	\centering
	\includegraphics[width=0.8\columnwidth]{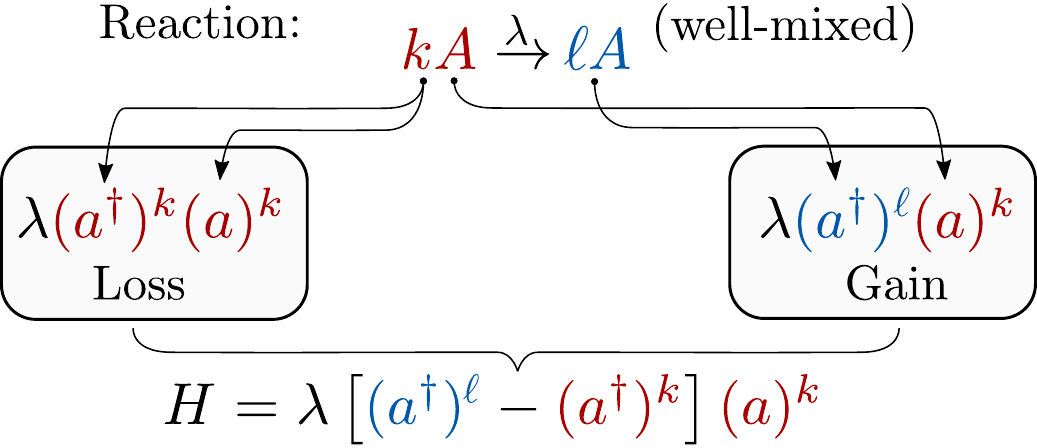}
\caption{Diagram illustrating how to construct the generator for a system with a general one species reaction with rate $\lambda$ in the well-mixed case.
\label{fig:genwellmixdiag} }	
\end{figure}

By itself, this term does not satisfy \eqref{infstoch}, as $\bra{\mathbf{1}} \lambda (a^\dagger)^\ell (a)^k  \neq 0$. To satisfy the conservation of probability, we should account for the \textit{loss} of probability of keeping the reactants in the system. This term is negative and diagonal: $-\lambda (a^\dagger)^k (a)^k$. The generator of the simple reaction \cref{singlespeciesreaction} is then:
\begin{equation}
    H = \lambda \left[ (a^\dagger)^\ell - (a^\dagger)^k \right] (a)^k\,,
\end{equation}
as displayed in \cref{fig:genwellmixdiag}.
In a general CRN, for each generic reaction~\eqref{eq:generic_reaction} we need to add the following terms, corresponding to the gain and loss of probabilities implied by the reaction:
\begin{itemize}
	\item[\textbf{Gain}] The term which implements the transition with a positive sign: $\lambda ({\mathbf{a}}^\dagger)^{\mathbf{n}} {\mathbf{a}}^{\mathbf{m}}$
	\item[\textbf{Loss}] The negative diagonal term implementing the loss of probability of keeping the reactants: $ - \lambda ({\mathbf{a}}^\dagger)^{\mathbf{m}} {\mathbf{a}}^{\mathbf{m}}$. 
\end{itemize}
Hence, to turn any CRN into a generator $H$, we can use the general formula
\begin{equation}\label{generalH}
	 H = \sum_\tau \lambda(\tau) \left[ ({\mathbf{a}}^\dagger)^{\mathbf{n}(\tau)} - ({\mathbf{a}}^\dagger)^{\mathbf{m}(\tau)} \right] {\mathbf{a}}^{\mathbf{m}(\tau)}\,,
\end{equation} 
where $\tau$ labels the distinct reactions and we are using the shorthand notation \cref{notation}. While the rates $\lambda(\tau)$ of the master equation and $r(\tau)$ of the rate equations are related to each other, they are not equal. The rate $\lambda(\tau)$ specifies the frequency of a single reaction, while the rate $r(\tau)$ specifies the frequency with which the concentrations of elements change. 

Any generator which satisfies the property of conserving non-negative probabilities under time-evolution is called \textit{infinitesimal stochastic}. These are the stochastic analogues for Hermitian operators in quantum mechanics. It is possible for an operator to be both Hermitian \textit{and} infinitesimal stochastic. In that case, the operator will define both a stochastic system (through the master equation) \textit{and} a quantum system (through the Schr\"odinger equation) and it is called a \textit{Dirichlet operator}. 

The formal solution of the master equation defines a \textit{one-parameter family semigroup} of linear operators $ U = e^{Ht}$. This operator is a \textit{stochastic} operator, which map probability vectors into probability vectors. Mathematically, such operators satisfy $\bra{\mathbf{1}} {U}(t) = \bra{\mathbf{1}}$. We have already encountered stochastic operators in \eqref{eigeneqFS}; any creation operator $\aidag$ is stochastic, since adding one particle does not change the total amount of probability, it merely maps the $n$ particle states to the $n+1$ particle states. This implies that whenever we are computing expectation values of some observables, due to the contraction with $\bra{\mathbf{1}}$ from the left, we can set all $\aidag \to 1$ after normal ordering\footnote{Normal ordering places creation operators to the left of annihilation operators by using the commutation relations \eqref{commutators}.}.

\subsection{Emergence of the rate equations}
\label{sec:emergencerateeqs}

In this section, we will connect the CME to the rate equations and see under what assumptions can we go from the microscopic stochastic theory to the macroscopic deterministic theory. This has been done rigorously in mathematical contexts using large copy number limits \cite{anderson2015stochastic,kurtz1970solutions,kurtz1971limit}. Here we want to show how this can be done with techniques using the second quantized representation.

The macroscopic variables are straightforward to define from the stochastic theory. The concentration of elements of a certain species is simply the expected number of elements of this species, divided by the system volume $V$, which we assume to be finite. So, in accordance with the notation above, we define:
\begin{equation}
	x_{i}(t) = \frac{1}{V}\vev{N_i}_t = \frac{1}{V} \bra{\mathbf{1}} N_i \ket{\rho(t)} \,.
\end{equation}
The next step is to work out the dynamical equations for $x_i(t)$, using the master equation \eqref{mastereqn}:
\begin{equation}\label{eq:commutator_with_H}
	\frac{d}{dt} x_{i} (t) 
    = \frac{1}{V} \bra{\mathbf{1}} N_i H \ket{\rho(t)}  = \frac{1}{V} \bra{\mathbf{1}} [N_i, H] \ket{\rho(t)}  \,.
\end{equation}
In the last equality, we have used the property of the generator $\bra{\mathbf{1}} H = 0$ to express the answer in terms of the commutator. Hence, the time evolution of the expectation value for the number operator (in fact, of any operator) is given by the expectation value of the commutator with the generator $H$.

For generic CRNs generated by \eqref{generalH}, the above relation implies that
\begin{equation}\label{ratefun_intermediate}
	\frac{d}{dt}x_i(t) = \sum_{\tau} \frac{\lambda(\tau)}{V} (n_i(\tau)- m_i(\tau)) \bra{\mathbf{1}} \mathbf{a}^\mathbf{m(\tau)} \ket{\rho(t)}\,.
\end{equation}
This is not yet the closed, multi-linear form of the rate equations \eqref{general_rateeqn}. To bring the right hand side to the desired form, some additional assumptions are needed. Let us focus on a single species for illustrative purposes. Equations \eqref{general_rateeqn} are derivable from the master equation when
\begin{equation}\label{mf_approx}
	 \bra{\mathbf{1}} a^m \ket{\rho(t)} - \bra{\mathbf{1}} a \ket{\rho(t)}^m = 0 \,.
\end{equation}
If this equation holds, we may replace $\bra{\mathbf{1}} a \ket{\rho(t)}$ by $V x(t)$ and we get
\begin{equation}\label{general_rateeqn2}
	\frac{dx_i(t)}{dt} = \sum_\tau \lambda(\tau)V^{\sum_i m_i(\tau) -1}  (n_i(\tau) - m_i(\tau))\mathbf{x}^{\mathbf{m}(\tau)}
\end{equation}
This is exactly equal to the rate equation \eqref{general_rateeqn}, if we identify $r(\tau) = V^{\sum_i m_i(\tau) -1} \lambda(\tau)$.

Equation \eqref{mf_approx} constitutes a mean-field approximation; it supposes that the $m$-th moment of the number operator is given by the $m$-th power of the expectation value. This is a good first approximation, but it is not exactly true when variances are large.
However, when the total number of elements is big, the law of large numbers and the \textit{central limit theorem} offer a helping hand. The law of large numbers roughly states that for $n$ \textit{independent and identically distributed} (iid) random variables $\{X_n\}$, the sample mean $S_n = \frac{1}{n} \sum_n X_n$ will converge to the mean $\mu$ of the distribution from which the $X_n$ are sampled, namely $\lim_{n \to \infty} S_n = \mu $. The central limit theorem states that in the same large $n$ limit, the distribution of the sample mean $S_n$ becomes Gaussian with standard deviation $\sigma/\sqrt{n}$, where $\sigma$ is the standard deviation of the distribution for the random variables. 
Or
\begin{equation}
	S_n \to \mu + \xi/\sqrt{n} \,,
\end{equation}
where $\xi$ is normally distributed $\mathcal{N}(0,\sigma^2)$. 
This in particular implies that the expectation value of any power of the sample mean $S_n^k$ will tend to the power of the mean $\mu^k$, with subleading corrections proportional to $\frac{1}{n}$. These corrections vanish in the large $n$ limit.

In our case, we can apply this when we assume that the number of elements of each species is very large, or likewise, if we are considering the dynamics of concentrations of species in a large volume expansion of the CRN. In chemistry this is called the van Kampen expansion, after \cite{van1976expansion}. In these situations, and assuming the elements are iid, we can approximate $\vev{a^k}$ as $\vev{a}^k$ and in this limit we recover the rate equations \eqref{general_rateeqn} from the master equation of the system. In other words, the leading contributions in a large $V$ expansion of \eqref{ratefun_intermediate} gives the rate equation \eqref{general_rateeqn2}, and the corrections to this equation vanish as $1/V$ in the limit $V \to \infty$.

\subsection{Generating functions}
\label{sec:wellm-genfuncs}

The probability vector $\ket{\rho(t)}$ gives a formal way of representing an infinite sequence of numbers, representing the probability of the system to have a certain copy number $\mathbf{n}$. In mathematics, and specifically in probability theory, another useful representation of infinite sequences is through a \textit{generating function}. The generating function encodes an infinite sequence of numbers ($a_n$) by treating them as coefficients of a formal power series of a variable $z$. If the infinite sequence $a_n$ has a finite sum, then the formal power series will converge and if we are lucky we may even obtain a closed form expression for the generating function in terms of analytic functions of the variable $z$. In general, we will see that it is possible to derive a partial differential equation for the probability generating function of a generic CRN. Afterwards, we will define the \textit{moment generating function} of an observable, where the series coefficients $a_n$ are not the probabilities $p_\mathbf{n}(t)$, but the different statistical moments of an observable: $\vev{O^n}$.

Formally, the generating function for an infinite series $a_n$ is defined as
\begin{equation}
	G(a_n; z) = \sum_n a_n z^n \,.
\end{equation}
This naturally generalizes to multiple variables. If we suppose there are $s$ species, then
\begin{equation}
	G(a_{\mathbf{n}}; \mathbf{z}) = \sum_{\mathbf{n}} a_{\mathbf{n}} \mathbf{z}^\mathbf{n} \,. 
\end{equation}
Here we make use of the notation introduced in \eqref{notation}. The generating function is now a formal power series in $s$ variables, where $s$ is the number of different species $s = |\mathcal{S}|$. To obtain the \textit{probability generating function} from the probability vector $\ket{\rho(t)}$, we can make use of the dual {Poisson state $\bra{\mathbf{z}}$ defined in \cref{dualcoherentdef}, together with its eigenvalue equation \cref{coherentstateEV}
\begin{equation}\label{Gzdefinition}
	G(\mathbf{z}, t) = \bracket{\mathbf{z}}{\rho(t)} = \sum_{\mathbf{n}} p_\mathbf{n}(t)  \mathbf{z}^\mathbf{n} \ .
\end{equation}
The probability generating function has a number of useful and interesting properties which we will highlight in the case of a single species. The generalization to multiple species is immediate.  
First, the normalization of $\ket{\rho(t)}$ is now translated to the property that the probability generating function evaluated at $z=1$ equals $1$
\begin{equation}
	G(z=1,t) = \bracket{1}{\rho(t)} = 1
\end{equation}
Then, if we wish to recover the probability mass function $p_n(t)$ giving the probability of having $n$ elements, we can do so by taking $n$ derivatives with respect to $z$, followed by evaluating the result at $z=0$
\begin{equation}
	p_n(t) = \left. \frac{1}{n!} \frac{d^n G(z, t)}{dz^n}  \right|_{z=0}
\end{equation}
The expectation value for the number of elements at time $t$ is also extracted by taking derivatives, but now evaluated at $z=1$
\begin{equation}
	\vev{N}_t = \left. \frac{d}{dz} G(z,t) \right|_{z=1} \,.
\end{equation}
If we wish to compute higher moments of the number operator, we can do so by the following relation
\begin{align}
    \vev{N^n}_t = \left. \left( z \frac{d}{dz} \right)^n G(z,t) \right|_{z=1} \, , 
    \label{eq:numOpMoments}
\end{align}
It should be understood here that one first performs the operation $\left( z \frac{d}{dz} \right)^k $ on the generating function and only afterwards set $z=1$.
This expression invites the identification of the number operator $N = a^\dagger a$ with $z \frac{d}{dz}$, such that creation operators correspond to multiplications with $z$ and annihilation operators are differentiation with respect to $z$. Indeed, speaking in terms of probability generating functions is completely equivalent to making the following identifications in the formalism developed in the last sections
\begin{equation}\label{polynomialrep}
\aidag \to z_i \,, \qquad \ai \to \frac{\partial}{\partial z_i} \,.
\end{equation}
As for example presented in \cite{Baez_2017}, this representation of the creation and annihilation operators satisfies the canonical commutation relations \eqref{commutators} and allows us to write the general master equation in terms of a partial differential equation for the probability generating function
\begin{equation}
	\partial_t G(\mathbf{z}, t) = \sum_\tau \lambda(\tau) (\mathbf{z}^{\mathbf{n}(\tau)} - \mathbf{z}^{\mathbf{m}(\tau)}) \partial_{\mathbf{z}}^{\mathbf{m}(\tau)} G(\mathbf{z}, t) \,,
\end{equation}
where $\partial_{\mathbf{z}}^{\mathbf{m}} = \prod_{i \in \mathcal{S}} \left(\frac{\partial}{\partial z_i} \right)^{m_i}$. Here are two simple examples which illustrate the computation of the probability generating function

\paragraph{The Poisson process $\emptyset \to A$.} The equation for the probability generating function is
\begin{equation}
	\partial_t G(z, t) = \lambda (z-1) G(z,t) \,.
\end{equation}
Together with the initial condition of having zero elements at $t=0$, which translates to $G(z,0) = 1$, this is easily solved as $G(z,t) = e^{\lambda t (z-1)}$, which is indeed the well-known result for the generating function of the Poisson process. 

\paragraph{The degradation reaction $A \to \emptyset$.} The differential equation for $G(z,t)$ is now
\begin{equation}\label{deathprocess}
	\partial_t G(z,t) = \lambda (1-z)\partial_z G(z,t) \,.
\end{equation}
One can solve this via the method of characteristics by finding a characteristic curve $c$ in terms of $z$ and $t$, such that $G(z,t) = f(c(z,t))$. We can do this by supposing that $z$ has some implicit time dependence, such that $G$ is a function of $z(t)$ and $t$ and the equation \eqref{deathprocess} splits into two equations
\begin{equation}
	\frac{d}{dt} G(z(t),t) = 0\,, \qquad \frac{d}{dt} z(t) = \lambda (z-1)\,. 
\end{equation}
The latter equation is solved by $z(t) = 1 + c e^{\lambda t}$, which introduces a constant $c$. Solving for $c$ gives
\begin{equation}
	c = (z-1)e^{-\lambda t} \,, \quad \text{such that: } \,\, G(z,t) = f(c(z,t))\,.
\end{equation}
The final step is to consider the initial conditions. If at $t=0$, there are $k$ particles present, then we must have $G(z, 0) = z^k$ and this then fixes the functional form of $f(c)$ to be $(1 + c(z,0))^k$ such that
\begin{equation}
	G(z,t) = (1 + e^{-\lambda t} (z-1))^k\,.
\end{equation}

Although it is generally a rare treat to find exact solutions for more involved CRNs, there are some noticeable examples in the literature (see \cite{jahnke2007solving} for a classical approach to solving monomolecular CRNs). In \cite{grima2012steady}, exact solutions of the generating function PDE for a stochastic model of a gene regulatory feedback loop are presented. See \cite{vastola2021solving,vastola2021analytic} for exact solutions of monomolecular and stochastic gene switching reaction networks in a second quantized formulation.

\subsubsection{Factorial moments and the inclusive product}
\label{sec:inclprod}
From the probability generating function, it is possible to compute the factorial moments $n_k(t)=\vev{N(N-1)...(N-k+1)}_t$ analogous to \cref{eq:numOpMoments}. Focusing again on a single species, the factorial moments are obtained as: 
\begin{align}
    n_k(t) = \left. \left(\frac{d}{dz} \right)^k G(z,t) \right|_{z=1} \,.
    \label{eq:numOpFacMoments}
\end{align}
This can be written as the expectation of the operator $N(N-1)...(N-k+1)=(a^\dagger)^k(a)^k$, that is $n_k(t)=\bra{1}a^k\ket{\rho(t)}$. Expressions of this kind already appeared in \cref{sec:emergencerateeqs}, where we invoked the mean-field approximation. Hence the factorial moments will be essential to capture higher-order moment closures.

In this light, it can sometimes be convenient to work with an `inclusive scalar product', which can be defined as 
\begin{align}
\langle m| n \rangle_{\rm in} =\langle m|e^{a^\dagger}e^{a}|n \rangle .
\label{eq:inclusiveProdDef}
\end{align}
The densities and factorial moments, originally given by $p_n(t) =\frac{1}{n!}\bra{0}a^n\ket{\rho(t)}$ and $n_k(t)=\bra{0}e^{a}a^k\ket{\rho(t)}$, can be expressed in terms of this `inclusive product' as
\begin{align}
p_n(t) &=\frac{1}{n!}\bra{0}e^{-a}a^n\ket{\rho(t)}_{\rm in} 
\label{eq:probIncl}\\
n_k(t)&=\bra{0}a^k\ket{\rho(t)}_{\rm in} \ .
\label{eq:momentsIncl}
\end{align}
In the space-dependent cases, the inclusive inner product will facilitate the computation of `inclusive densities', giving the probability of finding \textit{at least} $n$  particles at given locations (see \cref{sec:inclusivedensity}). This makes the inclusive inner product a convenient language in the thermodynamic limit. Nevertheless, using the inner product \cref{innerprod} and the exact microscopic master equation remains fully sufficient to obtain thermodynamic-limit results and systematic finite-$n$ corrections (e.g. via the van Kampen expansion). As shown in \cite{grassberger1980fock} and Appendix A of \cite{peliti1985path}, the inclusive inner product can be obtained by the shift $a_i^\dagger\mapsto a_i^\dagger+1$, which illustrates the mathematical equivalence and complementarity of the two formulations. The choice between them is therefore one of calculational convenience and in this paper we will mainly use the inner product \cref{innerprod}.

\subsubsection{Moment generating functions}

The \textit{moment generating function} for the observable $O$ is defined as the generating function for the series $a_n = \vev{O^n}/n!$. This is equivalent to the expectation value of $e^{z \, O}$, where $z$ is the variable of the power series. In the second quantized framework
\begin{equation}
	M_{O}(z,t) = \vev{e^{z \, O}}_t = \bra{\mathbf{1}} e^{z\, O} \ket{\rho(t)} \,.
\end{equation} 
For instance, if the observable is the number operator for species $i$, then we can express its moment generating function as
\begin{equation}
	M_{N_i} (z_i,t) = \bra{\mathbf{1}} e^{z_i N_i} \ket{\rho(t)}
\end{equation}
We can also construct the moment generating function for number operators of all species, which will then depend on a vector of variables $\mathbf{z}$
\begin{equation}\label{MN1}
	M_{\mathbf{N}} (\mathbf{z},t) = \bra{\mathbf{1}} e^{\mathbf{z} \cdot \mathbf{N}} \ket{\rho(t)}\,,
\end{equation}
where $\mathbf{z}\cdot \mathbf{N} = \sum_{i\in \mathcal{S}} z_i N_i$. If we now use the fact that $\ket{\rho(t)} = e^{Ht} \ket{\rho(0)} $ and \eqref{generalH} for the generic generator of a CRN, then, by commuting the $N_i$'s though $H$ using \eqref{commutators}, we may write
\begin{align}\label{tilted}
		M_{\mathbf{N}} (\mathbf{z},t) & = \bra{\mathbf{1}} e^{\mathbf{z}\cdot \mathbf{N} }  e^{H t} \ket{\rho(0)} \\
  & = \bra{\mathbf{1}} e^{\tilde{H}(\mathbf{z}) t} e^{\mathbf{s}\cdot \mathbf{N} } \ket{\rho(0)}\,,
\end{align}
with
\begin{equation}
		\tilde H(\mathbf{z}) = \sum_\tau \lambda(\tau) \left[ e^{\mathbf{z} \cdot (\mathbf{n}(\tau)- \mathbf{m}(\tau))}  (\mathbf{a}^\dagger)^{\mathbf{n}(\tau)} - (\mathbf{a}^\dagger)^{\mathbf{m}(\tau)} \right] \mathbf{a}^{\mathbf{m}(\tau)}\,.
\end{equation}
The moment generating function is related to the exponential of a so-called \textit{tilted} generator $\tilde{H}(\mathbf{z})$ (sometimes also called \textit{deformed} or \textit{biased} generator), which is easily obtained from the Markov generator $H$ by replacing the creation operators and annihilation operators as:
\begin{equation}
	\aidag \to e^{z_i} \aidag\,, \qquad \ai \to e^{-z_i} \ai \,.
\end{equation}
This procedure allows one to immediately write down the moment generating function of the number operators as the exponential of the tilted generator. For example, for the Poisson process $\emptyset \to A$, the tilted generator is $\tilde{H}(z) = \alpha(e^z a^\dagger - 1)$. When starting with zero elements, we have $e^{z N} \ket{\rho(0)} = \ket{0}$, such that we immediately obtain the moment generating function
\begin{equation}
    M_{N}(z, t) = \bra{\mathbf{1}} e^{\alpha t ( e^z a^\dagger -1)} \ket{0} = e^{\alpha t (e^z - 1)} \,.
\end{equation}
This is the well-known expression for the moment generating function of the Poisson point process with mean $\alpha t$.

\subsection{The large deviation principle}

We saw in the last section how the moment generating function for the number operator is given by the exponential of a tilted generator. A similar tilted generator can also be defined for another observable, called the \textit{dynamical activity} $K$. This observable is not easily expressible in terms of creation and annihilation operators, because it is a trajectory dependent observable. The dynamical activity counts the total number of reactions which have taken place up to a given time $t$. Since the reactions occur by a random process, the dynamical activity $K$ is a random variable.

To obtain the moment generating function $M_K(s,t)$ for the dynamical activity $K$, following \cite{garrahan2007dynamical,garrahan2009first}, we have to adapt our formalism slightly. We not only track the probabilities $\ket{\rho(t)}$, but we define the vector $\ket{\rho(t,k)}$, which gives the probability of having a certain combination of copy numbers at time $t$ \textit{and} having seen $k$ reactions since $t=0$. Then, the moment generating function for the dynamical activity becomes
\begin{equation}
	M_K(z,t) = \vev{e^{z K}}_t = \sum_{k=0}^\infty \bra{\mathbf{1}} e^{z k} \ket{\rho(t,k)} = \bracket{\mathbf{1}}{\tilde \rho(t,z)}
\end{equation}
Here we have defined the tilted probability vector $\ket{\tilde \rho(t,z)} = \sum_k e^{z k} \ket{\rho(t,k)}$ as the (discrete) Laplace transform of $\ket{\rho(t,k)}$. It is possible to show that for a general CRN, the tilted probability vector obeys the following (tilted) master equation
\begin{equation}\label{tiltedmaster}
	\partial_t \ket{\tilde \rho(t,z)} = \tilde{H}(z) \ket{\tilde \rho(t,z)} \,,
\end{equation}
with
\begin{equation}\label{tiltedgenerator}
	\tilde H(z) = \sum_\tau \lambda(\tau) \left[ e^{z}  (\mathbf{a}^\dagger)^{\mathbf{n}(\tau)} - (\mathbf{a}^\dagger)^{\mathbf{m}(\tau)} \right] \mathbf{a}^{\mathbf{m}(\tau)}\,.
\end{equation}
So the tilted generator for the dynamical activity is obtained by multiplying the gain terms (which generate reactions) with $e^z$. The formal solution to \eqref{tiltedmaster} is now $\ket{\tilde \rho(t,z)} = e^{\tilde{H}(z) t} \ket{\rho(0)} $, assuming that there has been no activity initially, so $K=0$ at $t=0$. This implies that the moment generating function is the exponential of the tilted generator:
\begin{equation}
	M_K(z,t) = \bra{\mathbf{1}} e^{\tilde{H}(z) t} \ket{\rho(0)} \,.
\end{equation}
We can think of this quantity, (or any moment generating function for that matter) as the analogue to a partition function in equilibrium statistical mechanics. Only here the observable is not energy with dual parameter inverse temperature ($\beta$), but it is the tilted generator, for which time serves as dual parameter.

A modern approach to statistical mechanics, which is rooted in probability theory, is \textit{large deviation theory}. We refer to the review \cite{touchette2009large} for a more extensive review on this subject. A review on large deviation theory in the context of the non-equilibrium dynamics of stochastic reaction-diffusion systems is provided in \cite{smith2011large}. In the context of this section, we are interested in the statistics of the dynamical activity at late times, as $t \to \infty$. The \textit{large deviation principle} states that the probability of seeing events with activity $k$ decays exponentially as
\begin{equation}
	P(k,t) \propto	 \exp(-t I(k/t) ) \,.
\end{equation}
Here $I(k)$ is the \textit{Cram\'er function} (also known as the \textit{rate function}). The Cram\'er function has the property that its zeros define the expected values of $k/t$, as for these values the probability $P(k,t)$ does not decay. For all other values of $k/t$, the Cram\'er function determines the late time decay rate of the probability of seeing these events, and hence it contains information on how probabilities for observing events deviating from the mean behave.

The Cram\'er function is the Legendre-Fenchel transform of the \textit{scaled cumulant generating function} (SCGF) $\Theta(z)$, or
\begin{equation}
	I(k) = \sup_z (k\, z -\Theta(z) )\,,
\end{equation}
where $\Theta(z)$ is defined as
\begin{equation}
	\Theta(z) = \lim_{t \to \infty} \frac{1}{t} \log M_K(z, t)\,.
\end{equation}
This definition is equivalent to stating that the moment generating function $M_K(z,t)$ behaves as
\begin{equation}
	M_K(z,t) \propto e^{ t \Theta(z)}\,,
\end{equation}
when $t\to \infty$. As the moment generating function is the exponential of the tilted generator \eqref{tiltedgenerator}, the leading contribution to $M_K(z,t)$ will come from the largest eigenvalue of $\tilde{H}(z)$. This argument relates the SCGF to the leading eigenvalue of the tilted generator as a function of $z$. Hence, to find the large deviation Cram\'er function, we need to compute the Legendre-Fenchel transform of the SCGF $\Theta(z)$, which can be found by optimizing for the leading eigenvalue of the tilted generator.

The large deviation principle is not always guaranteed to exist, just as the leading eigenvectors of $\tilde{H}(z)$ are not always unique. In some cases, the eigenvalue decomposition of $\tilde{H}(z)$ does not exist (as not all matrices are diagonalizable). Generally, a CRN (or Markov process in general) with absorbing states are going to cause problems, since the dynamics can get stuck in absorbing states, which makes the late time properties sensitive to the initial conditions. However, for systems which are \textit{ergodic}, such that every microscopic configuration can be reached from any other configuration by a series of transitions, the leading eigenvalue of $\tilde{H}$ is unique and these methods are very powerful. 

For further reading on the relationship between large deviation theory and CRNs we refer to the book \cite{feng2006large} and the reviews \cite{smith2011large,assaf2017wkb}. \cite{freidlin1985limit} deals with limit theorems for large deviations in the reaction-diffusion setting and \cite{dykman1994large} connects large fluctuations to optimal paths in chemical kinetics. A sample path Large Deviation Principle for a large class of CRNs was discussed in \cite{agazzi2018large}. Dynamical Large Deviations in the large volume limit are discussed in \cite{lazarescu2019large}, where it is shown that CRNs with multiple steady states generically undergo a first-order dynamical phase transitions in the vicinity of zero tilting parameter. 

\subsection{Applications and further advances}
The Fock space approach to CRNs and stochastic dynamics has found applications in a diverse range of disciplines. Here we list a selection of references which make use of these methods in the setting where populations are considered to be homogeneously mixed. Models with (discrete or continuous) spatial degrees of freedom are discussed in the following sections. 
An application to the Malthus-Verhulst process which essentially represents logistic growth, are discussed in \cite{moloney2006functional}. Other results in population dynamics
are found in \cite{santos2015fock}, which provides exact solutions of stochastic Michelson-Mentis enzyme kinetics, and \cite{de2022fock} treats stochastic finite-size SIR epidemic models. Analytical solutions of biological models of gene transcription involving bursting and gene switching are described in \cite{vastola2021analytic}.
An application of these methods to stochastic game theory on a population level was discussed in \cite{armas2023}, where a risk averse utility measure based on the moment generating function was shown to lead to cooperative behaviour in population games. Further applications in an economic setting were made in \cite{bleher2020orders,bleher2024algebraic}, where stochastic models of limit order books are described by creation and annihilation operators of orders. 
In physics these methods were used to uncover the statistics of electron-hole avalanches in semiconductors exposed to an electric field \cite{windischhofer2021statistics} and to describe neutron population under the influence of a feedback mechanism in nuclear reactors as a stochastic field theory in \cite{dechenaux2022percolation}.
Models of stochastic graph-rewriting leveraging a Fock space description were presented in \cite{behr2016stochastic}.

The second quantized approach to stochastic mechanics has also been used to further our theoretical understanding of CMEs and their properties. A one-parameter family of extensions of the second quantized stochastic mechanics is related to orthogonal Hermite and Charlier polynomials \cite{ohkubo2012one}. In \cite{baez2013noether} a Noether theorem for stochastic systems was derived using Fock space methods and an alternate proof of the Anderson-Craciun-Kurtz (ACK) theorem was presented in \cite{baez2015quantum}. A generalization of the ACK theorem to a class of CRNs without zero deficiency or weak reversibility was presented in \cite{hirono2023squeezing}, making use of squeezed Poisson states in Fock space. The link between dynamical equations for the probability generating functions, the exponential moment generating functions, the factorial moment generating functions  and results in mathematical combinatorics was made more explicit in \cite{behr2017combinatorics}. Equations of motion for the moments of the density operator at all orders for stochastic CRNs at general deficiency were derived in \cite{smith2017flows}.  The connection between the second quantized formalism for ecosystem dynamics and mean-field biochemical reaction networks was discussed in \cite{degiuli2022}.
Work on an action functional gradient descent algorithm for finding escape paths (i.e. least improbable or first passage paths) in stochastic CRN with multiple fixed points is presented in \cite{gagrani2023action}. Recently, a Doi-Peliti path integral formulation of a large class of compartmental models was discussed in \cite{visco2025effective}, where a dimensional reduction technique to study the critical behavior of the model was proposed.

\section{Incorporating space: field theories for reaction-diffusion}
\label{sec:continuum}

In the last section we described the dynamics of CRNs at two levels: the rate equations for the deterministic dynamics of concentrations and the CME for the stochastic dynamics of copy numbers. 
Here we relax the well-mixed assumption and introduce spatial dependence into the CME. 
To do so, we first discretize the diffusion process as reactions that add/remove particles in neighboring voxels in space, which enable us to apply the mathematical methods from the last section.
The limit of decreasing voxel size will lead to a continuous statistical field theory for \textit{reaction-diffusion} systems. Now, instead of having a finite and constant probability of reacting with one-another, particles rely on diffusion to be brought into the vicinity of one another. These are also called \textit{diffusion-influenced} reactions. In chemistry, it means the solution should not be stirred, but the reactants are depending on \textit{Brownian motion} to be brought within interaction range. In population biology, we may think of members of a species performing a \textit{random walk} through an environment in order to find each other. 

The second quantized formulation of interacting particle systems, such as reaction-diffusion models, was introduced in a seminal paper by Massao Doi in 1974 \cite{doi1976second} and thus referred to as the Doi formulation. In essence, it is equivalent to the formalism described in \cref{sec:Fockspace}, now with creation and annihilation operators depending on the spatial location of the particles: $a^\dagger(x), a(x)$. As such, the commutation relations in \eqref{commutators} are modified to give a Dirac delta function: $[a(x), a^\dagger(y)] = \delta(x-y)$, and the $n$ particle state becomes a symmetrized function of all the particles positions. The formalism was later adapted into a path integral representation by Peliti in \cite{peliti1985path}, where it is obtained using a continuum limit on a lattice discretization of the CRN. We will present here the path integral formulation, as it connects more closely to the last section. We revisit the Doi formulation in more detail in \cref{sec:deltabasis}, when we recover it by choosing delta functions as a basis for the formulation of section \ref{sec:stochMecRD}. There we also show how it can be used to obtain alternative representations and emerging models.

We will first briefly review the relation between random walks and the diffusion equation and formulate this in terms of the second quantized description of the last section. Then, following \cite{goldenfeld1984kinetics,peliti1985path}, we will turn the CME of the last section into a field theory for reaction-diffusion systems. We derive the propagator of the reaction-free theory and note that this is nothing else than the Green's function for the diffusion equation. Finally, we briefly comment on the perturbative expansion and renormalization group methods for reaction-diffusion systems. The aim is to give a pedagogical introduction to the construction of a reaction-diffusion path integral (also known as the Doi-Peliti path integral). For a deeper dive into this subject, we refer to the reviews \cite{tauber2005applications,chow2015path}, the book \cite{tauber2014critical} and \cite{cardy1996renormalisation,cardy1998field}.

\subsection{Random walks and the emergence of diffusion}
\label{randomwalks}
The microscopic diffusion process can be modeled as a random walk on a regular lattice by means of a master equation. More details on this exemplary subject can be found in \cite{spitzer1964principles, hughes1995random, rednerbook}. Here, we focus on introducing the Fock space description of the random walk and its continuous limit to the diffusion equation. To this end, we need a regular, $d$-dimensional lattice with lattice spacing $h$ and suppose that a particle has equal probability of hopping to any neighboring site. Importantly, each walker's hopping probability is independent of the presence of any other walker or of its own past trajectory. This means that we can treat the walkers as independent and identically distributed (iid) and that the random walk allows for a Markovian description. 

Let us consider random walkers of a single species $A$. The probability vector $\ket{\rho(t)}$ will now be expanded into a Fock space for the lattice occupation numbers $\{n\}$, representing the number of walkers $n_i$ per lattice site $i$, such that: $\ket{\rho(t)} = \sum_{\{n\}} p_{\{n\}}(t) \prod_i (a^{\dagger}_i)^{n_i}\ket{0}$. Hence, now we suppose that $a_i$, and $a^{\dagger}_i$ are the annihilation and creation operators for species $A$, but at lattice site $i$. The random walk master equation is then:
\begin{align}
\begin{split}
	\partial_t \ket{\rho(t)} 
  &  = - \frac{D}{h^2} \sum_{(i \, j)}  \big( \aidag - a^\dagger_j )(\ai - a_j ) \ket{\rho(t)}  \\
 &	\equiv H_0(D) \ket{\rho(0)} \,. \label{diffusiongenerator}
 \end{split}
\end{align}
Here $(i \, j)$ is shorthand notation for nearest neighboring sites. The hopping rate is set (with some amount of hindsight) to $\frac{D}{h^2}$, where $D$ will become the diffusion constant in the continuous $h \to 0$ limit. Let us take this continuous limit for the observable counting the local density of particles at site $i$, defined as:
\begin{equation}
    \rho_i(t)  = \frac{1}{V} \vev{N_i}_t = \frac{1}{V} \bra{\mathbf{1}} a_i \ket{\rho(t)}
\end{equation}
From \eqref{eq:commutator_with_H}, together with \eqref{diffusiongenerator}, it is not hard to work out that for the one-dimensional lattice
\begin{equation}\label{rwmaster}
	\partial_t \rho_i(t) = \frac{D}{h^2} \left[ \rho_{i+1}(t) + \rho_{i-1}(t) - 2 \rho_i(t) \right] 
\end{equation}
where $\rho_{i\pm1}(t)$ denotes the local density at the neighboring lattice sites. Replacing $\rho_i(t) \to \rho(x,t)$, such that $\rho_{i\pm 1}(t) \to \rho(x \pm h,t)$, we can perform the limit $h \to 0$ to obtain the 1D diffusion equation
\begin{equation}
	\partial_t \rho(x,t) = D \partial_x^2 \rho(x,t) \,.
\end{equation}
A generalization of the above argument to arbitrary dimensions results in
\begin{equation}\label{diffusioneqn}
	\partial_t \rho(\mathbf{x},t) = D \nabla^2 \rho(\mathbf{x},t)\,,
\end{equation}
where $\nabla^2 = \vec{\nabla} \cdot \vec{\nabla}$ is the Laplacian. The solution to this equation for a normalized density with an initial distribution localized at $x=0$ can be found by Fourier transformation and reads:
\begin{equation}\label{normaldistribution}
	\rho(\mathbf{x},t) = \frac{1}{(4\pi D t)^{d/2}} e^{- \frac{x^2}{4Dt}}\,.
\end{equation}
The density behaves as a (multivariate) normal distribution with vanishing mean and variance growing linearly in time as $2D t$. This implies that, while the expected position of the walker does not change in time, its mean squared displacement grows linearly with $t$, allowing the walker to explore a sphere around the origin of radius $\sqrt{Dt}$. The hopping rate is constant, however, so the number of sites visited grows linear in time. This implies that the density of sites visited within the exploration radius is proportional to $\phi \sim t/t^{d/2} = t^{1-d/2}$. From this heuristic argument, one is led to the conclusion that the number of spatial dimensions has an important effect on the property of the random walk, namely the question if the walker can explore all of space if given infinite time. If $d<2$, $\phi$ diverges as $t \to \infty$ and so each site is visited infinitely often. In this regime the random walk is certain to return to its starting point, a feature called \textit{recurrence}. When $d>2$, $\phi$ decreases with time and hence there will be points which are never visited. In this case, there is non-zero probability for the random walk to never return to its starting position and the random walk is called \textit{transient}. The marginal case where $d=2$ turns out to be recurrent, although a more careful analysis is required to show that the density of visited sites diverges logarithmically \cite{spitzer1964principles}. 

The density satisfying \eqref{diffusioneqn} does not have to be normalized (it is in general a concentration), and may satisfy other boundary conditions or initial conditions as those leading to \eqref{normaldistribution}. For example, constant Dirichlet boundaries imply contact with an external reservoir \cite{del2018grand,kostre2021coupling}. In some contexts, it might be useful to write \cref{diffusiongenerator} from a particle tracking perspective \cite{isaacson2008relationship}, i.e. in terms of positions of particles on the lattice instead of number of particles at a site.

One of the main advantages of this approach is that the generalization to multiple species is simple, although notationally more cumbersome. Now there is a set of occupation numbers for each species $s \in \mathcal{S}$. Let us denote the total set of $|\mathcal{S}|$ occupation numbers $\{n\}_{\mathcal{S}}$. Then  $\ket{\rho(t)}$ is expanded in the basis
\begin{equation}
	|\{n\}_{\mathcal{S}} \rangle = \prod_i \prod_{s \in \mathcal{S}} (a_{i,s}^\dagger)^{n_{i,s}}\ket{0} \,.
\end{equation}
where here $a_{i,s}^\dagger$ is the creation operator for species $s$ at lattice site $i$ and $n_{i,s}$ is the number of particles of species $s$ at lattice site $i$. Fortunately, the random walk process for each species is independent of any other species, and so the multi-species diffusion generator $H_0(\{D\}_{\mathcal{S}})$ is just the sum of the single species diffusion generators, where each species $s$ may have an independent diffusion constant $D_s$
\begin{equation}\label{multispecies_diffusion}
	H_0(\{D\}_\mathcal{S}) = \sum_{s\in \mathcal{S}} H_0(D_s)\,.
\end{equation}
In applications, one could have other forms of diffusion such as anomalous diffusion \cite{bouchaud1990anomalous}, non-Laplacian diffusion or interacting diffusion through pair potential interactions \cite{dibak2019diffusion}. Asymmetric diffusion, where particles have a preference for diffusing in a certain direction, is treated in \cite{sasamoto1998exact}. Related to this is the asymmetric exclusion process, which combines asymmetric diffusion with a particle exclusion principle at each site \cite{liggett1975ergodic,schutz2001exactly}.
In \cite{araujo2020revisiting} a Fock space formalism for L\'evy flights was introduced, which agrees with L\'evy distributions in the continuous limit. When the underlying geometry is not a regular lattice, but rather a complex network, the diffusion generator \eqref{diffusiongenerator} becomes the graph Laplacian; the generator for random walks on a graph \cite{burioni2005random}. More involved stochastic models on complex networks, such as spreading models, can also be formulated using an operator formalism, as was done in \cite{merbis2021exact,merbis2023emergent}.

\subsection{The Doi-Peliti path integral} 
\label{sec:firstSFTRD}

We will now turn to the construction of the Doi-Peliti path integral, first discussed in the context of the birth-death processes ($A \to 2A$ and $A \to \emptyset$) in \cite{peliti1985path}. 
The construction supposes that particles become point-particles in the continuum limit and reactions only take place when reactants are found at the same location in space. This has issues from a mathematical perspective: two point particles would encounter each other at exactly the same location in space with probability zero. However, this is a common physical approximation as one usually ends up working with discretizations or with densities instead of individual particles. In \cref{sec:stochMecRD,sec:specialapplications}, we will discuss more general stochastic field theories, which can handle space dependent reaction rate functions. 

As before, we restrict the presentation to a single species general reaction $k A \to \ell A$. The generalization to multiple species is not difficult and follows analogously.
The reaction-diffusion generator $H$ consists of the diffusion generator $H_0(D)$ \eqref{diffusiongenerator} and the reaction generator $H_I$, given in \eqref{generalH}. We further assume that only particles at the same lattice location can react. This implies that:
\begin{align}\label{reactiondiffusionH}
	H & = H_0(D) + H_I(\{\lambda\}) \,, \nonumber \\
	& = - \frac{D}{h^2} \sum_{(i\,j )}  \big( \aidag - a^\dagger_j )(\ai -a_j ) \nonumber \\
 & \qquad + \sum_{i,\tau} \lambda(\tau) \left[ (\aidag)^{\ell(\tau)} - (\aidag)^{k(\tau)} \right] \ai^{k(\tau)}\,.
\end{align}
Somewhat suggestively we have named the diffusive part of the generator $H_0$ and the reactive part $H_I$. This is because we can treat $H_0$ non-perturbatively, as the `free theory' for reaction-diffusion systems, but we should treat the total action perturbatively with respect to the non-linearities induced by the reactions in $H_I$.

To proceed, it is convenient to introduce Poisson states defined in terms of a complex valued variable $\eta$ as:
\begin{equation}
	\ket{\eta} = e^{-\frac{1}{2} |\eta|^2} e^{\eta a^\dagger} \ket{0} \,, \quad \bra{\eta} = \bra{0} e^{ \bar{\eta} a } e^{-\frac12 |\eta|^2} \,.
\end{equation}
Here the bar denotes complex conjugation. The overlap between two of these states can be shown to give:
\begin{equation}\label{overlap}
	\bracket{\eta_1}{\eta_2} = e^{- \bar{\eta}_1 (\eta_1 -\eta_2 ) }e^{\frac12 |\eta_1|^2 -\frac12 |\eta_2|^2}  \,.
\end{equation}
The complex valued Poisson states define a resolution of the identity operator as:
\begin{equation}
	\mathds{1} = \int \frac{\mathrm{d}^2\eta}{\pi} \ket{\eta} \bra{\eta}\,,
\end{equation}
or, defining one Poisson state for each lattice site $i$:
\begin{equation}\label{resofid}
	\mathds{1} = \int \prod_i \left( \frac{\mathrm{d}^2\eta_i}{\pi}\right) \ket{\eeta} \bra{\eeta}\,,
\end{equation}
where $\mathrm{d}^2\eta_i = \mathrm{d} \mathrm{Re}(\eta_i)\mathrm{d} \mathrm{Im}(\eta_i)$ and  $\ket{\eeta}  = \ket{\eta_1} \otimes \ket{\eta_2} \otimes \ldots$.

We now proceed to define the path-integral for the expectation value of any observable $A$. First, we write the expectation value in terms of the formal solution of the master equation as $\vev{A}_{t_f} = \bra{\mathbf{1}} A e^{H t_f} \ket{\rho(0)}$. Then, the time evolution induced by the generator $H$ is divided into infinitely many slices of infinitesimal size
\begin{equation}
	e^{H t_f} = \lim_{\Delta t \to 0} \left(e^{H \Delta t} \right)^{t_f/\Delta t} = \lim_{\Delta t \to 0 } \left(1 + \Delta t H\right)^{t_f/\Delta t}
    \label{eq:propslices}
\end{equation}
For a fixed $\Delta t$ and assuming equally spaced slices, the term inside the limit reduces to $\left(1 + \Delta t H\right)^N$ with $N$ the number of slices (which depends on $\Delta t$). Now, one can insert resolutions of the identity operator \eqref{resofid} in between each of the infinitesimally thin slices of $e^{H\Delta t}$.
Let $t$ label the time slices at $\Delta t, 2 \Delta t, \cdots , t_f$, such that the Poisson state parameters $\eeta$ also obtain a $t$-label and the expectation value becomes:
\begin{widetext}
\begin{equation}\label{pathintegralstep1}
	\vev{A}_{t_f} = \mathcal{N}^{-1} \lim_{\Delta t \to 0} \int \bigg(\prod_{i,t} d^2 \eta_{i, t} \bigg) \bra{\mathbf{1}}A\ket{\eeta_{t_f}}\left[ \prod_{ t =\Delta t}^{t_f} \bra{\eeta_t}(1 + \Delta t H ) \ket{\eeta_{t-\Delta t}} \right]\bracket{\eeta_0}{\rho(0)}\,.
 \end{equation}
\end{widetext}
Here $\mathcal{N}$ is a normalization factor, which can be determined in the end by requiring that if $A$ is the identity operator, the path integral should give unity. Lets discuss the terms appearing in this expression one by one.

First, inside \cref{pathintegralstep1}, the operator $A$ can be made to depend only on annihilation operators $a_i$, because any operator $A(\aidag, \ai)$ can be brought to a normal ordered form, where all creation operators are to the left of the annihilation operators. By the property $\bra{\mathbf{1}}\aidag = \bra{\mathbf{1}}$, the creation operators can then be set to one: $\aidag \to 1$. Moreover, since the Poisson state is an eigenvector of the annihilation operator $\ai\ket{\eeta} = \eta_i \ket{\eeta}$, we have that $\bra{\mathbf{1}}A\ket{\eeta_{t_f}} =  \bracket{\mathbf{1}}{\eeta_{t_f}} A(1, \eeta_{t_f})$, or: we can simply replace all annihilation operators $\ai$ in $A$ with the complex variables $\eta_{i,t_f}$. The inner product $\bracket{\mathbf{1}}{\eeta_{t_f}}$ gives $\prod_i \exp(-\frac12|\eta_{i,t_f}|^2 + \eta_{i,t_f} )$ by equation \eqref{overlap}.

The terms in the square brackets in \cref{pathintegralstep1} induce a similar replacement as with $A$, but now for $H$. Indeed, here we may write
\begin{align}\label{fieldthyderivation1}
&	\bra{\eeta_t}(1 + \Delta t H ) \ket{\eeta_{t-\Delta t}}  \\
& \qquad = \bracket{\eeta_t}{\eeta_{t-\Delta t}} (1 + \Delta t H(\bar{\eeta}_{t}, \eeta_{t-\Delta t}) ) \nonumber \\
&	\qquad  \simeq   \bracket{\eeta_t}{\eeta_{t-\Delta t}} \exp( \Delta t H(\bar{\eeta}_{t}, \eeta_{t-\Delta t}) ) \,.  \nonumber
\end{align}
where $H(\bar{\eeta}_{t}, \eeta_{t-\Delta t}) $ is given by $H$ with creation operators replaced as $\aidag \to \bar{\eta}_{i,t} $ and annihilation operators replaced by $\ai \to \eta_{i,t-\Delta t}$. In the last line above we have restored the original exponential form of $H\Delta t$. This is related to a subtle point (see section 3.3 of \cite{tauber2005applications}), where we are actually  swapping a sum (in the definition of the exponential operator) with doing an integral (in this case integrating over the complex numbers $\mathrm{d}^2\eta$ ). Such a procedure does not always neatly work out, so we should keep in the back of our heads the fact that we are actually \textit{always} defining $e^{Ht}$ in terms of its power series. So the actual explicit computation of the final object implicitly implies a power series, which can be cast into the form of a perturbative series in $\lambda$, restoring the proper order of summation and integration. 

Using \eqref{overlap}, the inner product in \eqref{fieldthyderivation1} is given as
\begin{align}\label{overlap2}
	 & \bracket{\eeta_t}{\eeta_{t-\Delta t}} = \prod_i  \bracket{\eta_{i,t}}{\eta_{i,t-\Delta t}} \\
  & \quad = e^{- \sum_i \bar{\eta}_{i,t} (\eta_{i,t}- \eta_{i, t -\Delta t})  }e^{\sum_i \frac12 |\eta_{i,t}|^2 - \frac12 |\eta_{i,t-\Delta t}|^2 } \,. \nonumber
\end{align}
The first exponential term in the second line can be expanded as a Taylor series in $\Delta t$ to give the time derivative of $\eta_{i,t}$ as leading term: it becomes $e^{- \sum_i \bar{\eta}_{i,t} \frac{d}{dt}\eta_{i,t}\Delta t + \mathcal{O}(\Delta t^2)  }$.
The terms in the second exponential will cancel when multiplying over all the $t$ slices, except for the initial term $-\frac12|\eta_{i,0}|^2$ and the final term $\frac12|\eta_{i,t}|$. This final term, however, cancels the factor of $-\frac12 |\eta_{i,t}|^2$ which came from the inner product of $\bracket{\mathbf{1}}{\eeta_t}$. 

Finally, the last term in \eqref{pathintegralstep1} concerns the initial conditions. We assume each lattice site is initially filled according to a Poisson distribution with on average $n_0$ particles. 
So the initial state becomes a Poisson state with mean $n_0$:  $\ket{\rho(0)} = \ket{n_0} \sim e^{\sum_i n_0 \aidag} \ket{0}$, where the normalization factor may be absorbed into $\mathcal{N}$.
The inner product with $\bra{\eeta_0}$ again has the effect that all creation operators $\aidag$ are replaced with $\bar{\eta}_i$ and so, when taken together with the factors of $\exp(-\frac12 |\eta_{i,0}|^2)$ left over from \eqref{overlap2}
\begin{equation}
	\bracket{\eeta_0}{\rho(0)}e^{- \sum_i \frac12 |\eta_{i,0}|^2} \sim e^{\sum_i \bar{\eta}_{i,0} (n_0 - \eta_{i,0}) }
\end{equation}
Putting everything back together and taking the limit $\Delta t \to 0$, the path integral becomes:
\begin{equation}
	\vev{A}_{t_f} = \mathcal{N}^{-1} \lim_{\Delta t \to 0} \int \bigg(\prod_{i,t} d^2 \eta_{i, t} \bigg) A(\eeta_{t_f})  e^{-S(\bar{\eeta}, \eeta)}  \,,
\end{equation}
with
\begin{align}
	S = & \sum_i \bigg( - \eta_{i,t_f} - \bar{\eta}_{i,0} (n_0 - \eta_{i,0})  \\ \nonumber
 & + \sum_{t =\Delta t}^{t_f} \Delta t \left(\bar{\eta}_{i,t}\frac{d}{dt}\eta_{i,t} - H(\bar{\eeta}_{t},\eeta_{t-\Delta t}) \right) \bigg)\,.
\end{align}
We can now take the $\Delta t \to 0$ limit and replace the $\sum_t \Delta t $ sum by an integral $\int_0^{t_f} \mathrm{d}t$. The $\eta$'s then become continuous functions of $t$ and the $\Delta t$ time difference in $H(\bar{\eeta}_{t},\eeta_{t-\Delta t})$ disappears, with the understanding that when it matters, we should think of the $\bar{\eta}(t)$ fields as following the $\eta(t)$ field in time. The $\Delta t \to 0$ limit turns the measure $\prod_{i,t} \mathrm{d}^2 \eta_{i, t} $ into functional differentials $\mathcal{D}\eta_{i} \mathcal{D}\bar{\eta}_i$  and we obtain
\begin{equation}\label{avevFT}
	\vev{A}_t = \mathcal{N}^{-1} \int \mathcal{D}\eta_{i} \mathcal{D}\bar{\eta}_i  A(\{\eta_{i}(t)\})  e^{-S(\bar{\eeta}, \eeta)}  \,.
\end{equation}
where $\mathcal{N} = \int \mathcal{D}\eta_{i} \mathcal{D}\bar{\eta}_i  e^{-S(\bar{\eeta}, \eeta)}$, such that choosing $A$ to be the identity gives $\vev{ \mathbb{1}}_t = 1$ at all times. The \textit{action} $S$ is given as:
\begin{align}
	S = & \sum_i \bigg( - \eta_{i}(t_f) - \bar{\eta}_{i}(0) (n_0 - \eta_{i}(0)) \\ \nonumber
 & + \int_0^{t_f} \mathrm{d}t \left(\bar{\eta}_{i}(t)\frac{d}{dt}\eta_{i}(t) - H(\bar{\eeta}_{t},\eeta_{t})  \right) \bigg)
\end{align}
What rests is to take the continuum space limit, and  discuss the meaning of the initial and final terms in the action. Like in the last section, the continuum limit is performed by taking the lattice spacing $h \to 0$, while keeping the action finite. This can be achieved with the scaling:
\begin{align}\label{fieldscaling}
&	 \eta_i(t) \to h^d \varphi(\mathbf{x},t)\,, \quad \bar{\eta}_i(t) \to \tilde{\varphi}(\mathbf{x},t)\,, \\
&  \sum_i \to \int h^{-d} \mathrm{d}^dx \,. \nonumber
\end{align}
The complex conjugate fields $\tilde{\varphi}$ and $\varphi$ are to be treated as independent field variables.
Performing the continuum limit in space gives the action
\begin{align}\label{Sfieldthy}
	S[\tilde{\varphi},\varphi] & = \int \mathrm{d}^dx \bigg( - \varphi(t_f) - \tilde{\varphi}(0) (n_0 - \varphi(0)) \\ 
 & +  \int_0^{t_f} \mathrm{d}t \left(\tilde{\varphi} (\partial_t - D\nabla^2) \varphi - H_I(\tilde{\varphi},\varphi) \right) \bigg)  \nonumber
\end{align}
Here $H_I(\tilde{\varphi},\varphi)$ represents the reaction generator $H_I$ with $\aidag \to \tilde{\varphi}(\mathbf{x},t)$ and $\ai \to \varphi(\mathbf{x},t)$. Additionally, the interaction rates $\lambda(\tau)$ appearing in \eqref{reactiondiffusionH} are scaled as $\lambda(\tau) = \lambda_0(\tau)/ h^{d(k(\tau)-1)}$ to keep the interaction terms finite as $h \to 0$. The result is:
\begin{equation}
	H_I(\tilde{\varphi},\varphi) = \sum_{\tau} \lambda_0(\tau) \left[ \tilde\varphi^{\ell(\tau)} - \tilde\varphi^{k(\tau)} \right] \varphi^{k(\tau)}\,.
 \label{eq:HintDoiPeliti}
\end{equation}
We have now turned the single species reaction network into a field theory for a pair of scalar fields $\tilde{\varphi}, \varphi$. The generalization to multiple species is immediate; each species $s \in \mathcal{S}$ has a scalar field associated to it. The diffusion term is diagonal in the sense that there is one term for each species, each with its own diffusion constant $D_s$. The interaction terms in $H_I$ may (and generally will) couple different species together. 
The upshot after this somewhat lengthy derivation, is that we arrive at a very simple prescription for turning any CRN with spatially independent reaction rates into a reaction-diffusion field theory: Replace the creation operators $\aidag$ by $\tilde{\varphi}_i$ fields and the annihilation operators $\ai$ by $\varphi_i$ fields and add a diffusion term for each species. 

Many authors use a slightly different formulation of the reaction-diffusion action, where the $\tilde{\varphi}$ fields are shifted by one: $\tilde{\varphi} \to 1 + \bar{\varphi}$. 
This field shift (sometimes called the Doi shift) traces back to the difference between the exclusive and inclusive definition of the inner product \cite{grassberger1980fock}, as explained in \cref{sec:inclprod}.
The flat state can be expressed as $\bra{\mathbf{1}} = \bra{0} e^{\sum_i \ai}$. It is possible to move the $e^{\sum_i \ai}$ to the right in the expectation value, commuting it through any operator it comes across its path. Since $e^{\ai} \aidag = (\aidag+1)e^{\ai}$, this has the net effect of shifting all creation operators by one: $\aidag \to \aidag +1$. In field theory language, this has the stated effect of the field shift $\tilde{\varphi} \to 1 + \bar{\varphi}$, under which the time derivative part gets an additional final and initial time contribution
\begin{equation}\label{shiftconsequence}
	\int_0^{t_f} dt \tilde{\varphi} \partial_t  \varphi = \varphi(t_f) - \varphi(0) + \int_0^{t_f} dt \bar{\varphi} \partial_t\varphi\,.
\end{equation}
The $\varphi(t_f)$ from this shift now cancels the $-\varphi(t_f)$ present in \eqref{Sfieldthy}, but it generates another $\varphi(0)$ initial term. This term is then canceled by performing the shift in the initial conditions. The $\bar{\varphi}(0)(n_0-\varphi(0))$ term remaining in the action \eqref{Sfieldthy} has the interpretation that $\bar{\varphi}(0)$ acts as a Lagrange multiplier field, enforcing the constraint $\varphi(0)= n_0$, and hence making sure the initial conditions are satisfied.

\subsection{Free propagator}
\label{sec:freeprop}

With the above prescription, any CRN is mappable to a field theory, describing the reaction process in the diffusion-influenced regime. This allows importing field theory methods to the field of reaction-diffusion systems. Most notably, these methods are perturbation theory (computing Feynman diagrams) and the renormalization group flow. We do not give an exhaustive overview of these methods; they are addressed in quantum field theory textbooks (such as \cite{peskin2018introduction,zinn2021quantum}) and in the context of reaction-diffusion systems in \cite{tauber2005applications,janssen2005field, tauber2014critical}. However, here and in the next section, we opt for a more pedagogical (and hence less deep-diving) presentation.

The first point of discussion is the free (purely diffusive) theory, where there are no reactions present in the system (i.e. all reaction rates are set to zero). The free theory is important, because it provides the elementary building block of the perturbative expansion in terms of Feynman diagrams, namely the \textit{propagator}. The propagator corresponds to the pair correlations between the fields $\bar{\varphi} $ and $\varphi$ and hence it represents mathematically the propagation of a $\bar{\varphi}$ field into a $\varphi$ field. In terms of diagrams, this is represented as a line (with time running from right to left), where a $\bar{\varphi}$ field is created at the right and propagates into a $\varphi$ field on the left. 
\begin{align}\label{twopointcorrelator}
\vev{\varphi(\mathbf{x},t) \bar\varphi(\mathbf{x}',t')} &= G(\mathbf{x}-\mathbf{x}' ,t-t') \\ \nonumber
& = 
	\begin{array}{c} 
		\begin{tikzpicture}
			\begin{feynman}
				\vertex (a) at (0,0) {$\varphi$};
				\vertex (b) at (2,0) {$\bar{\varphi}$};
				\diagram*{(a) -- [plain,line width=1] (b)};
			\end{feynman} 
		\end{tikzpicture}  
	\end{array}
\end{align}
We can obtain the propagator from the bilinear part of the action, which is in this case the Green's function of the diffusion equation\footnote{For monomolecular reactions the propagator generically receives contributions from bilinear terms in the reaction part of the action and/or from loop diagrams.}:
\begin{equation}\label{Greensfunction}
	(\partial_t - D \nabla_{\mathbf{x}}^2) G(\mathbf{x}-\mathbf{x}',t-t') = \delta^{d}(\mathbf{x}-\mathbf{x}')\delta(t-t')
\end{equation}
The Green's function is most easily obtained in momentum space after performing a Fourier transform:
\begin{equation}\label{fouriertsfm}
	G(\mathbf{p}, \omega) = \int d^dx \int dt e^{- i \mathbf{p}\cdot (\mathbf{x}- \mathbf{x}')} e^{i \omega (t-t')} G(\mathbf{x}-\mathbf{x}',t-t')
\end{equation}
Equation \eqref{Greensfunction} leads to the momentum space propagator
\begin{equation}\label{GFmomentum}
	G(\mathbf{p},\omega) = \frac{1}{-i\omega + D p^2} \,.
\end{equation}
It is convenient to invert the temporal Fourier transform and work in terms of $G(\mathbf{p},t-t') = \int d\omega/(2\pi) G(\mathbf{p},\omega) e^{-i \omega (t-t')}$. Performing the $\omega$ integral using Cauchy's residue theorem leads to 
\begin{equation}\label{propagator}
	G(\mathbf{p},t-t') =e^{-Dp^2 (t-t')} \Theta(t-t') \,. 
\end{equation}
Here $\Theta(t-t')$ is the Heaviside step function. Physically, the presence of this function represents causality: only earlier $\bar{\varphi}$ fields are connected to later $\varphi$ fields. 

The causality requirement has two important consequences: any term in the action where $\varphi$ fields appear without an earlier $\bar\varphi$ field will vanish when averaged with the statistical weight $e^{-S_0}$. Hence, as observables are only expressed in terms of $\varphi$ fields, the $\bar{\varphi}(0)\varphi(0)$ term in the initial conditions can be safely dropped from the action, since pairing the $\varphi$ fields in the observable with $\bar{\varphi}(0)$ would leave an unpaired $\varphi(0)$ field in the action. Secondly, it justifies an earlier hidden assumption where we extended the time domain from $[0,t_f]$ to the entire real time axis (including negative $t$) when performing the Fourier transform \eqref{fouriertsfm}. 

With these last thoughts in mind, the final result for the path integral representation of any (normal ordered) observable $A$ can be written as
\begin{equation}
	\vev{A}_t = \frac{1}{\mathcal{N}} \int \mathcal{D}\bar{\varphi} \mathcal{D} \varphi \; A(\varphi) e^{- S[\bar{\varphi},\varphi]}
\end{equation}
with the action (after applying the Doi-shift):
\begin{align}
\label{eq:pi_action} \nonumber 
	S[\bar{\varphi},\varphi] = \int d^d x \bigg\{  & \int dt \left[\bar{\varphi} (\partial_t - D \nabla^2) \varphi - H_I(1+\bar{\varphi}, \varphi) \right]  \\ & - \bar{\varphi} n_0    \bigg\}\,. 
\end{align}
The Euler-Lagrange equations for this action correspond to the rate equations obtained from the mean-field approximation discussed in \cref{sec:emergencerateeqs}, supplemented with a diffusion term. This action can be used as a starting point to explore perturbative corrections to the rate equations, in order to see where and when fluctuation become important. This leads to the construction of Feynman rules to compute contributions from Feynman diagrams and to the renormalization group for reaction-diffusion systems.

\subsection{Perturbation theory and the renormalization group}
\label{sec:perturbandrenormalize}
The renormalization group (RG) is a systematic procedure which can be applied to any field theory (quantum, classical, statistical or stochastic) to understand the impact of fluctuations and identify and compute universal properties of the system. Generically, a momentum-space cutoff $\kappa$ is introduced to regulate short-distance (ultraviolet: UV) singularities. These singularities are an artifact of the continuum limit taken above; because we have taken the lattice spacing to go to zero, we essentially require particles to be infinitesimally close to each other to react. The introduction of a momentum scale $\kappa$ makes the parameters of the theory (i.e. the reaction rates) $\kappa$ dependent. One can then define renormalized effective parameters, which absorb the short-distance singularities. This allows us to derive how the parameters depend on the momentum scale of the theory. Scale invariance is achieved when the effective (renormalized) parameters do not change with decreasing momentum scale and at this point the UV scaling properties translate to the long distance infrared (IR) regime. 
Here we review briefly the perturbative expansion and the RG method at the hand of the field theory for single species reaction $k A \to \ell A$ with $\ell < k$, which was discussed in detail in \cite{lee1994renormalization}. 

The field theory action, after the Doi shift $\tilde\varphi = 1 +\bar{\varphi}$ is
\begin{align}
\label{eq:reaction_action}
	S[\bar{\varphi}, \varphi] = \int d^d x \bigg\{ & \int_0^{t_f} dt \; \left[ \bar{\varphi} (\partial_t - D \nabla^2 ) \varphi + \sum_{i=1}^k \lambda_{i} \bar{\varphi}^i \varphi^k \right] \nonumber \\
 & - n_0 \bar \varphi(0) \bigg\} \,.
\end{align}
The coupling constants $\lambda_i$ are related to the original reaction rate $\lambda_0$ of the $kA \to \ell A$ field theory as
\begin{subequations}\label{lambdamap}
	\begin{align}
		\lambda_i & = \lambda_0 \left[\binom{k}{i} -  \binom{\ell}{i}   \right]\,,  & i \leq \ell  \\
		\lambda_i & =  \lambda_0 \binom{k}{i} \,, & i > \ell \,.
	\end{align}
\end{subequations}
In order to compute any observable, we treat the diffusive part of the action non-perturbatively, and expand the path integral in the reaction parameters $\lambda_i$. For instance, if our observable is the particle density, then $A = \varphi$ (recall the number operator is $\aidag \ai$ and we could set $\aidag = 1$ in the expectation value), then the path integral may be expanded in three exponents as
\begin{align}\label{pertcomputation}
	\vev{\varphi(\mathbf{x},t)}  = \int  \frac{\mathcal{D}\bar{\varphi} \mathcal{D} \varphi}{\mathcal{N}} \; \varphi(\mathbf{x}, t) e^{-S_0} e^{-S_I(\lambda_i)} e^{\int d^dy \; n_0 \bar \varphi(\mathbf{y}, 0)} 
\end{align}
The first exponent is the free, diffusive action, which we will use as statistical weight for the computation of the expectation value. The second term represents the reaction interaction, which we expand as a power series in $\lambda_i$. The last exponent is the implementation of the initial conditions, which effectively generates $\bar\varphi$ fields at $t=0$. The effect of computing everything with $S_0$ as statistical weight means that we can use the propagator \eqref{twopointcorrelator} to connect $\varphi$ fields to earlier $\bar{\varphi}$ fields. One only gets contributions from terms where pairs can be made with an equal number of $\varphi$ and $\bar \varphi$ fields, and the propagator imposes causality: later $\bar{\varphi}$ fields contracted with earlier $\varphi$ fields give vanishing contributions.

\subsubsection{Feynman rules and Feynman diagrams}

Each term in the expansion of $e^{-S_I(\lambda_i)}$ corresponds to a specific \textit{Feynman diagram}: a graphical representation of its contribution to the path integral. The rules for constructing these diagrams and mapping them to a formula are generically called \textit{Feynman rules}. 
It is often practically easier to compute the Feynman diagrams in the momentum-time representation, by Fourier transforming over the spatial variables, such that we may use the expression \eqref{propagator} for the propagator. 
The Feynman rules for constructing and computing the diagrams are:
\begin{itemize}
	\item Each line of the Feynman diagram represents a propagator $(2\pi)^d \delta^d(\mathbf{p}+\mathbf{p}')G(\mathbf{p},t_2-t_1)$, connecting an earlier $\bar{\varphi}(\mathbf{p}',t_1)$ field to a $\varphi(\mathbf{p},t_2)$ field.  
	\item Each vertex represents $k$ incoming $\varphi$ fields (coming from the right) and $i$ outgoing $\bar{\varphi}$ fields (going to the left), multiplied by $ - \lambda_i$
	\item Each external end point (labeled by $\times$), represents the initial insertion of $\bar{\varphi}$ fields, multiplied by $n_0$
	\item At each vertex, momentum conservation holds: multiply by $(2\pi)^d \delta( \sum_i \mathbf{p}_i)$ where $i$ labels all momenta connected to that vertex
	\item Each diagram represents an integration over all momenta $\int \prod_i \frac{d^dp_i}{(2\pi)^d}$ and over all time points of the inserted vertices from times  $[0,t]$.
	\item Each diagram is multiplied by a symmetry factor, which is the number of ways to relabel propagators without changing the diagram\footnote{This is a slightly different convention as is usual in standard quantum field theory (see for instance \cite{peskin2018introduction}), where the convention is to divide the coupling constants $\lambda_k$ of order $k$ by $1/k!$ and then one should \textit{divide} by the symmetry factor instead of multiply.}
\end{itemize}
Using these Feynman rules and the careful expansion of the path integral \eqref{pertcomputation}, the expectation value of the density receives the following diagrammatic contributions in the exemplary case where $k=2$:
\begin{align} \label{perturbativeexpansion}
	&
	\begin{array}{c} 
		\begin{tikzpicture}
			\begin{feynman}
				\vertex (a) at (0,0) {};
				\vertex (b) at (1,0) {};
				\diagram*{(a) -- [plain,line width=0.5,insertion=1] (b)};
			\end{feynman} 
		\end{tikzpicture}  
	\end{array}
	+ 
	\begin{array}{c} 
		\begin{tikzpicture}
			\begin{feynman}
				\vertex (a) at (0,0) {};
				\vertex (b) at (1,0.35) {};
				\vertex (c) at (1,-0.35) {};
				\vertex (l) at (0.5,0);
				\diagram*{(a) -- [plain,line width=0.5](l),
					(l) -- [plain,line width=0.5,insertion=1] (b),
					(l) -- [plain,line width=0.5,insertion=1] (c) };
			\end{feynman} 
		\end{tikzpicture} 
	\end{array}
	+
	\begin{array}{c} 
		\begin{tikzpicture}
			\begin{feynman}
				\vertex (a) at (0,0) {};
				\vertex (b) at (1,0.4) {};
				\vertex (c) at (1,-0.4) {};
				\vertex (l) at (0.5,0);
				\vertex (l2) at (0.75,-0.2);
				\vertex (d) at (1,0.1) {};
				\diagram*{(a) -- [plain,line width=0.5](l),
					(l) -- [plain,line width=0.5,insertion=1](b),
					(l) -- [plain,line width=0.5,insertion=1] (c),
					(l2) --  [plain,line width=0.5,insertion=1] (d)};
			\end{feynman} 
		\end{tikzpicture} 
	\end{array}
	+ \ldots \\
	& +
	\begin{array}{c} 
		\begin{tikzpicture}
			\begin{feynman}
				\vertex (a) at (0,0) {};
				\vertex (b) at (2,0.35) {};
				\vertex (c) at (2,-0.35) {};
				\vertex (l) at (0.5,0);
				\vertex (l2) at (1.5,0);
				\diagram*{(a) -- [plain,line width=0.5](l),
					(l2) -- [plain,line width=0.5,insertion=1](b),
					(l2) -- [plain,line width=0.5,insertion=1] (c) };
				\draw[plain] (l) arc [start angle=140, end angle=40, radius=.65];
				\draw[plain] (l2) arc [start angle=-40, end angle=-140, radius=.65];
			\end{feynman} 
		\end{tikzpicture}  
	\end{array} 
	+
	\begin{array}{c} 
		\begin{tikzpicture}
			\begin{feynman}
				\vertex (a) at (0,0) {};
				\vertex (b) at (2,0.4) {};
				\vertex (c) at (2,-0.4) {};
				\vertex (d) at (2,0.1) {};
				\vertex (l) at (0.5,0);
				\vertex (l2) at (1.5,0);
				\vertex (l3) at (1.75,-0.2);
				\diagram*{(a) -- [plain,line width=0.5](l),
					(l2) -- [plain,line width=0.5,insertion=1](b),
					(l2) -- [plain,line width=0.5,insertion=1] (c),
					(l3) --  [plain,line width=0.5,insertion=1] (d) };
				\draw[plain] (l) arc [start angle=140, end angle=40, radius=0.65];
				\draw[plain] (l2) arc [start angle=-40, end angle=-140, radius=0.65];
			\end{feynman} 
		\end{tikzpicture}  
	\end{array} 
	+
	\begin{array}{c} 
		\begin{tikzpicture}
			\begin{feynman}
				\vertex (a) at (0,0) {};
				\vertex (b) at (2,0.35) {};
				\vertex (c) at (2,-0.35) {};
				\vertex (d) at (2,-0.75) {};
				\vertex (l) at (0.5,0);
				\vertex (l2) at (1.5,0);
				\vertex (l3) at (1,-0.235);
				\diagram*{(a) -- [plain,line width=0.5](l),
					(l2) -- [plain,line width=0.5,insertion=1](b),
					(l2) -- [plain,line width=0.5,insertion=1] (c),
					(l3) --  [plain,line width=0.5,insertion=1] (d) };
				\draw[plain] (l) arc [start angle=140, end angle=40, radius=0.65];
				\draw[plain] (l2) arc [start angle=-40, end angle=-140, radius=0.65];
			\end{feynman} 
		\end{tikzpicture}  
	\end{array} 
	+ \ldots \nonumber \\
	& +
	\begin{array}{c} 
		\begin{tikzpicture}
			\begin{feynman}
				\vertex (a) at (0,0) {};
				\vertex (b) at (3,0.35) {};
				\vertex (c) at (3,-0.35) {};
				\vertex (l) at (0.5,0);
				\vertex (l2) at (1.5,0);
				\vertex (l3) at (2.5,0);
				\diagram*{(a) -- [plain,line width=0.5](l),
					(l3) -- [plain,line width=0.5,insertion=1](b),
					(l3) -- [plain,line width=0.5,insertion=1] (c) };
				\draw[plain] (l) arc [start angle=140, end angle=40, radius=0.65];
				\draw[plain] (l2) arc [start angle=-40, end angle=-140, radius=0.65];
				\draw[plain] (l2) arc [start angle=140, end angle=40, radius=0.65];
				\draw[plain] (l3) arc [start angle=-40, end angle=-140, radius=0.65];
			\end{feynman} 
		\end{tikzpicture}  
	\end{array} + \ldots
	\nonumber
\end{align}
The contribution from each diagram can be computed using the Feynman rules. For instance, the first two tree diagrams give:
\begin{align}
    \begin{array}{c} 
		\begin{tikzpicture}
			\begin{feynman}
				\vertex (a) at (0,0) {};
				\vertex (b) at (1,0) {};
				\diagram*{(a) -- [plain,line width=0.5,insertion=1] (b)};
			\end{feynman} 
		\end{tikzpicture}  
	\end{array}  & =
 \int d^d y \; n_0 G(\mathbf{x}-\mathbf{y},t)  = n_0 \label{eq:diagrams_1} \\
 \begin{array}{c} 
		\begin{tikzpicture}
			\begin{feynman}
				\vertex (a) at (0,0) {};
				\vertex (b) at (1,0.35) {};
				\vertex (c) at (1,-0.35) {};
				\vertex (l) at (0.5,0);
				\diagram*{(a) -- [plain,line width=0.5](l),
					(l) -- [plain,line width=0.5,insertion=1](b),
					(l) -- [plain,line width=0.5,insertion=1] (c) };
			\end{feynman} 
		\end{tikzpicture} 
	\end{array} & =   \int_0^{t} dt'\;  G(0,t-t')(-\lambda_1)  G(0, t')^2 n_0^2  \label{treelevel}  \\ \nonumber
 & = -\lambda_1 n_0^2 t 
\end{align}
The first \textit{loop} diagram in the perturbative expansion \eqref{perturbativeexpansion} gives an effective correction to the tree level contribution of the vertex with two incoming fields and one outgoing field.
Working this out following the Feynman rules gives:
\begin{align}\label{loopintegral}
	& I_{\rm loop} =  \int_0^{t} dt'' \int_0^{t''} dt' G(0,t -t'') \lambda_1 \\ \nonumber
 &  \times \int \frac{d^d p}{(2\pi)^d} 2 G(\mathbf{p},t''-t') G(-\mathbf{p},t''-t') \, \lambda_2 G(0,t')^2 n_0^2\,,
\end{align}
Using the expression for the diffusion propagator \eqref{propagator}, the integrals evaluate to:
\begin{equation}\label{loopintegral2}
	I_{\rm loop} = \frac{8 \lambda_1 \lambda_2 n_0^2}{(8\pi D)^{d/2}} \frac{t^{2-d/2}}{(2-d)(4-d)} \,.
\end{equation}
Compared to the \textit{tree-level} diagram computed in \eqref{treelevel}, we see that the one-loop contribution effectively adds a term to the $\lambda_1$ coupling constant which scales as $(\lambda_2/D^{d/2}) t^{1-d/2}$. This contribution diverges at late times $t \to \infty$ when the dimensionality of space is lower than a \textit{critical dimension}: $d < d_c = 2$ (while for $d>2$ it diverges as $t \to 0$). At $d = d_c$, the effective coupling goes as $(\lambda_2/D) \ln(Dt)$, which diverges both at small and large $t$'s; it diverges both in the IR and the UV. The UV (short times, short distances) singularities can be cured by introducing a small distance cut-off. This is reasonable, as originally the theory contained a lattice spacing, or particles usually have a finite interaction radius, setting a smallest distance scale in the theory. The IR divergences are more serious. These imply that at late times, the loop contributions start to become increasingly important and hence the naive power series in $\lambda_i$ does not make sense anymore. 

Conversely, for $d>d_c=2$, the corrections from loop diagrams become less important when $t \to \infty$. So in this regime, perturbation theory is applicable and the overall scaling behavior of parameters in the theory are not affected by loop corrections. This implies that only the tree diagrams are sufficient to compute the late time behaviour of the system. Fortunately, the contributions of the tree diagrams can be resumed by realizing the following. If we represent the sum over all tree diagrams as a triangle vertex (and call its contribution $a_{\rm tree}(t)$), then the following relation holds for general $k$:
\begin{equation}
	\begin{array}{c} 
		\begin{tikzpicture}
			\begin{feynman}
				\vertex (a) at (0,0) {};
				\vertex (b) at (1,0) {};
				\diagram*{(b) -- [plain, with arrow=0] (a)};
			\end{feynman} 
		\end{tikzpicture}  
	\end{array}
	=
	\begin{array}{c} 
		\begin{tikzpicture}
			\begin{feynman}
				\vertex (a) at (0,0) {};
				\vertex (b) at (1,0) {};
				\diagram*{(a) -- [plain,insertion=1] (b)};
			\end{feynman} 
		\end{tikzpicture}  
	\end{array}
	+
	\begin{array}{c} 
		\begin{tikzpicture}
			\begin{feynman}
				\vertex (a) at (0,0) {};
				\vertex (b) at (1,0.5) {};
				\vertex (c) at (1,-0.5) {};
				\vertex (l) at (0.5,0);
				\vertex (dots) at (1,0.1) {$\vdots \; k$};
				\diagram*{(a) -- [plain] (l),
					(b) -- [plain, with arrow = 0] (l),
					(c) -- [plain, with arrow = 0] (l) };
			\end{feynman} 
		\end{tikzpicture}  
	\end{array}
\end{equation}
or, expressed in formula:
\begin{equation}
	a_{\rm tree}(t) = n_0 - \lambda_1 \int_0^t dt' \; (a_{\rm tree}(t'))^k
\end{equation}
Taking the time-derivative on both sides and using that $\lambda_1 = (k-\ell)\lambda_0$ (see equation \eqref{lambdamap}), we obtain exactly the rate equation \cref{singlespeciesrateeqn} for the density of $A$ particles. Hence, we have shown that the tree level diagrams lead to the mean-field result. The loop diagrams are responsible for corrections to the mean-field result, which become increasingly important when increasing the length and time scales at or below the critical dimension. Alternatively, we can think of the tree-level result as giving the leading contributions to the large volume (van Kampen) expansion, whereas loop diagram contributions are responsible for higher orders in this expansion \cite{calisto1993omega,thomas2014system}.

\subsubsection{Renormalization group}

The idea of the renormalization group is to identify how physical observables, such as densities, fluctuations or other correlation functions, scale with time, momentum or length scale transformations. To this end, we should first identify the UV singularities, stemming from the short distance (large wave number) contributions to the loop integrals. We can then absorb these UV singularities into a renormalized coupling which will depend explicitly on a \textit{momentum scale} $\kappa$. Associated to this, there is a typical length scale $\kappa^{-1}$ and a typical time scale $t_0 = 1/(D\kappa^2)$. 

The first step is to identify the primitive UV divergences as $\kappa \to \infty$; which components of the diagrams are responsible for the divergences? To figure this out, we can perform a power-counting argument on the vertices. The action has to be dimensionless and as stated above
\begin{equation}
	[x] = \kappa^{-1} \,, \qquad [t] = \kappa^{-2} \,.
\end{equation}
In the continuum limit of the previous subsection we have used the scaling \eqref{fieldscaling}, which imply that
\begin{equation}
	[\bar\varphi] = \kappa^0\,, \qquad [\varphi] = \kappa^d\,.
\end{equation}
All this taken together implies that the diffusive part of the action $\int d^dx\int dt \bar{\varphi}(\partial_t - D \nabla^2) \varphi$ is dimensionless, as it should be. In order for the interaction terms to be dimensionless, the vertices $\lambda_i \bar{\varphi}^i \varphi^k$ impose a scaling of the coupling constant as
\begin{equation}\label{lambdascaling}
	[\lambda_i] = \kappa^{2-(k-1)d} 
\end{equation} 
Now, a loop correction to the vertex  $\lambda_i \bar{\varphi}^i \varphi^k$ would generate a contribution proportional to $\lambda_i \lambda_k$ (see \eqref{loopintegral2} for $i=1, k=2$). In order for this contribution to scale as the tree level contribution $\lambda_i$, the momentum integral over the loop must be of dimension $[\lambda_k]^{-1} = \kappa^{(k-1)d-2}$. Hence, if $(k-1)d-2 >0$, or equivalently: if $d \geq d_c = 2/(k-1)$, the loops will give a contribution which diverges as $\kappa \to \infty$ and the vertex should be renormalized. Conversely, if $d < d_c$, IR singularities appear as $\kappa \to 0$. At the critical dimension $d=d_c$ the loop diagrams carry logarithmic UV and IR divergences.

Using the primitive divergencies, we can renormalize the coupling constants by absorbing the UV singularities. For the reaction $kA \to \ell A$ the procedure is relatively simple, as the whole perturbative series can be resummed \cite{lee1994renormalization}. As noted above, the propagator does not receive any loop corrections, hence this leaves just the vertex functions $\Gamma^{(m,k)}$, representing all diagrams with $k$ incoming and $m$ outgoing legs. Without considering the propagators for the external lines, the vertex function represents the sum over all possible loop diagrams with the specified external lines. Graphically, for $m=1$ and $k=3$, the contributions are:
\begin{align}
	\begin{array}{c} 
		\begin{tikzpicture}
			\begin{feynman}
				\vertex (a) at (0,0) {};
				\vertex (b) at (1,0.35) {};
				\vertex (c) at (1,-0.35) {};
				\vertex (d) at (1,0) {};
				\vertex[circle, draw, fill=black] (l) at (0.5,0) {};
				\diagram*{(a) -- [plain] (l),
					(l) -- [plain](b),
					(l) -- [plain] (c),
					(l) -- [plain] (d) };
			\end{feynman} 
		\end{tikzpicture}  
	\end{array}
	& = 
	\begin{array}{c} 
		\begin{tikzpicture}
			\begin{feynman}
				\vertex (a) at (0,0) {};
				\vertex (b) at (1,0.35) {};
				\vertex (c) at (1,-0.35) {};
				\vertex (d) at (1,0) {};
				\vertex[circle,draw, fill=black] (l) at (0.5,0);
				\diagram*{(a) -- [plain](l),
					(l) -- [plain](b),
					(l) -- [plain] (c),
					(l) -- [plain] (d)};
			\end{feynman} 
		\end{tikzpicture}  
	\end{array}
	+
	\begin{array}{c} 
		\begin{tikzpicture}
			\begin{feynman}
				\vertex (a) at (0,0) {};
				\vertex (b) at (2,0.35) {};
				\vertex (c) at (2,-0.35) {};
				\vertex (d) at (2,0) {};
				\vertex (l) at (0.5,0);
				\vertex (l2) at (1.5,0);
				\diagram*{(a) -- [plain](l),
					(l2) -- [plain](b),
					(l2) -- [plain] (c),
					(l2) -- [plain] (l),
					(l2) -- [plain] (d) };
				\draw[plain] (l) arc [start angle=140, end angle=40, radius=0.65];
				\draw[plain] (l2) arc [start angle=-40, end angle=-140, radius=0.65];
			\end{feynman} 
		\end{tikzpicture}  
	\end{array}  \\  \nonumber
	& \qquad + 
	\begin{array}{c} 
		\begin{tikzpicture}
			\begin{feynman}
				\vertex (a) at (0,0) {};
				\vertex (b) at (3,0.35) {};
				\vertex (c) at (3,-0.35) {};
				\vertex (l) at (0.5,0);
				\vertex (d) at (3,0) {};
				\vertex (l2) at (1.5,0);
				\vertex (l3) at (2.5,0);
				\diagram*{(a) -- [plain](l),
					(l3) -- [plain](b),
					(l3) -- [plain] (c),
					(l) -- [plain] (d) };
				\draw[plain] (l) arc [start angle=140, end angle=40, radius=0.65];
				\draw[plain] (l2) arc [start angle=-40, end angle=-140, radius=0.65];
				\draw[plain] (l2) arc [start angle=140, end angle=40, radius=0.65];
				\draw[plain] (l3) arc [start angle=-40, end angle=-140, radius=0.65];
			\end{feynman} 
		\end{tikzpicture}  
	\end{array} + \ldots
\end{align}
Using the Feynman rules, this diagram can be expressed in terms of a nested set of convolutions of the single loop diagram, which can be solved using the Laplace transform \cite{lee1994renormalization}. 
After some computation, this ultimately leads to the definition of a renormalized coupling constant $g_R = Z_g g_0$ with $g_0 = (\lambda/D)\kappa^{-2 (d_c-d)/d_c}$ and $Z_g^{-1} =  1 + g_0 k! k^{-d/2} (4\pi)^{-d/d_c} \Gamma( \tfrac{d_c-d}{d_c})$, which tells us how the effective reaction rates change as a function of the momentum scale at which we probe the theory. It is a function of the momentum scale $\kappa$ and its associated $\beta$-\textit{function} shows how it changes with momentum:
\begin{align}
	\beta_g(g_R) & = \kappa \frac{d}{d\kappa} g_R  \\ \nonumber
 & = 2 g_R \left[-\frac{\epsilon}{d_c} + B_k \Gamma \left(1+\frac{\epsilon}{d_c} \right) g_R \right]
\end{align}
Here $\epsilon = d_c -d$ and $B_k = k! k^{-d/2} (4\pi)^{-d/d_c} $
This $\beta$-function, computed to all orders in perturbation theory, is finite for $\epsilon \to 0$ when expressed in terms of renormalized quantities. When it vanishes, the theory is manifestly scale invariant, which happens when $g_R = 0$ (the trivial fixed point, corresponding to pure diffusion with no reactions), or at the non-trivial value, when $g_R = g^*_R$ with:
\begin{equation}\label{gRstar}
	g^*_R = [B_k \Gamma(\epsilon/d_c)]^{-1}
\end{equation}
This is of order $\epsilon$. By using the definitions of $g_R$ and $g_0$, together with \eqref{gRstar}, one can expand the bare coupling $g_0$ as: 
\begin{equation}\label{couplinginverse}
	g_0 = \frac{g_R}{1-g_R/g^*_R} = g_R + \frac{g_R^2}{g_R^*} + \ldots
\end{equation}
Using this relation, any perturbative expansion in $g_0$ can be exchanged for one in the renormalized coupling. 
Effectively, this can then be used to turn the perturbative expansion for observables into an expansion in $\epsilon$, which is the small parameter of the theory. This is done through the computation of \textit{Callan-Symanzik} (CS) equations, which determine how the density, density correlations or any other correlation function scales with the momentum scale $\kappa$. The details for this procedure are found in \cite{lee1994renormalization,tauber2005applications}, and we cannot treat it in full glory here. However, the computation goes according to the following rough lines. First, one determines the CS equation of the observable of interest, by requiring that the bare observable is independent of the renormalization scale $\kappa$ and so $\kappa \frac{d}{d\kappa} O(t, n_0,D,\lambda) = 0$. This defines a partial differential equation for the renormalized observable $O(t, n_0,D,\kappa, g_R)$, whose scaling behaviour can be determined via the method of characteristics. This gives characteristic equations for the renormalized coupling $g_R$ and densities $n_0$, which determine how the effective couplings change with time. This \textit{running} of the coupling constant can in this case be solved exactly, showing that at late times, the theory \textit{flows} to the non-trivial fixed point $g^*_R$ and hence becomes scale-invariant when $t\to\infty$. 

To solve the CS equations, one still needs a known value for the observable for some value of the running parameters, which one can compute perturbatively in $g_0$. The CS equations and \eqref{couplinginverse} then allow one to write the perturbative expansion for $g_0$ in terms of an expansion in terms of $\epsilon$. Moreover, as the renormalized running coupling flows to the fixed point $g_R^*$, universal behavior emerges in the asymptotic regime, where for instance the density scales as $n \sim A_k t^{-d/2}$ for $d<d_c$ and as $A_k (\ln t/t)^{1/(k-1)}$ when $d = d_c$ \cite{lee1994renormalization}. These scaling exponents do not depend on the value of $\ell$, nor on initial conditions, and hence they are universal.

\subsection{Applications of the Doi-Peliti path integral}

We close this section with discussing some application of the Doi-Peliti path integral to different types of reactions and universality in the diffusion-influenced reaction systems in more generality. Without being completely exhaustive, we aim to give a balanced overview of the past and present works in this broad field. Early applications of renormalization of the Doi-Peliti path integral focussed mainly on \textit{relaxational models}, where the system shows universal behaviour in its approach to an absorbing state, or on models containing a \textit{dynamical phase transition} between an absorbing state and an active regime. An example of a relaxational model is given by the coagulation reaction $ A + A \to A$. Simulations in $d=2$ and $d=1$ reveal that the density of coagulation clusters scales as $(t/ \log(t))^{-1}$ in $d=2 =d_c$ and as $t^{-1/2}$ in $d=1$ \cite{kang1984fluctuation}. This scaling behavior was confirmed from a renormalization group computation in \cite{peliti1986renormalisation}, moreover the coagulation reaction was shown to belong to the same universality class as the pairwise annihilation $A+ A \to \emptyset$. \cite{lee1994renormalization} generalized this to the single species reactions $k A \to \ell A$ with $k>\ell$, which was shown to be in the same universality class as $k A \to \emptyset$. In this case, $d_c = \frac{2}{k-1}$ (as shown above) and the density scales as $t^{-d/2}$ for $d < d_c$ and as $(\log{t}/ t )^{1/(k-1)}$ for $d = d_c$. \cite{winkler2012validity} studied the validity of the law of mass action (LMA) for the 3D coagulation process using RG methods and report a violation of the LMA in the nonequilibrium force driving the reaction kinetics.

In the bi-molecular version of the coagulation process $A + B \to 0 $ \cite{kang1984scaling,kang1985fluctuation}, or $A + B \to C$ \cite{ovchinnikov1978role,toussaint1983particle}, fluctuations slow the decay rate of the density compared to the mean-field prediction. For $d<4$ the densities scale as $C_d \sqrt{n_0} t^{d/4}$, where $n_0$ is the initial density, assumed to be uniformly distributed \cite{bramson1991spatial}. RG methods were used in \cite{lee1995renormalization} to confirm and extend these results and to determine that the upper critical dimension is in fact $d_c = 2$ when the two species are initially segregated \cite{lee1994scaling}, showing asymptotic sensitivity to initial conditions in this system. Generalizations to multiple ($q$) species pair annihilation were made in  \cite{deloubriere2002multispecies,hilhorst2004segregation}, where it was shown that for $d=1$ the system segregates into single-species domains and the density decays with exponent $\alpha = (q-1)/(2q)$, while for $d\geq2$ and $q>2$ the asymptotic decay is in the single-species pair annihilation universality class.

Universal behavior also occurs near continuous dynamical phase transitions separating an active state (with non-zero density as $t\to \infty$) from an absorbing (inactive) state (see \cite{hinrichsen2000non} for a comprehensive review). 
Many different models exhibit universal scaling relations close to the phase transition belonging to the universality class of \textit{directed percolation} (DP). Examples of models within this class are Schl\"ogl's models \cite{janssen1981nonequilibrium,grassberger1981phase}, kinetic aggregation models such as the Eden model and Diffusion Limited Aggregation (DLA) \cite{parisi1985field,peliti1985field, elderfield1985field}, branching and annihilating random walks with odd offsprings \cite{cardy1998field} or spreading processes such as the contact process \cite{deroulers2004field}, although there are many exceptions and subtleties\footnote{Most notably, the theory of branching and annihilating random walks with even offsprings (see \cite{cardy1996theory,cardy1998field}), tricritical Directed Percolation \cite{grassberger1981phase,ohtsuki1987nonequilibrium,janssen2005survival} and the diffusive epidemic process (see \cite{van1998wilson,tarpin2017nonperturbative,polovnikov2022subdiffusive}), which seems to be related to DP with a conservation law \cite{janssen2016directed}.}. In directed percolation, bonds are placed at random in a regular lattice with probability $p$, which can be traversed in only one direction. When increasing $p$, a percolating cluster appears and the size and correlation lengths of this cluster satisfies power-law scaling near the percolation threshold. The scaling exponents are universal; they are found to be the same for all reaction-diffusion systems within the DP universality class. The critical dimension for DP is $d_c =4$. Field theory methods and the renormalization group has been applied to compute the critical exponents in an expansion in $\epsilon = d_c -d$ in \cite{janssen2005field}. Moreover, the \textit{effective} field theory for the DP universality class is the \textit{Reggeon field theory} \cite{grassberger1979reggeon,cardy1980directed,janssen1981nonequilibrium}. Under the RG group, irrelevant higher-order coupling constants will not contribute to the long-distance behavior of the theory and the effective action becomes equivalent for many different microscopic models.

Beyond the study of universal properties in reaction-diffusion systems, applications of the Doi-Peliti formalism have appeared in various scientific contexts. In cellular biology and biochemistry, the formalism was used to connect microscopic models of cancer cell migration to macroscopic PDEs, which had previously only been validated experimentally \cite{deroulers2009modeling}. Stochastic models of gene expression and gene switches, which lie at the heart of understanding the stability of cell types, have also been formulated in terms of the second quantized stochastic mechanics \cite{sasai2003stochastic}. Building on this identification, in \cite{walczak2005self} a broad range of stochastic gene switch models was solved exactly and \cite{lan2006variational} discussed similar methods for solving stochastic signaling in enzymatic cascades, relevant for cellular signaling networks. \cite{zhang2014stem} generalized these models to explore the effects of DNA binding with protein synthesis and degradation in large genetic networks, leading to approximation methods which determine the most probable transition paths and stochastic switching rates between steady states in the epigenetic landscape, emphasizing the role of fluctuation in stem cell differentiation.

Use cases of the formalism relating to population dynamical models have led to various applications where fluctuations are important drivers in the dynamical system. 
\cite{hernandez2004clustering} show how fluctuations lead to a finite-wavelength instability responsible for cluster formation in diffusive birth-death models (Brownian Bugs).
The role and importance of fluctuations in persistent spatial ecological pattern formation was elucidated using a comparison between mean-field and a Doi-Peliti formulation of stochastic Levin-Segel predator-prey type models \cite{butler2009robust,butler2011fluctuation}. Stochastic Lotka-Volterra predator-prey models were shown to display instabilities which drive the system into a limit cycle \cite{bettelheim2001quantum}. This limit-cycle behaviour was shown to be quasi-stationary and erratic using field theory methods \cite{mobilia2007phase,butler2009predator,tauber2012population}. Furthermore, including spatial structure and stochasticity induces a continuous phase transition for predator extinction, with the critical properties of the directed percolation universality class \cite{mobilia2006fluctuations}. This link between ecosystems at the edge of extinction and the DP universality class has led to a surprising application of stochastic ecological models to describe the onset of turbulance in \cite{shih2016ecological}.

A Doi-Peliti stochastic field theory for neural activity was presented in \cite{buice2007field}, for which the corrections to mean-field lead to a simple version of the Wilson-Cowan equations of interacting excitatory and inhibitory neurons. Furthermore, the model contains a dynamical phase transition of the DP universality class. The scaling laws found in measurements of neocortical activity, in various functional conditions of the brain, are consistent with the existence of DP and related phase transitions at a critical point \cite{buice2009statistical}. Truncated equations describing the higher-order statistics in neural activity were derived for these types of models in \cite{buice2010systematic} and in \cite{bressloff2010stochastic} by utilizing the Van Kampen system-size expansion (see also \cite{buice2013dynamic} for a discussion on finite-size effects).
Field theory methods to describe neural avalanches using branching processes were studied in \cite{garcia2018field} and used in \cite{pausch2020time} to model time-dependent branching leading to oscillating neural avalanches.

Further applications of the mathematical formalism are found in the field of kinetically constrained models (KCM), where restrictions are placed on the allowed hopping transitions of particles on a lattice \cite{ritort2003glassy,schulz1997analytical}. Often these models have trivial (non-interacting) steady-states, but interesting slow dynamics, including a first-order dynamical phase transition between active and inactive phases \cite{garrahan2007dynamical,garrahan2009first}. RG calculations on a field theory for glass-forming liquids derived from a KCM has been shown to belong to the DP universality class for $d>2$ in \cite{whitelam2004dynamic}. 
A statistical field theory approach, equivalent to the Doi-Peliti path integral method, has been used to compute higher moments in the Kuramoto model in \cite{buice2007correlations}. 
RG and field theory method have also successfully been applied to study disordered elastic manifolds, the depinning transition, stochastic avalanches, and sandpiles (see \cite{wiese2022theory} for an extensive review).
RG methods on (concealed) voter models have been used to show these models belong to the DP universality class \cite{garcia2020concealed}.
Applications to models of active matter include non-equilibrium run-and-tumble dynamics, inspired by the motion of bacteria \cite{thompson2011lattice,garcia2021run}, active Ornstein-Uhlenbeck processes \cite{bothe2021doi} and flocking models \cite{scandolo2023active}. An RG analysis for a model of swarming behaviour was performed in \cite{cavagna2023natural}, where they reported the emergence of a novel fixed point and matched critical exponents to experimental data.

Besides the wide range of applications of the formalism, connections with other theoretical approaches have been discussed by various authors. In \cite{droz1994equivalence}, the equivalence between the Doi-Peliti path integral and the Poisson representation of the master equation due to \cite{gardiner1977poisson} was proven. The connection between the Doi-Peliti formalism and the Langevin equation is also well studied (see for example \cite{cardy1996renormalisation}), mainly for bimolecular reactions (of the type $A+A \to \emptyset$). In this case, the Langevin equation for the (complex) field $\phi$ contains an imaginary multiplicative noise term; a peculiarities which has been studied by many authors \cite{howard1997real,munoz1998nature,deloubriere2002imaginary,gredat2011imaginary,benitez2016langevin}. It is, however, important to note that the field $\phi$ is not directly representing the particle densities. An alternative formulation in terms of the density field can be obtained by performing a Cole-Hopf transformation on the fields \cite{tauber2005applications,andreanov2006field,itakura2010two}, and the corresponding path integral connects to the Martin-Siggia-Rose-De Dominicis-Janssen representation for stochastics PDEs \cite{lefevre2007dynamics}. Alternatively, effective stochastic equations of motions (with real valued noise) may be derived by a coarse-graining procedure of the microscopic lattice model \cite{wiese2016coherent}.

Further theoretical advances have been made by formulating fermionic versions of Fock space methods, which may account for systems with particle exclusion principles \cite{rudavets1993phase} (see also \cite{duarte2020fock}). In \cite{greenman_2018} the path integral formalism is generalized to parafermions, to describe stochastic models with finite maximum occupation numbers per site.
Path-integral representations for stochastic hybrid systems, where continuous, deterministically evolving variables are coupled to discrete variables subject to Markovian dynamics, are constructed in \cite{bressloff2014path} and connected to a large deviation action principle. An approach based on dressing the propagator before summing the vertex functions can reduce the number of diagrams in the perturbative expansion which in some cases leads to accurate approximations for the dynamics \cite{harsh2023accurate}.
Equations for the entropy production using a Doi-Peliti formalism were derived in \cite{pruessner2022field,zhang2023entropy} and applied to active particle systems.
Non-equilibrium work relations (the Jarzynski and Crooks relations) were derived using Fock space methods recently in  \cite{baish2024fock}.

\section{The basis-independent representation}
\label{sec:stochMecRD}
In this section, we present a general field theory representation for reaction-diffusion systems, which handles space-dependent reaction rates and generalizes the creation and annihilation operators to act on regions in space. It is constructed by characterizing the copy number representation and the creation and annihilation operators in terms of an arbitrary unspecified basis, following \cite{del2021probabilistic}. By choosing a specific basis (which can also be the basis of a subspace or a degenerate basis), one recovers other field theory formulations as special cases, such the Doi representation (see \cref{sec:deltabasis}), as well as formulations for numerical discretizations (see \cref{sec:RDMEgalerkin,sec:pelitipathint}). Thus, it supplies a unifying mathematical framework to connect different field representations at different scales (\cref{fig:maindiag}), useful for multiscale modeling.   

To introduce the theory, we first generalize the copy number representation from \cref{sec:Fockspace} to continuous space in a basis-independent form. Then, we generalize the creation and annihilation operators, followed by the derivation of the generators for general one species and bimolecular reaction-diffusion processes. Finally, we introduce the Galerkin discretization and extend the path integral formulation to this setting. Additional details are covered in \cite{del2021probabilistic,delRazo2}, although the bra-ket notation is solely covered in detail here.

\subsection{Copy number representation in continuous space}
\label{sec:opendiff}

In \cref{randomwalks}, we introduced a lattice space discretization to handle diffusion, where the copy number representation
\begin{align}
\ket{\mathbf{n}} = \prod_{i=1}^N ( a_i^\dagger)^{n_i} \ket{0},
\label{eq:copynumberLattice}
\end{align}
represents the state with $n_i$ particles in lattice site $i$, where $N$ is the total number of lattice sites. The creation operator $ a_i^\dagger$ adds a particle at $i$-th site in the lattice. 
In the continuum limit, the lattice site converges to a point in space, and the creation and annihilation operators, $a^{\dagger}(x)$ and $a(x)$, create or annihilate a particle at location $x$ \cite{doi1976second}.  
Implicitly, we have chosen a set of basis functions for the creation and annihilation operators: delta functions, which act point-wise. However, we might prefer another basis, for example, one which creates particles according to a specified probability density function, or one which implements a specific spatial discretization. This leads to the following copy number representation for the continuous case: 
\begin{align}
	\ket{\mathbf{n}} = \prod_{\alpha \in S} ( a^\dagger\{u_\alpha\}))^{n_\alpha} \ket{0}, 
	\label{eq:copynumRepcont}
\end{align}
where $\ket{\mathbf{n}}=\ket{n_1,n_2...,n_m,...}$ and $n_\alpha\in\mathbb{N}$ is the number of particles created with basis index $\alpha$. The factor $a^\dagger\{u_\alpha\}$ represents the creation operator for one particle at position $x\in\mathbb{X}$ in a bounded domain $\mathbb{X}$,  with probability density $u_\alpha(x)$. These densities are not arbitrary. The set $\{u_\alpha(x)\}_{\alpha \in S}$ ---where $S$ is a countable set (e.g. $\alpha=1,\dots,\infty$)--- corresponds to a normalized basis for the space of integrable functions $L^1(\mathbb{X})$; the space where probability density functions for the position of a single particle reside in. From now on, if further unspecified, all sums and products over $\alpha$ or $\beta$ are over the set $S$. The representation from \cref{eq:copynumRepcont} depends on the positions $x$ of each particle through the bases $u_\alpha(x)$. 

\begin{figure}
    \centering
    \includegraphics[width=\columnwidth]{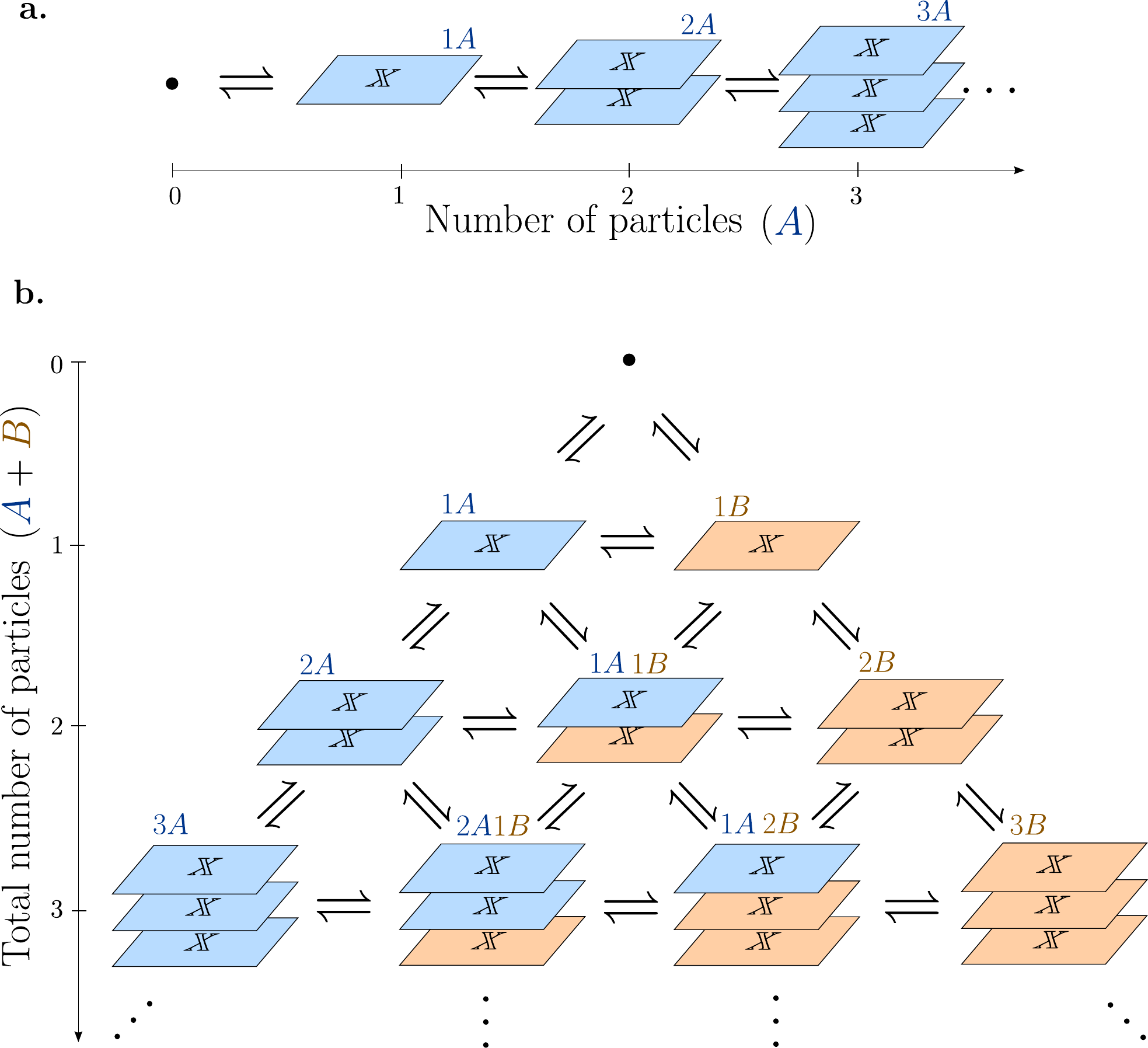}
    \caption{Structure of the phase space of the chemical diffusion master equation. \textbf{a.} Phase space for a system with one chemical species $A$, consisting of discrete sets of continuous diffusion domains $\mathbb{X}$ (e.g. a bounded region within $\mathbb{R}^3$), one for each particle in the current configuration. \textbf{b.} Analogous phase space but for a system with two chemical species $A$ and $B$. The arrows denote possible transitions between first neighbors. Reactions can in general transition between any two subsets in the phase space (not depicted). Figure adapted from \cite{del2024open}.}
    \label{fig:CDMEphasespace}
\end{figure}

The annihilation operators $a\{u_\beta^*\}$ now remove a particle with a rate function $u_\beta^*$---generally a bounded function---where the set $\{u^*_\beta(x)\}_{\beta \in S}$ corresponds to the dual set induced by the basis $\{u_\alpha(x)\}_{\alpha \in S}$, given an orthonormality condition. From now on, when the argument inside the creation or annihilation operator is a basis or its dual set, we will use the compact notation $a^\dagger_{\alpha} = a^\dagger\{u_\alpha\}$ and $a_{\beta} = a\{u_\beta^*\}$.\footnote{Although the creation/annihilation operators could take input from functions in different spaces, they are both linear maps from the Fock space (\cref{eq:fcokspacedef}) onto itself.}

Consider now a system of many stochastic particles of the same species in a finite domain $\mathbb{X}$. The configuration of the system is given by the numbers of particles and their positions. The interactions between particles are capable of not only changing the spatial configuration of the particles but also the number of particles. The probability distribution $\rho$ of such a system is
\begin{align}
	\rho=\left( \rho_0, \rho_1(x^{(1)}), \rho_2(x^{(2)}), \dots, \rho_n(x^{(n)}), \dots \right),
	\label{eq:prodistfamily}
\end{align}
where $\rho_n$ is the probability density of finding $n$ particles at positions $x^{(n)}=(x_1^{(n)},\dots, x_n^{(n)})$ with $x_i^{(n)}$ the position of the $i$th particle in the ensemble of $n$ particles\footnote{In general, these densities depend on time and can be written as $\rho_n(x^{(n)},t)= p_n(x^{(n)},t|N(t)=n) \mathbb{P}[N(t)=n]$, with $p_n$ the conditional density and $\mathbb{P}[N(t)=n]$ the probability of having $N$ particles at time $t$. This implies $\mathbb{P}[N(t)=n]=\int_{\mathbb{X}^n}\rho_n(x^{(n)},t) dx^{(n)}$.}. If the dimension is clear from context, we will use the simplified notation: $\mb{x}=x^{(n)}$ and $x_i = x_i^{(n)}$. We further assume the particles are statistically indistinguishable from each other, so these densities are symmetric in its arguments. The normalization condition is (analogous to \cref{eq:normwellmix}):
\begin{align}
	\rho_0 + \sum_{n=1}^\infty \int_{\mathbb{X}^n} d\mb{x} \, \rho_n(\mb{x}) = 1 \,,
	\label{eq:densitynorma}
\end{align}
with each $\mb{x}\in\mathbb{X}^n$. In general, $\rho$ will depend on time and live in a phase space with a complex structure, as depicted in \cref{fig:CDMEphasespace}. We can express the state of the system, $\ket{\rho}$ as a linear combination of Fock space basis elements $\ket{\mathbf{n}}$, weighted by the probabilities $p_{\mathbf{n}}$ of each configuration
\begin{align} 
	\label{eq:densitybasisexp}
	\ket{\rho} =& \sum_{\mathbf{n}}p_{\mathbf{n}} \ket{\mathbf{n}},
\end{align}
where the sum is over all possible combinations of $\mathbf{n}$. The relation between the copy number representation and the classical $n$ particle densities is given by 
\begin{align}
\ket{\rho_m(\mb{x})} &= \sum_{\mathbf{n}\in \mathbb{M}_m} p_{\mathbf{n}} \ket{\mathbf{n}},
\label{eq:connectiondenscopy}
\end{align} 
where $\displaystyle \mathbb{M}_m=\big\{\mathbf{n}=(n_1,n_2,...):\sum_{\alpha \in S} n_\alpha=m\big\}$
and thus
\begin{align}
	\ket{\rho} = \sum_{m\geq 0}\sum_{\mathbf{n}\in \mathbb{M}_m} p_{\mathbf{n}} \ket{\mathbf{n}}=\sum_{m\geq 0}\ket{\rho_m(\mb{x})},
 \label{eq:ketrhoexpan}
\end{align}
which is equivalent to \cref{eq:densitybasisexp}. 

We can now more precisely define what we mean by Fock space in this context. The densities $\rho_n(x^{(n)})$ reside in a symmetrized $L^1(\mathbb{X})^{\otimes n}$ space, which we denote by $W_{n}$. Then, the Fock space is:
\begin{align}
F:=\bigoplus_{n=0}^\infty W_n=\{\rho=(\rho_0,\rho_1,\dots):\rho_n \in W_n \, | \forall n \},
\label{eq:fcokspacedef}
\end{align}
In bra-ket notation, the direct sum can be understood as a sum over the standard basis $\ket{e_m}=(0, \dots, 0, 1 , 0, \dots)$ ($1$ in the $m$th entry and zero elsewhere): $\ket{\rho} \in F$ can be expanded as $\ket{\rho}=\sum_n \rho_n \ket{e_n}$. The distributions $\rho$ are assumed normalized following \cref{eq:densitynorma}. 
In this setting $F$ is not necessarily a Hilbert space, but it still holds similar properties as formally shown in \cite{del2021probabilistic}. It has been suggested this can be rigorously addressed using measure theory approaches \cite{kolokoltsov2010nonlinear} or rigged Hilbert spaces \cite{grassberger1980fock}. 

A rigged Hilbert space (or Gelfand triple) extends Hilbert spaces to include generalized functions like distributions  \cite{bohm1989dirac,gadella2002unified,de2005role}. It consists of three nested spaces, written as $\Phi(\mathbb{X}) \subset L^2(\mathbb{X}) \subset \Phi^*(\mathbb{X})$ for the one-particle space $\mathbb{X}$. Although tempting, it is not possible to form a Gelfland triple rigorously with $\Phi$ and $\Phi^*$ given by $L^\infty$ and $L^1$.
A possible solution is to set $\Phi$ as the Schwartz space of rapidly decreasing smooth functions (``nice'' test functions including probability densities) and $\Phi^*$ as the space of tempered distributions. 
By constructing the Fock space in terms of such Gelfand triple, one could rigorously define the spectral properties of unbounded operators and extend the formalism to accommodate generalized eigenstates.
However, to the best of our knowledge, this has not been addressed in detail in the context of Fock space for reaction-diffusion, and thus is beyond the scope of this review. 

\subsection{Creation and annihilation operators}
\label{sec:generalbasisCA}

In general, the creation operator adds a particle with probability density function $w:=w(x)$, and the annihilation operator removes a particle with a rate function $f:=f(x)$. For convenience, we express explicitly the action of the operators on $\ket{\rho_n}$ here\footnote{To write them in terms of probability densities (not kets), we apply the standard basis: $\ket{\rho_m(\mb{x})} = \rho_m(\mb{x})\ket{e_m}$}: 
\begin{align}
\label{eqs:apamdensities}
    a^\dagger \{w\} \ket{\rho_n} &= \frac{1}{n+1}\sum_{j=1}^{n+1} w(x_j) \ \ket{\rho_n(\mb{x}_{\setminus \{j\}})} \,, \quad\mb{x}\in\mathbb{X}^{n+1},\\ 
    a \{ f\} \ket{\rho_n} &= n \int_\mathbb{X} f(y) \ \ket{\rho_n(y, \mb{x})} \ dy \,, \quad \mb{x}\in\mathbb{X}^{n-1}, \nonumber
\end{align}
where $x_j$ correspond to the $j$th entry of the tuple $\mb{x}$ and $\mb{x}_{\setminus \{j\}}$ means that the entry with index $j$ is excluded. Here $w$ is an arbitrary probability density for one particle and $f$ is an arbitrary rate function; they are not necessarily elements of the basis and its dual. In particular, when applied to basis elements of the copy number representation $\ket{\mathbf{n}}$ (or their duals), we obtain them in their basis-independent form (analogous to \cref{removing_ni,adding_ni}):
\begin{align}
\begin{aligned}
    a^\dagger_{\alpha'} \ket{\mathbf{n}} &= \prod_{\alpha} (a^\dagger_\alpha)^{n_\alpha + \delta_{\alpha\alpha'}} \ \ket{0}, \\
	a_{\beta} \ket{\mathbf{n}} &= n_\beta \prod_{\alpha} (a^\dagger_\alpha)^{n_\alpha - \delta_{\alpha\beta}} \ \ket{0}.
\end{aligned}
\label{eq:a+a-kets}  
\end{align} 
Naturally if $n_\beta=0$, then $a_{\beta} \ket{\mathbf{n}}=0$. 
These creation and annihilation operators satisfy the following commutation relations, which is a generalization of \cref{commutators}:
\begin{align}
\begin{aligned}
	\left[a \{ f\} , a^\dagger \{ w\}\right] &= \langle f,w \rangle \mathcal{I}\,, \\
	\left[a^\dagger \{ w\} , a^\dagger \{ \nu\}\right] &= 
	\Big[a \{ f\} , a \{ g\}\Big] = 0 \,.
\end{aligned}
\label{eq:commutrelRD}
\end{align}
Here $\mathcal{I}$ is the identity operator, and the dual pairing (a generalization of the inner product) is defined as:
\begin{align}
    \langle f,w \rangle = \int_{\mathbb{X}} f(x)w(x) dx \,.
    \label{eq:dualpairing}
\end{align}
This definition is analogous for components in higher dimensions (integral over $\mathbb{X}^n$) or no integral if $n=0$. Through this pairing, we can obtain the dual set of the basis. For the basis set $\{u_\alpha (x)\}$, its dual set $\{u_\beta^* (x)\}$ satisfies the following orthonormality condition: 
\begin{align}
\langle u_b^*,u_\alpha\rangle =\delta_{\alpha, \beta} \, .
\label{eq:basisortho}
\end{align}
Together with the first commutation relation from \cref{eq:commutrelRD}, this relation yields the commutator in terms of basis elements $\left[a_\beta , a^\dagger_\alpha \right] = \delta_{\alpha, \beta}$. Moreover, similarly to \cref{vacuum,vacuum2}, the vacuum state satisfies the following relations:
\begin{align}
	a \{ f\}\ket{0}=0\,, \ \quad \bra{0}a^\dagger \{ w\} =0\,, \ \quad \ \bracket{0}{0} = 1\,,
	\label{eq:vacuumRD}
\end{align}
which also apply in the special case where $f$ is $u_\alpha$ and $w$ is $u_\beta^*$. We can now generalize the copy number representation from \cref{eq:copynumRepcont} to include its dual, as in \eqref{vacuum2}:
\begin{align}
	\ket{\mathbf{n}} = \prod_{\alpha} (a^\dagger_\alpha)^{n_\alpha} \ket{0} ,\quad 
	\bra{\mathbf{n}} = \bra{0} \frac{1}{\mb{n}!}\prod_{\beta}  ( a_\beta)^{n_\beta}
	\label{eq:cprepstates}  
\end{align}
with $\mb{n}!=\prod_{\beta}n_\beta !$. One can intuitively understand the $\mb{n}!$ as a normalizing factor to take into account all the equivalent ways of removing $n$ particles of the same species from a symmetric density. Combining \cref{eq:cprepstates} with the properties from \cref{eq:commutrelRD,eq:vacuumRD} and the orthogonality of the basis, we can show the orthogonality of the Fock space basis in terms of bras and kets
\begin{align}
	\bracket{\mathbf{n}}{\mathbf{m}} = \prod_\alpha\ \delta_{m_\alpha, n_\alpha} =  \delta_{\mb{n},\mb{m}}.
        \label{eq:orthonormalityRD}
\end{align}

The Poisson states \cref{coherentdef} and their duals \cref{dualcoherentdef} are
\begin{subequations}\label{eq:coherentstatesBasis}
    \begin{align}
	\ket{\mathbf{z}} & = \exp \left(\sum_\alpha z_\alpha a^\dagger_\alpha  \right) \ket{0}\,, \\
	\bra{\mathbf{z}} & = \bra{0} \exp \left(\sum_\beta z_\beta a_\beta \right)\,,
    \end{align}
\end{subequations}
such that $a_{\beta} \ket{\mathbf{z}} = z_\beta \ket{\mathbf{z}}$ and $\bra{\mathbf{z}} a^\dagger_\alpha = z_\alpha \bra{\mathbf{z}}$, i.e. they are eigenstates of the creation and annihilation operators, respectively. This can be shown by direct substitution and use of the commutation relations \cref{eq:commutrelRD}.

In practical computations, we can again use the Poisson state $\bra{\mb{1}}$ to calculate expectations; it corresponds to $z_\alpha=1$ for all $\alpha$'s: $\bra{\mb{1}} = \bra{0} \exp \left(\sum_\alpha a_\alpha  \right)$. Other properties are also analogous to the expressions for the well-mixed case, like the number operators
\begin{align}
	\mathcal{N}_\alpha = a^\dagger_\alpha a_\alpha \qquad \mathcal{N} = \sum_\alpha a^\dagger_\alpha a_\alpha,
	\label{eq:numpartops_basis}
\end{align}
such that $\mathcal{N}_\alpha \ket{\rho} = n_\alpha \ket{\rho}$ and $\mathcal{N} \ket{\rho} = \left(\sum_\alpha n_\alpha \right)\ket{\rho}$. One can further obtain the generating functions analogously to \cref{sec:wellm-genfuncs}.

\subsection{The chemical diffusion master equation}
The chemical diffusion master equation (CDME) is the master equation for reaction-diffusion processes.\footnote{We use this name since the name reaction-diffusion master equation is normally reserved for discrete space settings (see \cref{fig:maindiag}).}
The CDME is the analog of \cref{mastereqn} but with continuous spatial dependence. To construct it, we write the generator of the underlying reaction-diffusion process as the sum of operators acting on the Fock space elements. For a system with one chemical species and $R$ reactions, the generator is
\begin{align}
	H = H_0(D) + \sum_{r=1}^R H_r(\lambda_r),
	\label{eq:CDMEgeneral}
\end{align}
where $H_0(D)$ is the diffusion generator and $H_r(\lambda_r)$ are the reaction generators of the $r$-th reaction with reaction rate function $\lambda_r$. Using this, we can write the CDME as
\begin{align}
	\partial_t \ket{\rho} = H\ket{\rho}.
\end{align}
The equation can be written out explicitly in a more standard form, as shown in \cref{sec:formdensities} and \cite{del2021probabilistic,delRazo2,doi1976second}. Partial or full versions of the CDME have been presented in previous works \cite{doi1976second,isaacson2022mean,isaacson2018unstructured}. The underlying process and the CDME itself have also been referred to as the Doi model, and in special cases, the volume reactivity or $\lambda-\rho$ model \cite{erban2009stochastic,smith2019spatial}. 

Some operators in \cref{eq:CDMEgeneral} conserve the number of particles of each species, such as diffusion generators, while others change the particle numbers, like reaction generators. Nonetheless, all operators are expressible in terms of creation and annihilation operators, as we will show shortly. In particular, we show how to expand the diffusion generator and the reaction generator for a general one species reaction and how it extends to reactions with multiple species.

\subsubsection{Particle number conserving operators}
\label{sec:partconservops}
Lets focus first on operators that do not change the number of particles. Consider an operator $A$ acting on a single particle ($A:L^1(\mathbb{X}) \to L^1(\mathbb{X})$); the action of this operator on the whole system (i.e. on the Fock space) can be expressed as an operator $F_A$, which has a representation in terms of creation and annihilation operators: 
\begin{align}
	F_A= \sum_{\alpha, \beta} \langle u_\alpha^*,A u_\beta\rangle a^\dagger_\alpha a_\beta.
	\label{eq:consPartExp}
\end{align}
For instance, the diffusion generator $D\nabla^2$ doesn't change the number of particles and acts on a single particle. Its action on the whole Fock space is expressed as the operator
\begin{align}
	H_0(D) = \sum_{\alpha, \beta} \langle u_\alpha^*,D\nabla^2 u_\beta\rangle a^\dagger_\alpha a_\beta,
	\label{eq:diffExp}
\end{align}
corresponding to the general form of \cref{eq:consPartExp}.

To intuitively understand this, we can think of a lattice discretization. Each term in \cref{eq:diffExp} represents the action of removing one particle from site $\beta$ and adding it into site $\alpha$ in the lattice. The quantity $\langle u_\alpha^*,\nabla^2 u_\beta\rangle$ denotes how much probability should be transferred between those voxels, and it will only be non-zero between neighboring voxels. To obtain the action of the operator on the whole system, we must sum over all possible site combinations $(\alpha,\beta)$.

We can also have operators $B$ that act on two particles at a time  ($B:L^1(\mathbb{X})^{\otimes 2} \to L^1(\mathbb{X})^{\otimes 2}$), such as diffusion with pair interactions through a potential. Assuming $B$ is symmetric under particle permutations, then its action in the whole system is:
\begin{align}
	F_B= \sum_{\aalpha, \bbeta \in S^2} \langle u^*_{\aalpha},B u_{\bbeta} \rangle a^\dagger_{\aalpha} a_{\bbeta},
	\label{eq:cons2PartExp}
\end{align}
with $u_{\bbeta}(\mb{x})=u_{\beta_1}(x_1)u_{\beta_2}(x_2)$, $u^*_{\aalpha}(\mb{x})=u^*_{\alpha_1}(x_1) u^*_{\alpha_2}(x_2)$ compact notation for the corresponding basis and its dual set;
$a^\dagger_{\aalpha}=a^\dagger_{\alpha_1}a^\dagger_{\alpha_2}$ and $a_{\bbeta}=a_{\beta_1}a_{\beta_2}$. We can use analogous expansions for operators $Z:L^1(\mathbb{X})^{\otimes m} \to L^1(\mathbb{X})^{\otimes m}$ acting on $m$ permutation symmetric particles at a time
\begin{align}
	F_Z= \sum_{\aalpha, \bbeta\in S^m} \langle u^*_{\aalpha},Z u_{\bbeta} \rangle a^\dagger_{\aalpha} a_{\bbeta},
	\label{eq:consmPartExp}
\end{align}	
where we generalized the compact notation to $\aalpha = (\alpha_1,...,\alpha_m) \in S^m$, $\bbeta = (\beta_1,...,\beta_m) \in S^m$ and $\mb{x} = (x_1,...,x_m) \in \mathbb{X}^m$ with 
\begin{align}
    \label{pre_this_notation}
    \begin{aligned}
    u_{\bbeta}(\mb{x}) &= u_{\beta_1}(x_1)  \cdots  u_{\beta_m}(x_m), \\ u^*_{\aalpha}(\mb{x}) &= u^*_{\alpha_1}(x_1)  \cdots  u^*_{\alpha_m}(x_m).
    \end{aligned}
\end{align}
and	$a^\dagger_{\aalpha}=a^\dagger_{\alpha_1}\cdots a^\dagger_{\alpha_n}$ and $a_{\bbeta} = a_{\beta_1}\cdots a_{\beta_m}$.

\subsubsection{Reaction generators}
\label{sec:reactopers}
Reaction events will in general change the number of particles. The reaction generator $H(\lambda)$ with reaction rate function $\lambda$ is split into two operators: 
\begin{align}\label{eq:reactiongenerator}
	H(\lambda)=\mathcal{G}-\mathcal{L}.
\end{align}
Analogous to the two terms in \cref{generalH}, the gain $\mathcal{G}$ quantifies how much probability is being ``gained'' by the current state from another state due to the reaction. Similarly, the loss quantifies how much probability is being ``lost'' by the current state, as it is required to ensure that total probability is conserved. The gain operator does not conserve particle numbers, while the loss does and thus belongs to the class of operators \eqref{eq:consPartExp}. To simplify the presentation, lets focus first on examples involving only one chemical species.

\paragraph{General one species reaction}
\label{sec:gen1SpecsBasInden}
Consider again the single species reaction $ kA \rightarrow \ell A$ of \eqref{singlespeciesreaction}, now with space-dependent reaction rate function $ \lambda (\mb{y};\mb{x})$, where $\mb{x}\in\mathbb{X}^k$ and $\mb{y}\in\mathbb{X}^\ell$ are the positions of the reactants and products respectively. As the particles are treated as indistinguishable, the reaction rate function is symmetric with respect to permutations of the reactants positions, as well as permutations of the products positions. 
To construct the reaction generator, we first write the loss and gain operators per reaction: $L:L^1(\mathbb{X})^{\otimes k} \to L^1(\mathbb{X})^{\otimes k}$ and $G:L^1(\mathbb{X})^{\otimes k} \to L^1(\mathbb{X})^{\otimes \ell}$, which act on $k$ particles at a time,
\begin{align} 
\begin{aligned}
    \bigl(L u_{\bbeta} \bigr)(\mb{x}) :=& u_{\bbeta} (\mb{x}) \int_{\mathbb{X}^\ell} d \mb{y} \; \lambda(\mb{y};\mb{x}) , \\
    \bigl(G u_{\bbeta} \bigr)(\mb{y}) :=& \int_{\mathbb{X}^k}  d\mb{x} \; \lambda(\mb{y};\mb{x}) u_{\bbeta}(\mb{x}),
\end{aligned}
\label{eq:propOperators_kl}
\end{align}
where $\bbeta \in S^k$, so $u_{\bbeta}$ corresponds to a $k$ particle space basis function. The loss and the gain are both probability fluxes, given by the product of the reaction rate function times a probability density of the reactants, $u_{\bbeta}$.
As the loss represents the loss of probability due to one reaction from the current state to any other state, it is integrated over all product positions, and it depends only on the positions of the reactants. The gain function signifies the probability gain due to one reaction into the current state and is hence integrated over the reactants positions, leaving a function of only the products locations. The total loss and gain operators over all possible reactions in the system is obtained in terms of creation and annihilation operators
\begin{align}
\begin{aligned}
    \mathcal{L} &= \frac{1}{k!} \sum_{\substack{\aalpha \in S^k \\ \bbeta \in S^k}} \langle u_{\aalpha}^*, L u_{\bbeta} \rangle a_{\aalpha}^\dagger a_{\bbeta} \ ,  \\
    \mathcal{G}  &= \frac{1}{k!} \sum_{\substack{\aalpha \in S^\ell \\ \bbeta \in S^k}} \langle u_{\aalpha}^*, G u_{\bbeta} \rangle a_{\aalpha}^\dagger a_{\bbeta} \ ,
\end{aligned}
\label{eq:fockpropOperators_kl}
\end{align}
where these operators now act on the Fock space. The reaction generator is $H(\lambda) = \mathcal{G}- \mathcal{L}$; the diagram in \cref{fig:genbasiindependiag} illustrates its construction. The gain is understood as removing $k$ particles and adding $\ell$ particles, weighted by the reaction rate function and summing over all possible combinations. If applied to $\ket{\rho_{n}}$, it yields the probability gain into the $(n-k+\ell)$--state. The loss applied to $\ket{\rho_n}$ quantifies how the probability to remain in the current state decreases, ensuring total probability is conserved. 

\begin{figure}
	\centering
	\includegraphics[width=\columnwidth]{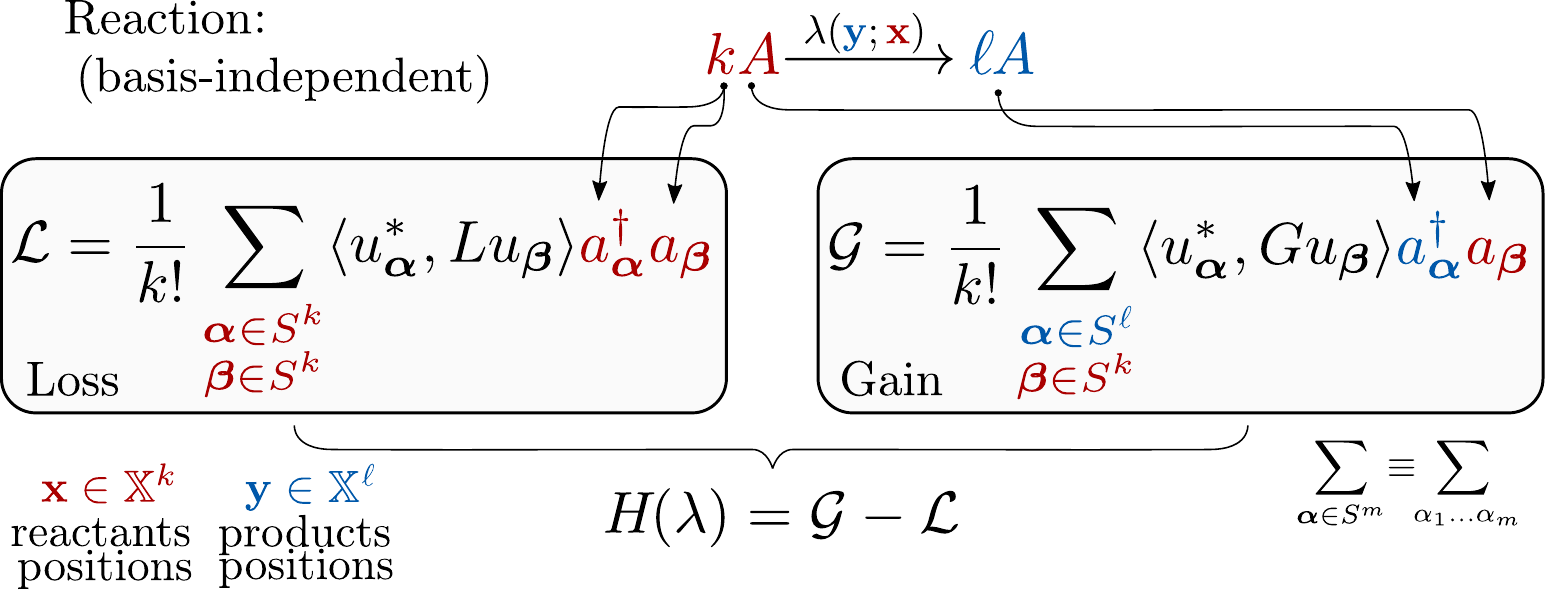}
	\caption{Diagram illustrating how to construct the generator for a system with a general one species reaction with reaction rate function $\lambda (\mb{y};\mb{x})$ for the basis-independent case. This case takes into account the spatial dynamics in continuous space, so it generalizes the diagram presented for the well-mixed case, (\cref{fig:genwellmixdiag}).
	\label{fig:genbasiindependiag} }	
\end{figure}

\paragraph{Degradation and creation reactions}
As special cases, we write the generators for degradation and creation reactions. Consider first a degradation reaction $A\rightarrow \emptyset$ with reaction rate function $\lambda_d(x)$. The loss and gain per reaction are special cases of \cref{eq:propOperators_kl}
\begin{align}
\begin{aligned}
	\left(L_d u \right) (x) & = u(x)\lambda_d(x), \\
	\left(G_d u \right) & = \int_\mathbb{X} u(y)\lambda_d(y) dy.
 \end{aligned}
\end{align}
Analogously, the loss and gain operators acting on the Fock space are special cases of \cref{eq:fockpropOperators_kl}, and thus the reaction generator due to this reaction is given by 
\begin{align}
	H_d(\lambda_d) = \sum_{\beta} \langle 1, G_d  u_\beta \rangle a_\beta - \sum_{\alpha, \beta} \langle u_\alpha^*,L_d  u_\beta \rangle a^\dagger_\alpha a_\beta .
	\label{eq:basisHd}
\end{align}
where $\langle 1, G_d  u_\beta \rangle =G_d  u_\beta$. Analogously, we can consider the creation reaction $\emptyset \rightarrow A$ with spatially-dependent rate $\lambda_c(x)$. The loss and gain operators per reaction are
\begin{align}
	L_c = \int_{\mathbb{X}} \lambda_c(y)dy, \qquad G_c = \lambda_c(y), 
\end{align}
such that the generator for the creation reaction is:
\begin{align}
	H_c(\lambda_c) = \sum_{\alpha} \langle u_\alpha^*, G_c 1 \rangle a^\dagger_\alpha - L_c.
    \label{eq:basisHc}
\end{align}

\subsection{Extension to multiple species}
Extending this formalism to systems with multiple chemical species is straightforward, but the notation can become cumbersome. Thus, instead of presenting the general case, we focus on a simple bimolecular reaction
\begin{equation}
	A+B\rightarrow C,
\end{equation} 
with reaction rate function $\lambda(y_c; x_a,x_b)$, where $x_a$ and $x_b$ are the locations of the reactants and $y_c$ is the location of the product. The key to write the generator is to use different creation and annihilation operators for each species.
Analogously to \cref{eq:propOperators_kl}, the loss and gain operators per reaction are given by
\begin{align}
	\big( L u_{\bbeta}\big)(x_a,x_b) &:= u_{\bbeta}(x_a,x_b) \int_\mathbb{X}\lambda(y_c;x_a,x_b)dy_c,
	\\
	\big(G  u_{\bbeta}\big)(y_c) &:= \int_{\mathbb{X}^2} u_{\bbeta}(x_a,x_b) \lambda(y_c;x_a,x_b)dx_a dx_b, \nonumber
\end{align}
with $u_{\bbeta}=u_{\beta_1}u_{\beta_2}$ to shorten notation. Analogous to \cref{eq:fockpropOperators_kl}, the total loss and gain terms are:
\begin{align}
    \mathcal{L} &= \frac{1}{2}
	\sum_{\substack{\alpha_1,\alpha_2\\ \beta_1,\beta_2 }} \left\langle u_{\alpha_1}^* u_{\alpha_2}^*, L u_{\bbeta} \right\rangle  a^\dagger_{\alpha_1} b^\dagger_{\alpha_2} a_{\beta_1} b_{\beta_2}, \notag\\
	\mathcal{G} &=\frac{1}{2}
	\sum_{\substack{\alpha_1\\ \beta_1,\beta_2 }} \left\langle u_{\alpha_1}^*, G  u_{\bbeta}\right\rangle c^\dagger_{\alpha_1} a_{\beta_1} b_{\beta_2},
\end{align} 
where $a^\dagger,b^\dagger,c^\dagger,a,b,c$ correspond to creation and annihilation of particles of type $A,B$ and $C$ correspondingly. Operators from different species commute, while those of like species follow  commutation relations (\cref{eq:commutrelRD}). The generator includes a diffusion term of each chemical species and the reaction generator $H(\lambda)$ is formed as \eqref{eq:reactiongenerator}
\begin{align}
	H = H_0(D_A) + H_0(D_B) + H_0(D_C) + H(\lambda)\,.
\end{align}
Here $D_A$, $D_B$ and $D_C$ are the corresponding diffusion coefficients. This case and others were studied in detail in \cite{delRazo2}, where it is also shown how to recover the standard integro-differential form as presented in full or partial form in the works \cite{doi1976second,erban2009stochastic,isaacson2022mean,isaacson2018unstructured,smith2019spatial}.

\subsection{Galerkin discretization: indicator functions as basis} \label{sec:RDMEgalerkin}

In this section, we apply a Galerkin discretization to the basis-independent formulation, recovering the lattice representation used in \cref{sec:stochMecRD}. We refer to this, and the generalization to arbitrary discretizations, as \textit{second quantized Galerkin discretizations} in \cref{fig:maindiag}. 
Galerkin discretizations reduce dimensionality to simplify the problem, while also serving three key purposes in this setting: to recover other field theory representations as special cases; to enable efficient numerical schemes that remain convergent regardless of the discretization; and to apply field theory methods at the discretization level, as we will show in \cref{sec:pelitipathint}.

To exemplify the discretization process, we apply a piecewise constant approximation. Consider a partition of the full space $\mathbb{X}$ with $M$ non-overlapping subsets denoted by $\mathbb{X}_i$ and $i=1,\dots,M$ (in a lattice, each $\mathbb{X}_i$ corresponds to a voxel). The indicator function $\mathbb{1}_{\mathbb{X}_i}(x)$ has value $1$ if $x\in\mathbb{X}_i$ and $0$ elsewhere. We use the normalized indicator functions as a basis of the discretized space
\begin{align}
\label{eq:ind_func_def}
	\xi_i=\frac{\mathbb{1}_{\mathbb{X}_i}}{||\mathbb{1}_{\mathbb{X}_i}||},
\end{align}
where $||\mathbb{1}_{\mathbb{X}_i}||$ is the volume of $\mathbb{X}_i$ and thus, its integral is always one. We denote $\xi_j^*$ as the dual of the basis $\xi_j$, which corresponds to $\xi_j^* = \mathbb{1}_{\mathbb{X}_j}$, so the orthonormality condition is satisfied $\langle \xi_i^*,\xi_j\rangle=\delta_{ij}$.

Based on \cref{sec:generalbasisCA}, we obtain the creation and annihilation operators as $a_i^\dagger=a^\dagger\{\xi_i\}$ and $a_j=a\{ \xi_j^*\}$, which correspond to creating or annihilating a particle in the region $\mathbb{X}_i$. The copy number representation follows from \cref{eq:cprepstates}, with the main difference that the basis is now finite:
\begin{align}
		\ket{\mathbf{n}} :&= \ket{n_1,...,n_M} =(a^\dagger_1)^{n_1} ... (a^\dagger_M)^{n_M} \ket{0}\,,
\end{align}
where there are $n_i$ particles in region $\mathbb{X}_i$. 
The action of the operators on these states follows from \cref{eq:a+a-kets}
\begin{align}
	\begin{split}
		a^\dagger_i\ket{\mathbf{n}} &= \ket{n_1,..,n_i+1,..,n_M}\,, \\
		a_j\ket{\mathbf{n}} &= n_j \ \ket{n_1,..,n_j-1,..,n_M} \,.
	\end{split}
	\label{eq:a+a-kets_discrete}
\end{align}  
These expressions are similar to \cref{adding_ni,removing_ni}, with the difference that here we do not distinguish the species, but its location in discretized space.
An expression for the discretized density $\ket{\hat{\rho}}$ is analogous to \cref{eq:densitybasisexp}
\begin{align}
	\ket{\hat{\rho}} = \sum_{\mathbf{n}} p_\mathbf{n} \ket{\mathbf{n}},
    \label{eq:rhofockGalerk}
\end{align}
where $p_\mathbf{n}$ is the probability of configuration $\mathbf{n}=(n_1,...,n_M)$.
Next, we want to discretize the generators that act on the density by projecting into the discretized subspace. We start with the diffusion generator. Using \cref{eq:diffExp}, we substitute the basis $u_\alpha$ with the basis of the subspace $\xi_i$ and similarly for $u_\beta^*$
\begin{align}
	H_0(D) = \sum_{i, j=1}^M D_{ij} a^\dagger_i a_j\,, 
\label{eq:diffExpdiscrete}
\end{align}
with $D_{ij}=\langle \xi_i^*, D\nabla^2  \xi_j\rangle$. 
Lets calculate this explicitly in one dimension, such that $||\mathbb{1}_{\mathbb{X}_i}||=h$, and apply it to a test function,  $\phi(x)$. For a fixed $i$, the sum over $j$ of these terms yields:
\begin{align}
	\sum_{j=1}^M  D_{ij}\phi(x) & = \sum_{j=1}^M \langle \xi_i^*, D \partial_x ^2  (\xi_j \phi(x))\rangle \nonumber \\
	& \approx \frac{D}{h^2}\left(\phi_{i+1} + \phi_{i-1}  - 2\phi_{i} \right) , \label{lapdiscrete}
\end{align}
where we approximated the derivatives with a piecewise constant discretization of the test function. Thus, in a 1D lattice, the coefficients will be $D_{ii}=-2D/h^2$, $D_{ij}=D/h^2$ for $j=i\pm 1$ and zero otherwise, similar to \cref{rwmaster}.  
Using this in \cref{eq:diffExpdiscrete}, one obtains the discretized diffusion generator of \cref{diffusiongenerator}.
This can be extended to higher dimensions and non-lattice discretizations. In general, $D_{ij}$ will be non-zero only for $i,j$ nearest neighbors and $D_{ii}=-\sum_{i\neq j}D_{ji}$, recovering the graph Laplacian of the discretization. 

Applying an analogous approach to the generators for degradation and creation reactions from \cref{eq:basisHd,eq:basisHc}, we obtain the discretized generators:
\begin{align}
H_d(\lambda_d)  &=\sum_{i=1}^M \lambda_d^i \left( 1 - a_i^{\dagger} \right)a_i\,, \\
H_c(\lambda_c) &=\sum_{i=1}^M \lambda_c^i \left( 
	a^{\dagger}_i - 1 \right) \,,
\end{align}
with $\lambda_d^i = \int_{\mathbb{X}_i} \lambda_d(x) \xi_i(x)dx$ and $\lambda_c^i =\int_{\mathbb{X}_i} \xi_i^*(x)\lambda_c(x) dx$.
The master equation for discretized reaction-diffusion process with creation and degradation reactions is then
\begin{align}
	\partial_t \ket{\hat{\rho}} = \left(H_0(D) + H_d(\lambda_d)  + H_c(\lambda_c) \right) \ket{\hat{\rho}}.
    \label{eq:theRDME}
\end{align}
To obtain this equation in terms of $p_{\mathbf{m}}$ explicitly, we multiply by $\bra{\mathbf{m}}$ on the left. From \cref{eq:orthonormalityRD,eq:rhofockGalerk}, we recognize $p_{\mathbf{m}}=\bra{\mathbf{m}} \hat{\rho} \rangle$, and we can calculate all the terms explicitly using \cref{eq:a+a-kets_discrete},
\begin{align}
\begin{split}
\partial_t & p_\mathbf{m} = \sum_{\substack{{i,j=1}\\i\neq j}}^M \left( D_{ij}  (m_j+1) p_{\mathbf{m}_{i-1,j+1}}   -  D_{ji} m_i p_\mathbf{m}  \right) \ +\\
&\sum_{i=1}^M \left[\lambda_d^i \left( 
(m_i+1)p_{\mathbf{m}_{i+1}} - m_i p_\mathbf{m}\right) 
+ \lambda_c^i \left( 
 p_{\mathbf{m}_{i-1}} - p_\mathbf{m}\right)\right],
\end{split}
   \label{eq:theRDMEexplicit}
\end{align}
where we used $p_{\mathbf{m}_{i\pm1}}:=p_{m_1,..,m_i\pm1,..,m_M}$ for compact notation. We refer to this type of equation as the reaction-diffusion master equation (RDME). In this work, the term RDME (\cref{fig:maindiag}) refers to any lattice-based or spatially discrete approximation of the CDME, such as \cref{eq:theRDMEexplicit} (or \cref{eq:theRDME} in the second quantized representation).

For bimolecular/non-linear reactions, it is well-known that phenomenological formulations of the RDME do not converge in the microscopic limit\footnote{In the literature, RDME often refers to these phenomenological/non-convergent formulations, while convergent ones are specifically referred to as convergent RDMEs.} (e.g. lattice spacing $\rightarrow$ 0), as discussed by \cite{fange2010stochastic,hellander2012reaction,isaacson2008relationship,isaacson2009reaction,isaacson2009reaction2}. The lack of convergence is due to the reaction rate function depending on the interparticle distance, resulting in the loss of non-linear reactions. This was first proved in \cite{isaacson2009reaction}, and together with \cite{hellander2012reaction}, these works identified the conditions under which the RDME more closely agrees with microscale particle-based models. However, it was the works \cite{isaacson2013convergent, isaacson2018unstructured} that first established a truly convergent RDME in the microscopic limit.
A huge advantage of the approach presented in this section is that one can systematically discretize complex reaction generators, while assuring convergent results and consistency between different discretizations (lattice/voxel size), including bimolecular reactions \cite{del2021probabilistic}. The resulting equations are also suitable for numerical computations, as implemented in many software packages (e.g. \cite{bartmanski2020stospa2,drawert2012urdme}). 

The field theory representation for the well-mixed case from \cref{sec:Fockspace} is recovered when choosing a basis covering the whole domain $\mathbb{X}$, e.g. $\mathbb{X}/||\mathbb{X}||$ (see \cref{fig:maindiag}). However, for non-linear reactions, the effective rates in the well-mixed theory do not match the spatial dependent rates. These relations are not trivial due to diffusion-influenced phenomena, see \cite{del2024open, doi1976stochastic} for a discussion on this topic.

Galerkin discretizations are not limited to lattices nor to this choice of basis functions. For instance, \cite{engblom2009simulation} pioneered deriving transport operators for the RDME in unstructured meshes and \cite{isaacson2018unstructured} extended it to convergent RDMEs. Such and other discretizations are also possible within this framework, including basis functions for unstructured meshes or a subset of basis functions that only covers the metastable regions instead of the whole space $\mathbb{X}$ (recovering the so-called spatio-temporal master equation \cite{winkelmann2016spatiotemporal}).
Formal continuum limits in the form of reaction-diffusion PDEs are obtained from the RDME in a multi-scale settings in \cite{arnold1980deterministic,blount1992law,egan2024macroscopic,kotelenez1988high}. 

\subsection{Path integrals for arbitrary discretizations} 
\label{sec:pelitipathint}

In this section, we will show how one can construct the path integral formalism based on a generic Galerkin discretization, leading directly to a discretized reaction-diffusion PDE. We first apply the Galerkin method from the previous section without fixing a specific basis, yielding the second quantized Galerkin description---and thus the RDME---in a basis-independent form. We use this to extend the path integral of  \cref{sec:firstSFTRD} and derive the mean-field equations for a generic discretization. Stochastic corrections can then be captured using diagrammatic techniques, analogous to \cref{sec:freeprop,sec:perturbandrenormalize}.

Consider the Galerkin discretization using an arbitrary finite basis $\{\xi_\alpha \}_{\alpha \in S}$, where $S$ is a finite set (e.g. \cref{eq:ind_func_def}). The loss and gain operators for the general one species reaction $kA \to \ell A$ are derived from \eqref{eq:fockpropOperators_kl}, where now $S$ is a finite set. They are expressed as:
\begin{align}
    \mathcal{G}  &= \sum_{\substack{\aalpha \in S^{\ell} \\ \bbeta \in S^k}} G^{(\ell,k)}_{\aalpha \bbeta} a_{\aalpha}^\dagger a_{\bbeta} , 
    \quad \text{with} \; G^{(\ell,k)}_{\aalpha \bbeta}=\frac{1}{k!} \langle \xi_{\aalpha}^*, G \xi_{\bbeta} \rangle , \\
    \mathcal{L}  &= \sum_{\substack{\aalpha \in S^k \\ \bbeta \in S^k}} L^{(k)}_{\aalpha \bbeta} \  a_{\aalpha}^\dagger a_{\bbeta},  
    \quad \ \text{with} \; \ L^{(k)}_{\aalpha \bbeta}=\frac{1}{k!} \langle \xi_{\aalpha}^*, L \xi_{\bbeta} \rangle , \nonumber
\label{eq:fockpropOperators_kl_disct}
\end{align}
where we introduced the compact notation $G^{(\ell,k)}_{\aalpha \bbeta}$ and $\ L^{(k)}_{\aalpha \bbeta}$. These expressions describe how the reaction rate function inside $G$ and $L$ (defined through \cref{eq:propOperators_kl}) is discretized for any chosen finite basis, ensuring consistency across different discretizations. The discretized reaction generator $H(\lambda) = \mathcal{G} - \mathcal{L}$ together with \cref{eq:diffExpdiscrete} constitutes the generator of the master equation for the second quantized Galerkin discretization (\cref{fig:maindiag}). The infinitesimal stochasticity of the reaction generator after the discretization can be expressed as 
\begin{equation}
   \label{eq:pcons_reac}
	0 = \bra{\mb{1}} H = \bra{\mb{1}} \sum_{\bbeta \in S^k} a_{\bbeta} \left( \sum_{\aalpha \in S^{\ell}} G^{(\ell,k)}_{\aalpha \bbeta} - \sum_{\aalpha \in S^k} L^{(k)}_{\aalpha \bbeta}\right), 
\end{equation}
which implies that for each $\bbeta \in S^k$ the expression in the parenthesis vanishes. Similarly for the diffusion generator 
\begin{equation}
\label{eq:pcons_diff}
	0 = \bra{\mb{1}} H_0(D) = \bra{\mb{1}} \sum_{\beta \in S} a_\beta \left( \sum_{\alpha \in S} D_{\alpha \beta} \right),
\end{equation}
implying that the finite matrix with elements $D_{\alpha \beta}$ should be infinitesimally stochastic.
Thus, from the above equations, the conservation of probability implies
\begin{align}
	0 &= \sum_{\aalpha \in S^{\ell}} G^{(\ell,k)}_{\aalpha \bbeta} - \sum_{\aalpha \in S^k} L^{(k)}_{\aalpha \bbeta}, \quad \text{and} \label{rtrt}\\
 \label{alle2}
	0 &= \sum_{\alpha \in S} D_{\alpha \beta} ,
\end{align}
since all vectors $\langle 1 | a_{\bbeta}$  in \cref{eq:pcons_reac} and $\langle 1 | a_{\beta}$ in \cref{eq:pcons_diff} are linearly independent. 
This will be used later to simplify the path integral action after performing the Doi shift. Similar to the previous derivation of the path integral, we introduce Poisson states for discretized space and a completeness relation. Let $\mb{z} \in \mathbb{C}^S$, then
\begin{align}
	& \ket{\mb{z}} = \exp\left( \sum_{\alpha \in S} z_\alpha a^\dagger_\alpha \right) \ket{\mb{0}}, \\ 
    & \frac{1}{\pi^{|S|}}\int_{\mathbb{C}^S} d\bar{\mb{z}}d\mb{z} \; e^{-\bar{\mb{z}}\mb{z}} \ket{\mb{z}}\bra{\mb{z}} = 1 \; ,
\end{align}
as in \eqref{eq:coherentstatesBasis} and where $|S|$ is the cardinality of $S$. From here on, the derivation of the path integral action follows \cref{sec:firstSFTRD}, the main difference being that space is already discretized, hence the continuum limit is only applied to the temporal discretization. Thus,  similar to \eqref{eq:pi_action}, the path integral action is
\begin{align}
    \begin{split}
     \label{eq:gal_pi}
	S[\bar{\pphi},\pphi] &= \int dt \; s(\bar{\pphi},\pphi,\del_t \pphi)\\ &=\int dt \; \bigl[\bar{\pphi}(\del_t -  \mathbb{D})\pphi - s_{\mathcal{R}}(\bar{\pphi},\pphi)\bigr],
 \end{split}
\end{align}
where $\mathbb{D}$ is the matrix/operator with elements $D_{\alpha \beta}$, $\bar{\pphi},\pphi : \mathbb{R} \to \mathbb{C}^S$, and $s_{\mathcal{R}}(\bar{\pphi},\pphi)$ is obtained from $H$ with the substitution 
\begin{align}
	a_\beta &\mapsto \phi_\beta,\\
	a^\dagger_\beta &\mapsto \bar{\phi}_\beta + 1.
\end{align}
The second line corresponds to the Doi shift.
In \cref{eq:gal_pi} we omitted  the boundary term in the last line of \cref{eq:pi_action}, since it does not contribute to the derivation of the mean-field equations.
The form of $s_{\mathcal{R}}(\bar{\pphi},\pphi)$ for the reaction $k A \to \ell A$ can be expressed in terms of elementary symmetric polynomials in $x_j$ for $j \in \{ 1,...,n \}$ 
\begin{align}
	e^{(n)}_q(\mb{x}) &:= \sum_{1 \leq j_1 <  \dots < j_q \leq n } x_{j_1} \dots x_{j_k}, \\
	e^{(n)}_0(\mb{x}) &= 1,
\end{align}
yielding 
\begin{align}
\begin{split}
	s_{\mathcal{R}}(\bar{\pphi},\pphi)
	&= \sum_{\substack{\aalpha \in S^{\ell}\\\bbeta \in S^k}} G^{(\ell,k)}_{\aalpha \bbeta} \sum_{q = 1}^{\ell} e^{(\ell)}_q(\bar{\phi}_{\aalpha}) \phi_{\bbeta} \\
	& \qquad -\sum_{\substack{\aalpha \in S^k \\ \bbeta \in S^k}} L^{(k)}_{\aalpha \bbeta} \sum_{q = 1}^k e^{(k)}_q(\bar{\phi}_{\aalpha}) \phi_{\bbeta},
 \end{split}
 \label{eq:actionGLkLk}
\end{align}
where the term proportional to $e^{(\ell)}_0 = e^{(k)}_0$ vanishes because of \eqref{rtrt}.
We define the \textit{symmetrically contracted} reaction generators to be:
\begin{align}
\label{eq:this_def}
\begin{split}
    G^{(\ell|q,k)}_{\aalpha \bbeta} &:= \frac{\ell !}{q!(\ell -q)!} \sum_{\alpha_{q+1},...,\alpha_{\ell}}G^{(\ell,k)}_{\aalpha,\alpha_{q+1},...,\alpha_{\ell};\bbeta},\\
    L^{(k|q)}_{\aalpha \bbeta} &:= \frac{k!}{q!(k-q)!} \sum_{\alpha_{q+1},...,\alpha_k}L^{(k)}_{\aalpha,\alpha_{q+1},...,\alpha_k;\bbeta}
 \end{split}
\end{align}
with $\aalpha \in S^q$. Note that for the gain $q\leq l$ and $G^{(\ell |\ell, k)}_{\aalpha \bbeta}\equiv G^{(\ell,k)}_{\aalpha \bbeta}$, and for the loss $q\leq k$ and $L^{(k|k)}_{\aalpha \bbeta}\equiv L^{(k)}_{\aalpha \bbeta}$. If $q=1$, we contract all indices except one, and we obtain $G^{(\ell | 1,k)}_{\alpha \bbeta}$ and $L^{(k|1)}_{\alpha \bbeta}$ with $\alpha \in S$. Recalling that the reaction generators are symmetric in permutations of $\aalpha$ and $\bbeta$, we can rewrite the terms in \cref{eq:actionGLkLk} as
\begin{align}
    \sum_{q=1}^{\ell} \sum_{\substack{\aalpha \in S^{\ell} \\ \bbeta \in S^k}} G_{\aalpha \bbeta}^{(\ell,k)} e^{(\ell)}_q(\bar{\phi}_{\aalpha})\phi_{\bbeta} &= \sum_{q=1}^{\ell}\sum_{\substack{\aalpha \in S^q\\ \bbeta \in S^k}} G_{\aalpha \bbeta}^{(\ell |q,k)}\bar{\phi}_{\aalpha}\phi_{\bbeta},\\
	\sum_{q=1}^k \sum_{\substack{\aalpha \in S^k \\ \bbeta \in S^k}} L_{\aalpha \bbeta}^{(k)} e^{(k)}_q(\bar{\phi}_{\aalpha})\phi_{\bbeta} &= \sum_{q=1}^k \sum_{\substack{\aalpha \in S^q \\ \bbeta \in S^k}} L_{\aalpha \bbeta}^{(k|q)}\bar{\phi}_{\aalpha}\phi_{\bbeta}.
\end{align}
This simplifies the expression of the reaction part of the action, as in \eqref{eq:reaction_action}, to
\begin{equation}
    \label{eq:esse_erre}
    s_{\mathcal{R}}(\bar{\pphi},\pphi) = \sum_{q=1}^{\ell}\sum_{\substack{\aalpha \in S^q \\ \bbeta \in S^k} }G_{\aalpha \bbeta}^{(\ell |q,k)}\bar{\phi}_{\aalpha} \phi_{\bbeta} - \sum_{q=1}^k \sum_{\substack{\aalpha \in S^q \\ \bbeta \in S^k}} L_{\aalpha \bbeta}^{(k|q)}\bar{\phi}_{\aalpha} \phi_{\bbeta}.
\end{equation}
This is the starting point for the perturbation theory in terms of Feynman diagrams prescribed in \cref{sec:perturbandrenormalize}, bringing together field theory techniques and the spatial discretizations required for numerical simulations. To derive the reaction-diffusion equation of the reaction above, we look at the extremum of the action given by:
\begin{equation}
	\frac{\delta}{\delta \pphi}S[\bar{\pphi},\pphi] = 0, \quad \frac{\delta}{\delta \bar{\pphi}}S[\bar{\pphi},\pphi] = 0,
\end{equation}
which yields the Euler-Lagrange equations of motion. 
All elementary symmetric polynomials for $k\neq0$ vanish at $\bar{\pphi} = \mb{0}$, which results from the conservation of probability enforced on the generator of the 
Markov process. We can then cast the reaction-diffusion equation as 
\begin{equation}
	\frac{\del}{\del\bar{\phi}_\alpha}s(\bar{\pphi},\pphi,\del_t \pphi) \bigg|_{\bar{\pphi} = \mb{0}} =0.
\end{equation}
The above equation involves only the terms proportional to $G_{\alpha ;\bbeta}^{(\ell|1,k)}$ and $L_{\alpha ;\bbeta}^{(k|1)}$ with $\alpha \in S$. Reorganizing the previous results, the reaction-diffusion equation is
\begin{equation}
    \label{pel_mfode}
	\del_t \phi_\alpha = \sum_{\beta \in S} D_{\alpha \beta} \phi_\beta + \sum_{\ggamma \in S^k} \biggl( G_{\alpha \ggamma}^{(\ell |1,k)} - L_{\alpha \ggamma}^{(k|1)} \biggr)\phi_{\ggamma} \,,
\end{equation}
where $G_{\alpha \ggamma}^{(\ell |1,k)}$ and $L_{\alpha \ggamma}^{(k |1)}$ are the contracted operators from \cref{eq:this_def} for $q=1$. This equation is determined by $D_{\alpha \beta}, G_{\alpha \bbeta}^{(\ell |1,k)}$ and $L_{\alpha \bbeta}^{(k|1)}$, which depend on the choice of basis, reaction rate function and diffusion generator. \Cref{pel_mfode} is effectively a discretization of a reaction-diffusion PDE. The field $\phi_\alpha$ satisfies the same equation as the concentrations on the basis function indexed by $\alpha$ (e.g. for a lattice, the concentration in voxel $\alpha$).\footnote{The fields and concentrations are related to one another by a Cole-Hopf transformation \cite{tauber2005applications}.} 

To summarize, the chosen basis corresponds to a specific discretization. Applying this basis, results in a master equation for the projected stochastic dynamics. Then we define a path integral action for the master equation. Its saddle points correspond to the mean-field behavior, yielding a discretized reaction-diffusion PDE for the concentrations (\cref{fig:maindiag}).

This description also enables the use of  diagrammatic techniques. Although these are likely not easy to handle analytically, they could be implemented numerically as contractions of tensors of the type appearing in the reaction part of the action. In simple cases, the propagator for the reaction under consideration is given by the diffusive (bilinear) part of the action.\footnote{In general, however, monomolecular reactions can lead to terms proportional to $\bar{\phi}_\alpha \phi_\beta$ in  the action and hence contribute to the propagator.}

In the case of the lattice discretization, the diffusive operator $D$ is precisely the lattice Laplacian as shown in \cref{lapdiscrete}. Laplacian operators on graphs are always diagonalizable, which implies $U D U^{-1} = \Lambda$ with $\Lambda$ diagonal. The propagator for a diagonalizable diffusion generator can then be written in terms of the eigenbasis as the time dependent matrix 
\begin{equation}
\label{eq:gal_prop}
		G(t-t') = \frac{1}{2 \pi} \int d\omega \; e^{i \omega (t-t')} U^{-1} (i\omega  - \mathbb{D} \Lambda)^{-1}U,
\end{equation}
necessary for diagrammatic calculations as in \cref{GFmomentum}. To perform such calculations, we need to reinstate the boundary terms omitted above (similar to \cref{eq:reaction_action}) accounting for the initial conditions. Expressions for the tree level quantities can be derived as in \cref{eq:diagrams_1,treelevel}. From such tree level quantities, more complicated diagrams can be constructed as convolution of smaller blocks. Once the process is discretized such convolutions will be translated to tensor contractions involving the propagator in \cref{eq:gal_prop}, without spoiling the applicability of the method. 

\section{Doi formulation and alternative representations}
\label{sec:specialapplications}
Starting from the basis-independent representation presented in the last section, one may obtain different field theory representations, tailored to fit the specific analytical, numerical or physical application in mind. We had shown in \cref{sec:RDMEgalerkin,sec:pelitipathint} how discretized models arise from an appropriate choice of basis functions, and here we will discuss in more detail the link with the formulation of \cite{doi1976second}. We use this to express the equations in terms of physical quantities, such as the inclusive densities, fluxes and stochastic concentrations, yielding alternative representations that are insightful for specific applications. Finally, we discuss the emergence of the macroscopic reaction-diffusion equations in the limit of large copy numbers. These approaches are also possible with the basis-independent representation; however, the Doi formulation is simpler in this context.\footnote{In contrast, the basis-independent representation is more convenient for numerical approaches and to bridge models.} 
Although many of these results are based on previous work \cite{doi1976second,doi1976stochastic,grassberger1980fock,delRazo2}, some parts have not been covered with the same detail.

\subsection{Doi formulation: Dirac deltas instead of basis}
\label{sec:deltabasis}
The Doi formulation (\cite{doi1976second}) is recovered as a special case of the formalism from \cref{sec:stochMecRD} by replacing the basis functions by Dirac deltas (\cref{fig:maindiag}).\footnote{Formally, the Dirac deltas do not constitute a basis of the underlying space, but they are a singular limit of an approximate basis, e.g. as the lattice spacing goes to zero in \cref{sec:RDMEgalerkin}.}
This procedure follows \cite{delRazo2}. The basis functions for the creation operators from \cref{sec:generalbasisCA} now become Dirac deltas: $u_\alpha(x)\rightarrow \delta_{y}(x)$, where $\delta_{y}(x) \equiv \delta(x-y)$. We use $y$ instead of $\alpha$ since the index variable of the basis is now continuous and not discrete. We do the same for $u_\beta^*(x)$, and we employ a short-hand notation for these operators: $a^\dagger(y):=a^\dagger(\{\delta_{y}(x)\}$ and $a(y):=a(\{\delta_{y}(x)\})$, which based on \cref{eqs:apamdensities} yields
\begin{subequations}	\label{eqs:creaannihopsdelta}
    \begin{align}
    a^\dagger(y)\ket{\rho_{n}} &=\frac{1}{n+1}\sum_{j=1}^{n+1} \delta(x_j-y) \ \ket{\rho_n(\mb{x}_{\setminus \{j\}})} \, , \quad \mb{x}\in\mathbb{X}^{n+1} \label{eqs:creaopsdelta}\\
    a(y)\ket{\rho_{n}} &=  n \, \ket{\rho_n(\mb{x},y)} \, , \quad \mb{x}\in\mathbb{X}^{n-1}  . \label{eqs:annihopsdelta}
	\end{align}
\end{subequations}
where $\mathbb{X}$ again corresponds to a bounded domian. The creation operator $a^\dagger(y)$ adds one particle at position $y$, and the annihilation operator $a(y)$ removes one particle at position $y$, both in a way that preserves the symmetry over permutations. The commutation relations from \cref{eq:commutrelRD} simplify to
\begin{align}
\begin{aligned}
\left[a(y_1), a^\dagger(y_2)\right]  &= \delta(y_1-y_2), \\
\left[a(y_1), a(y_2)\right] &= \left[a^\dagger(y_1), a^\dagger(y_2)\right] = 0,
\end{aligned}
\label{eq:a+a-CommutRels}
\end{align}
which one can also prove independently by direct substitution of \cref{eqs:creaannihopsdelta}. The vacuum state $\ket{0}$ satisfies \cref{eq:vacuumRD}, i.e.
\begin{align}
	\bra{0}a^\dagger(y) = 0, \qquad a(y) \ket{0}=0, \qquad \bracket{0}{ 0}=1.
	\label{eqs:vacuumprops}
\end{align}
Analogously, the copy number representation from \cref{eq:copynumRepcont} and its dual can be written as
\begin{align}
	\ket{x^n} = a^\dagger(x^{(n)}) \ket{0} , \qquad \bra{x^{(n)}}= \bra{0} \frac{1}{n!}a(x^{(n)}),
	\label{eq:braketdef} 
\end{align}
where $a^\dagger(x^{(n)})=a^\dagger(x_1)\dots a^\dagger(x_n)$ and likewise for $a(x^{(n)})$. Throughout this section, it will be convenient to use the following notation interchangeably 
\begin{align}
a^\dagger(x^{(n)}) \equiv a^\dagger(x_1)\dots a^\dagger(x_n)\equiv  a^\dagger(\mb{x}),
\end{align}
whenever the dimension of $\mb{x}$ is clear from context. The ket $\ket{x^n}$ represents the state of the system with $n$ particles at positions $x^{(n)}$, and the bra $\bra{x^n}$ represents its dual.\footnote{In the original work by Doi \cite{doi1976second}, the factorial term was included in the normalization, not in the definition of the dual.} The creation and annihilation operators act on the states $\ket{x^n}$ as
\begin{align}
\begin{aligned}
a^\dagger(y)\ket{x^n} &= a^\dagger(y) a^\dagger(x^{(n)}) \ket{0} \\ 
a(y) \ket{x^n} &= \sum_{j=1}^n \delta(y-x_j) a^\dagger(x^{(n)}_{\setminus \{j\}})\ket{0},
\end{aligned}
\end{align}
where $x_{j}$ is the $j$th component of $x^{(n)}$ and $x^{(n)}_{\setminus \{j\}}$ means the $j$th component is excluded from $x^{(n)}$. Using the vacuum properties \cref{eqs:vacuumprops}, the commutation relations \cref{eq:a+a-CommutRels} 
and the symmetry of the densities, we can show by direct substitution that the states are orthonormal (analogous to \cref{eq:orthonormalityRD})
\begin{align}
	\bracket{y^m}{ x^n} = \delta_{m,n} \ \delta(x^{(n)}-y^{(m)}).
	\label{eq:orthogonality}
\end{align}
We can now write the state of the whole system as a linear combination of all the possible states $\ket{y^m}$, weighted by the probability density $\rho_m(y^{(m)})$ of observing any of these states. This is denoted as
\begin{align}
	\ket{\rho} = \sum_{m \geq 0}\ket{\rho_m} =\sum_{m \geq 0} \int_{\mathbb{X}^m} dy^{(m)} \ \rho_m(y^{(m)}) \ \ket{y^m},
	\label{eq:DoidenistyKets}
\end{align}
which is the analog of \cref{eq:ketrhoexpan}, where the sum over the discrete basis must become an integral over continuous space when using deltas. Using orthonormality \eqref{eq:orthogonality}, we can apply $\bra{x^n}$ on the left to obtain
\begin{align}
	\rho_n(x^{(n)}) = \bracket{x^n}{ \rho},
	\label{eq:rhoninner}
\end{align}
which establishes a precise connection between the Fock space vector and the probability densities.

The Poisson states and their duals are obtained from \cref{eq:coherentstatesBasis} and lead to:
\begin{align}
\begin{aligned}
\ket{z} &= \exp \left( \int_{\mathbb{X}} dx \ z(x) a^\dagger (x)  \right) \ket{0}\,, \\
\bra{z} &= \bra{0} \exp \left( \int_{\mathbb{X}} dx \ z(x) a(x) \right)\,.
\end{aligned}
\label{eq:coherentstatesDoi}
\end{align}
Once again they are eigenstates of the creation and annihilation operators: $a(x) \ket{z}=z(x)\ket{z}$ and $\bra{z}a^\dagger(x)=\bra{z}z(x)$. Analogous to \cref{flat_def}, expectations are calculated with the flat state corresponding to $z=1$,
\begin{align}
\bracket{1}{\dots} = \sum_{m\geq 0} \int_{\mathbb{X}^m}dx^{(m)} \cdots \ ,
\label{eq:flatstateDoi}
\end{align}
satisfying $\bra{1}a^\dagger(x)=\bra{1}$ and $\bracket{1}{\rho} = 1$.

In analogy to \cref{sec:partconservops}, generic operators are given in terms of creation and annihilation operators. A single particle operator $A$ is obtained by substituting the basis functions in \cref{eq:consPartExp} with delta functions and replacing the sum over basis indices by integrals, yielding
\begin{align}
	F_A= \int_{\mathbb{X}^2} dx dy \ a^\dagger(x) \tilde A(x;y) a(y) \,,
\label{eq:generalConservingPartExp2}
\end{align}
where $\tilde A(x;y)=\langle \delta_x,A \delta_y\rangle$. 
More generally for an operator $Z$ acting on $k$ particles at a time, the representation is
\begin{align}
F_Z= \int_{\mathbb{X}^k \times \mathbb{X}^k} d\mb{x} d\mb{y} \ a^\dagger(\mb{x}) \tilde Z(\mb{x};\mb{y}) a(\mb{y}),
\end{align}
where $\mb{x},\mb{y}\in \mathbb{X}^k$ and $\tilde Z(\mb{x};\mb{y})=\langle\delta_{\mb{x}},Z\delta_{\mb{y}} \rangle$, analogous to \cref{eq:consmPartExp}.
For diagonal operators $Z$, it simplifies to
\begin{align}
	F_Z= \int_{\mathbb{X}^k} d\mb{x} \ a^\dagger(\mb{x}) \tilde Z(\mb{x}) a(\mb{x}).
\end{align}
When $\tilde Z$ is a differential operator, the proper way to interpret this equation is in the distributional sense, see \cite{delRazo2}.
Analogous results hold for the reaction generators from \cref{sec:reactopers}, as we will explore in \cref{sec:genOneSpeciesReacDeltas}. 

\subsubsection{Additional relations based on commutators}
If is often helpful for calculations to compute more relations based on commutators. We will need these expressions for the flux formulation in \cref{sec:doifluxes}. Using the commutation relations \cref{eq:a+a-CommutRels}, along with \cref{eqs:vacuumprops}, we can show the following relations hold:
\begin{align}
\begin{aligned}
\left[a(y), a^\dagger(x^{(n)})\right] &= \sum_{i=1}^n \delta(y-x_i)a^\dagger(x^{(n)}_{\setminus \{i\}}), \\ \left[a(y^{(n)}), a^\dagger(x)\right] &= \sum_{i=1}^n \delta(x-y_i)a(y^{(n)}_{\setminus \{i\}}).
\end{aligned}
\label{eq:a+a-CommutRels2}
\end{align}
Based on these relations, we can further compute the following
\begin{align}
    \left[ a(y^{(n)}),  a^\dagger(x^{(n)}) \right] \ket{0} 
    &=n! \ \delta(y^{(n)} - x^{(n)}) \, \ket{0}, 
    \label{eq:a+a-CommutRels3} \\
    \bra{0}\left[a(y^{(n)}), a^\dagger(x^{(n)})\right]  
    &=\bra{0} \, n! \ \delta(y^{(n)} - x^{(n)}) .
  \label{eq:a+a-CommutRels3alt}
\end{align}
To simplify notation, from now onward, we will denote sums over ordered indices as 
\begin{align}
    \sum_{\bm{\nu}\in S^{(m)}_n}\equiv \sum_{{1\leq \nu_1<\dots<\nu_m\leq n}}, 
    \label{eq:compactnot_ordersum}
\end{align}
with $\bm{\nu}=\nu_1,\dots , \nu_m$ and $S_n$ corresponds to the countable set $\{1,\dots,n\}$. The parenthesis in $(m)$ is to note the $m$ tuples are ordered. 
With this notation in mind, we can generalize these relations for arbitrary $n$ and $m$:
\begin{align}
&\left[  a(y^{(m)}), a^\dagger(x^{(n)})\right] \ket{0}  \label{eq:a+a-CommutRels4}\\
    &=\begin{cases}
    m! \displaystyle {\sum_{\bm{\nu}\in S^{(m)}_n}} \delta\left(y^{(m)}-x^{(n)}_{\bm{\nu}} \right) a^\dagger\left(x^{(n)}_{\setminus\{\bm{\nu} \}}\right)\ket{0} \quad \text{if} \quad n\geq m \\
     0 \quad \text{otherwise}
    \end{cases} \notag
\end{align}
and
\begin{align}
    &\bra{0} \left[a(y^{(m)}), a^\dagger(x^{(n)})\right]   \label{eq:a+a-CommutRels5} \\
    &=\begin{cases}
    \bra{0} \ n! \displaystyle{\sum_{\bm{\nu}\in S^{(n)}_m}}\delta \left( x^{(n)}-y^{(m)}_{\bm{\nu}}\right)a\left(y^{(m)}_{\setminus\{\bm{\nu} \}}\right) \quad \text{if} \quad m\geq n \\
    0 \quad \text{otherwise},
    \end{cases} \notag
\end{align}
where we used that $x^{(n)}_{\bm{\nu}}$ picks the $\bm{\nu}$ indices of $x^{(n)}$ and $x^{(n)}_{\setminus\{\bm{\nu} \}}$ excludes the entries with indices $\bm{\nu}$ from $x^{(n)}$.
The number of terms in an ordered sum over $\bm{\nu}\in S^{(m)}_n$ is simply the binomial coefficient $n!/((n-m)!m!)$. 

\subsection{General one species reaction using deltas}
\label{sec:genOneSpeciesReacDeltas}
We now revisit in Doi's formulation the general one species reaction $kA\rightarrow \ell A$ with reaction rate function  $\lambda(\mb{y};\mb{x})$, where  $\mb{x} \in \mathbb{X}^k$ and $\mb{y} \in \mathbb{X}^\ell$ are the reactants and products positions, respectively. 
The diffusion generator is expanded using \cref{eq:generalConservingPartExp2}. After the integration over the Dirac deltas (in distributional sense), it simplifies to
\begin{align}
H_0(D) = \int_{\mathbb{X}} dx \ a^\dagger(x) \,D\nabla_x^2 \, a(x) \ .
\label{eq:diffdeltasexp}
\end{align}
For the reaction generator $H(\lambda)=\mathcal{G}-\mathcal{L}$, we again substitute the basis by delta functions and sums over indices by integrals; applying this to \cref{eq:fockpropOperators_kl} results in
\begin{align}
	\mathcal{L} &= \frac{1}{k!}\int_{\mathbb{X}^k \times \mathbb{X}^k} d\mb{x} d\mb{z} \
	a^\dagger(\mb{z}) \,\tilde L(\mb{z}; \mb{x}) \, a(\mb{x}) \,, \label{eq:lossperdelta}
	\\
	\mathcal{G} &= \frac{1}{k!}\int_{\mathbb{X}^k \times \mathbb{X}^\ell} d\mb{x} d\mb{y} \
	a^\dagger(\mb{y}) \, \tilde G(\mb{y}; \mb{x}) \, a(\mb{x}) \, . \label{eq:gainperdelta}
\end{align}
where $\mb{z},\mb{x} \in \mathbb{X}^k$ and $\mb{y} \in \mathbb{X}^\ell$.
The coefficient functions are calculated from the definitions of the loss and gain operators per reaction in \cref{eq:propOperators_kl},
\begin{align}
	\tilde L(\mb{z}; \mb{x})  &:= \langle \delta_{\mb{z}},L\delta_{\mb{x}} \rangle
	= \delta(\mb{z} - \mb{x}) \int_{\mathbb{X}^\ell} d\mb{y} \, \lambda(\mb{y};\mb{x}) \,, \label{tildeL}\\
	\tilde G(\mb{y}; \mb{x}) &:= \langle \delta_{\mb{x}},G\delta_{\mb{y}} \rangle
	= \lambda(\mb{y}; \mb{x}) \,,
\end{align}
These equations together with \cref{eq:lossperdelta,eq:gainperdelta} yield the loss and gain explicitly in terms of the reaction rates
\begin{align}
	\mathcal{L} &= \frac{1}{k!}\int_{\mathbb{X}^k \times \mathbb{X}^\ell} d\mb{x} d\mb{y} \
	a^\dagger(\mb{x}) \,\lambda(\mb{y};\mb{x}) \, a(\mb{x}) \,, 
	\label{eq:lossdoigeneralonespecies}\\
	\mathcal{G} &= \frac{1}{k!}\int_{\mathbb{X}^k \times \mathbb{X}^\ell} d\mb{x} d\mb{y} \
	a^\dagger(\mb{y}) \, \lambda(\mb{y}; \mb{x}) \, a(\mb{x}) \,,
	\label{eq:gaindoigeneralonespecies}
\end{align}
corresponding to the results of \cite{doi1976second}. The CDME is then $\partial_t \ket{\rho} = H \ket{\rho}$ with the generator $H = H_0(D) + H(\lambda)$. The construction of the generator for the reaction part is illustrated in \cref{fig:gendoidiag}. 

\begin{figure}
	\centering
	\includegraphics[width=\columnwidth]{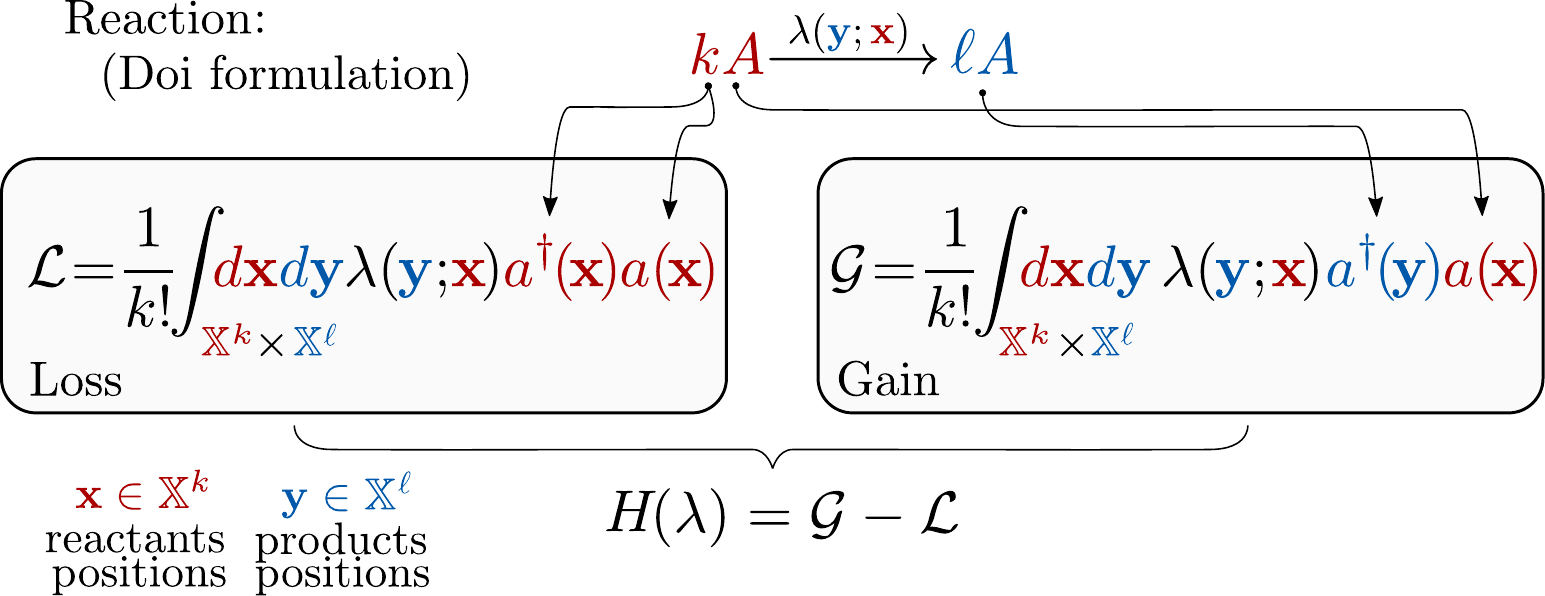}
	\caption{Diagram illustrating the construction of the generator for a system with a general one species reaction with reaction rate function $\lambda (\mb{y};\mb{x})$ in the Doi formulation. This case takes into account the spatial dynamics in continuous space, and it is a special case of the basis-independent case presented in \cref{fig:genbasiindependiag}.
		\label{fig:gendoidiag} }	
\end{figure}

\subsection{Representation in terms of scalar probability densities}
\label{sec:formdensities}
Some readers might be more familiar with the CDME in terms of the scalar densities $\rho_n(x^{(n)})$. Using \cref{eq:rhoninner}, this formulation is recovered by multiplying $\partial_t \ket{\rho} = H \ket{\rho}$ with $\bra{x^n}$ from the left, effectively projecting $\ket{\rho}$ to its $n$-particle subspace:
\begin{align}
 \partial_t \rho_n(x^{(n)}) =  \bra{x^n} H \ket{\rho}.
 \label{eq:densityformulationCDME}
\end{align}
The right hand side can be written in terms of $\rho_n(x^{(n)})$, but it requires careful consideration. We first substitute the copy number representation of $\ket{\rho}$ (\cref{eq:DoidenistyKets}),
\begin{align}
    \bra{x^n} H \ket{\rho}  = \sum_{m \geq 0} \int_{\mathbb{X}^m}  \ \bra{x^n} H  \ket{y^m}  \ \rho_m(y^{(m)}) dy^{(m)},
    \label{eq:bradiff2den}
\end{align}
We can further split this into the diffusion and reaction components: $\bra{x^n} H \ket{y^m} = \bra{x^n} H_0(D) \ket{y^m} + \bra{x^n} H(\lambda) \ket{y^m}$. The diffusion part can be simplified by using the expansion of the diffusion coefficient (\cref{eq:diffdeltasexp}) and of the bras and kets (\cref{eq:braketdef}),
\begin{align}
&\bra{x^n} H_0(D) \ket{y^m}   \\ \nonumber
&=\frac{1}{n!}\int_{\mathbb{X}} dz \ \bra{0} a(x^{(n)})  \ a^\dagger(z) \,D\nabla_{z}^2 \, a(z)   \  a^\dagger(y^{(m)})\ket{0} \\ \nonumber
&=  \sum_{i=1}^n \delta_{n,m } \,D\nabla_{x_i}^2 \,   \delta(y^{(n)} - x^{(n)}) \ ,
\end{align}
where we used \cref{eq:a+a-CommutRels2} to commute $a^\dagger(z)$ to the left and $a(z)$ to the right, along with the orthogonality of the resulting states \cref{eq:orthogonality}.
Incorporating this into the form of \cref{eq:bradiff2den}, we obtain
\begin{align}
    \bra{x^n} H_0(D)\ket{\rho}  &=  \sum_{i=1}^n D\nabla_{x_i}^2 \rho_n(x^{(n)}),
\end{align}
Similar expressions are obtained analogously for the reaction generator $H(\lambda)=\mathcal{G}-\mathcal{L}$. We first calculate
\begin{align}
\bracket{x^n}{\ \mathcal{G} \ |y^m} &=  \label{eq:generalfluxxnym} \\ 
 \delta_{n,m-k+\ell} &
	\frac{m!\ \ell !}{n! \ k!} \sum_{\bm{\nu}\in S_n^{(\ell)}}
    \lambda\left(x^{(n)}_{\bm{\nu}}; y^{(m)}_{[k]}\right)\delta\left( y^{(m)}_{\setminus [k]} - x^{(n)}_{\setminus \{\bm{\nu}\}}\right),
    \notag
\end{align}
and 
\begin{align}
    \bra{x^n} \ \mathcal{L} \ \ket{y^m} &= \label{eq:generalfluxxnym_loss} \\  \delta_{n,m} & \sum_{\bm{\nu}\in S_n^{(k)}}  \int_{\mathbb{X}^\ell}  \lambda\left(\mb{z}; x^{(n)}_{\bm{\nu}}\right)\delta(y^{(n)}-x^{(n)}) d\mb{z}.
    \notag
\end{align}
where $[k]=1,\dots,k$, the ordered sums follows \cref{eq:compactnot_ordersum} and we have used the relations \cref{eq:a+a-CommutRels4,eq:a+a-CommutRels5}. Based on \cref{eq:bradiff2den}, we use these expressions to obtain the final form
\begin{align}
&\bracket{x^n}{\ \mathcal{G} \ |\rho} = \frac{(n-\Delta)!\ \ell !}{n! \ k!}   \\ &\qquad \times \int_{\mathbb{X}^k}   \sum_{\bm{\nu}\in S_n^{(\ell)}}
\lambda\left(x^{(n)}_{\bm{\nu}}; \mb{z}\right)\rho_{n-\Delta}(\mb{z},x^{(n)}_{\setminus\{\bm{\nu}\}}) \, d\mb{z} \notag 
\end{align}
with $\Delta:=\ell-k$, and similarly for the loss
\begin{align}
\bracket{x^n}{\ \mathcal{L} \ |\rho} &= \sum_{\bm{\nu} \in S_n^{(k)}} \rho_n(x^n) \int_{\mathbb{X}^\ell}  \lambda\left(\mb{z}; x^{(n)}_{\bm{\nu}}\right) d\mb{z}.
\end{align}
Now all the terms in \cref{eq:densityformulationCDME} are expressed uniquely in terms of the scalar probabiliy densities:
\begin{align}
 \partial_t \rho_n(x^{(n)}) =  \bra{x^n} H_0(D) \ket{\rho} + \bra{x^n} \mathcal{G} \ket{\rho} - \bra{x^n} \mathcal{L} \ket{\rho}.
 \label{eq:formdensitiesFinal}
\end{align}
The results obtained are consistent previous formulations of the general one species reaction \cite{delRazo2}, and the methodology yields the same equations as \cite{doi1976second,isaacson2022mean,lanconelli2024analysis} for other reaction systems. 

\subsection{Fluxes and many-particle propensities}
\label{sec:doifluxes}
It can be useful for practitioners to write the CDME in terms of probability fluxes. For instance, flux formulations for well-mixed systems are essential to study nonequilibrium processes and their driving forces \cite{qian2003stoichiometric,qian2006open,beard2007relationship} as well as stochastic thermodynamics in general \cite{hong2021stochastic,van2013stochastic}. 

Consider again the general one species chemical reaction ($kA\rightarrow \ell A$ with $\Delta:=\ell-k$). We define two fluxes: the influx $J^+_{n}$ from any particle state into the current state with $n$ particles, and the outflux $J^-_{n}$ from the current $n$ particle state to any other state. These can be written explicitly using the gain and loss operators:
\begin{align}
    \label{eq:generalfluxrhonrhom}
    J^+_{n} (x^{(n)})&=\bracket{x^n}{\ \mathcal{G} \ |\rho} \\
    J^-_{n} (x^{(n)})&=\bracket{x^n}{\ \mathcal{L} \ |\rho}
    \label{eq:generalfluxrhonrhom_m}
\end{align}
which correspond to the two terms in the CDME from \cref{eq:formdensitiesFinal}. We have already computed these explicitly in \cref{sec:formdensities}. However, we can rewrite the fluxes in terms of the many-particle propensities $\Lambda_n$ introduced in \cite{delRazo2}, 
\begin{align}
    \Lambda_{n-\Delta}\left(x^{(n)};y^{(n-\Delta)}\right) := \bracket{x^n}{\ \mathcal{G} \ |y^{n-\Delta}}.
	\label{eq:manypartprop_def}
\end{align}
This was calculated in \cref{eq:generalfluxxnym} (and only contributes when $m=n-\Delta$), and it denotes the total transfer of probability density from the $n-\Delta$ to the $n$ state per unit time for all the possible reactions of this type. The influx and outflux can be expressed as
\begin{align}
J^+_{n} (x^{(n)}) &= \int_{\mathbb{X}^{n-\Delta}}
\Lambda_{n-\Delta}\left(x^{(n)};\mb{z}\right) \rho_{n-\Delta}(\mb{z}) d\mb{z} \label{eq:gainmanyprop}\\
J^-_{n} (x^{(n)}) &= \rho_n(x^{(n)}) \int_{\mathbb{X}^{n+\Delta}}
\Lambda_{n}\left(\mb{z};x^{(n)}\right) d\mb{z},
\label{eq:lossmanyprop}
\end{align}
yielding an alternative way to write the CDME, similarly to \cite{doi1976second,isaacson2022mean}. The first equation is shown by expanding \cref{eq:generalfluxrhonrhom} in the form of \cref{eq:bradiff2den}. Similarly, although not as trivial, the second one is equal to \cref{eq:generalfluxrhonrhom_m}, which can be proven by direct substitution of \cref{eq:manypartprop_def} and \cref{eq:generalfluxxnym_loss,eq:generalfluxxnym}.

For a system of $R$ reactions involving one species, one can write the total fluxes as
$J^+_{n} (x^{(n)}) =\sum_{r=1}^R  \bracket{x^n}{ \mathcal{G}^{(r)}  |\rho}$ and
$J^-_{n} (x^{(n)}) =\sum_{r=1}^R  \bracket{x^n}{  \mathcal{L}^{(r)}  |\rho}$.
Alternatively, one can also write them in terms of the propensity functions
$\Lambda_{m}\left(x^{(n)};y^{(m)}\right) := \sum_{r=1}^R \bracket{x^n}{\ \mathcal{G}^{(r)} \ |y^{m}},
$
as
\begin{align}
    J^+_{n} (x^{(n)}) &= \sum_{\substack{m\geq 0 \\ m\neq n}}\int_{\mathbb{X}^{m}}
    \Lambda_{m}\left(x^{(n)};\mb{z}\right) \rho_{m}(\mb{z}) d\mb{z} \\
    J^-_{n} (x^{(n)}) &= \sum_{\substack{m\geq 0 \\ m\neq n}} \rho_n(x^{(n)}) \int_{\mathbb{X}^{m}}
    \Lambda_{n}\left(\mb{z};x^{(n)}\right) d\mb{z},
\end{align}
The general equation with multiple species will have the same form, but the notation needs to keep track of the species numbers and positions for each species.

\subsection{Inclusive densities representation \label{sec:inclusivedensity}}
The work \cite{grassberger1980fock} introduces the \textit{inclusive densities} as a convenient representation for large systems. Similarly to \cref{sec:inclprod}, it facilitates the calculation of factorial moments.
We begin by recognizing that the particle number operator in the Doi formulation requires an additional integration
\begin{align}
    \mathcal{N} = \int_{\mathbb{X}} dx \ a^\dagger (x) a(x), \quad  \mathcal{N}_\Omega = \int_{\Omega} dx \  a^\dagger (x) a(x),
    \label{eq:partNumOpSpace}
\end{align}
such that $\mathcal{N} \ket{\rho}$ yields the number of particles in the whole domain $\mathbb{X}$ and $\mathcal{N}_\Omega \ket{\rho}$ in a subdomain $\Omega$, analogous to \cref{eq:numpartops_basis}. 
The corresponding operator for the factorial moments is then given by $\mathcal{N}(\mathcal{N} -1)\dots (\mathcal{N} -k+1)= \int_{\mathbb{X}^k} dx^k \ a^\dagger (x^{(k)}) a(x^{(k)})$. The factorial moments $n_k(t)$ correspond to the expectation with respect to the distribution $\ket{\rho}$, such that $n_k(t) = \bra{1} \int_{\mathbb{X}^k} dx^k \ a^\dagger (x^{(k)}) a(x^{(k)})\ket{\rho}$, which simplifies to
\begin{align}
    n_k(t) = \int_{\mathbb{X}^k} \rho^{\rm{in}}_k(x^{(k)}) \ dx^{(k)}
\end{align}
where $\rho^{\rm{in}}_k(x^{(k)})=\bra{1} a(x^{(k)})\ket{\rho}$ are the inclusive densities:
\begin{align}
    \rho^{\rm{in}}_k(x^{(k)}) &= \sum_{m\geq k} \frac{m!}{(m-k)!}\int_{\mathbb{X}^{m-k}} \rho_m(\bm{z},x^{(k)}) \ d\bm{z} \ .
    \label{eq:inclusivedensDef}
\end{align} 
Hence the inclusive densities give the probability of finding \textit{at least} $k$ particles at positions $x^{(k)}$. This naming convention follows high-energy physics, as pointed out in \cite{grassberger1980fock}, where the densities $\rho_n(x^{(n)})$ are referred to as \textit{exclusive densities}. 

Similarly, the inclusive product \cref{eq:inclusiveProdDef} can be generalized to $\bracket{y^m}{x^n}_{\rm{in}} = \bra{y^m}e^{\int dz a^\dagger(z)}e^{\int dz a(z)}\ket{x^n}$. This enables spatial analogous expressions to \cref{eq:probIncl,eq:momentsIncl},
\begin{align}
\rho_n(x^{(n)}) &= \bra{0} e^{-\int dz a(z)} a(x^{(n)})  \ket{\rho}_{\rm{in}}\\
\rho^{\rm{in}}_k(x^{(k)}) &= \bra{0} a(x^{(k)})\ket{\rho}_{\rm{in}} \ ,
\label{eq:momentsInclSpace}
\end{align}
which simplify the calculation of the inclusive densities.
Inclusive densities are a natural representation to obtain the factorial moments and thus the macroscopic behavior of the system. For instance, in \cref{sec:emerRD}, \cref{eq:bracketConc} corresponds to \cref{eq:inclusivedensDef} with $k=1$. Although in the current presentation, the advantages of inclusive densities may not be so evident, they become especially useful when calculating higher moments.

\subsection{Stochastic concentrations}
\label{sec:stochconcentrations}

The family of densities $\rho$ (\cref{eq:prodistfamily}) yields the probability of finding $n$ particles at locations $x^{(n)}$ for all $n$. Alternatively, suppose we want to know the distribution of finding $n$ particles in a delimited region in space at time $t$, or more precisely the (stochastic) concentration of particles at a point $(x,t)$ in space and time\footnote{Mathematically, the stochastic concentrations formulation is related to measure-valued stochastic processes \cite{dawson1993measure}, as shown in \cite{isaacson2022mean}.}. \Cref{eq:partNumOpSpace} suggests that the density/concentration operator is
\begin{align}
	\mathcal{C}_x = a^\dagger (x) a(x),
\end{align}
when applying it to $\ket{\rho}$ from \cref{eq:DoidenistyKets}, it yields the stochastic concentration 
\begin{align}
	\label{eq:stochconc}
		 \mathcal{C}_x \ket{\rho} & =  \\ \nonumber
		& \sum_{m\geq 1}m  \int_{\mathbb{X}^{m-1}} dz^{(m-1)} \rho_m(x,z^{(m-1)})  a^\dagger (x)\ket{z^{m-1}},
\end{align}
where we used \cref{eq:a+a-CommutRels2}. 
We can also apply $\mathcal{C}_x$ to the diffusion generator to compute
\begin{align}
[\mathcal{C}_x, H_0(D)]= D \left[a^\dagger(x) \nabla_x^2 a(x) - (\nabla_x^2 a^\dagger(x))  a(x)\right]\,.
\label{eq:commutdensopdiff}
\end{align}
which can be proven with the commutation relations \cref{eq:a+a-CommutRels} and by integration by parts.
One can now apply the density operator to an arbitrary CDME $\partial_t \ket{\rho} = H \ket{\rho}$ to yield the equation for the stochastic concentrations as
\begin{align} \label{eq:concentrations}
    \partial_t \mathcal{C}_x \ket{\rho}
    & = \left[\mathcal{C}_x, H\right]\ket{\rho} + H\mathcal{C}_x \ket{\rho},
\end{align}
written in terms of the commutator. As we see shortly in \cref{sec:emerRD}, macroscopic reaction-diffusion equations for the concentrations can be recovered by taking the expectation value of the above equation, along with a mean-field approximation. In general, \cref{eq:concentrations} could also be used to study fluctuations and higher-order moments of the concentration field.

One can apply the same procedure with the particle number operator from \cref{eq:partNumOpSpace}. Its commutator with diffusion is $\left[\mathcal{N}, H_0(D)\right] = 0$, which means the diffusion process alone does not change the total number of particles. Applying $\mathcal{N}$ to the CDME removes the spatial dependence and yields the CME from \cref{sec:mastereqn}. However, the relation between the CDME and CME rates is in general not trivial for systems with non-linear reactions, due to the influence of diffusion in the reaction process \cite{doi1976stochastic,del2024open}.

Following this procedure, we can apply any operator of interest to the master equation and directly obtain equations for the fluctuating observable with considerable less effort than with standard methods. Then, as we will now discuss, we can take the expectation value (or compute higher-order moments) of the desired observable to obtain macro or mesoscopic equations, which highlights the effectiveness of the second quantized representation. 

\subsection{Emergence of macroscopic reaction-diffusion equations}
\label{sec:emerRD}
In the same spirit as in \cref{sec:emergencerateeqs}, we obtain the mean-field equations for the concentrations with space dependence, see \cref{fig:maindiag}. To do so, we derive an equation for the expected concentration at $x$, $c(x) = \bra{1} \mathcal{C}_x \ket{\rho}=\bra{1} a(x) \ket{\rho}$ as may be verified by explicit contraction of \cref{eq:stochconc} with the flat state $ \bra{1}$ from \cref{eq:flatstateDoi}:
\begin{align}
\bra{1}\mathcal{C}_x \ket{\rho} 
&= \sum_{m\geq 1} m \int_{\mathbb{X}^{m-1}} d\mb{z} \, \rho_m(x,\mb{z}), 
\label{eq:bracketConc}
\end{align}
which is equivalent to the inclusive density of \cref{eq:inclusivedensDef} for $k=1$.
This equation hints at two possible pathways to calculate the equation for the concentrations: via the left hand side using Fock space or via the right hand side using the scalar probability densities. We will demonstrate the former one here in the case of a non-linear bimolecular reaction-diffusion system: mutual annihilation $A+A\rightarrow \emptyset$ with reaction rate function $\lambda(\mb{z})$, where $\mb{z}=(z_1,z_2)$ are the positions of the two reactants. 

We start with the master equation for the diffusion generator $\partial_t \ket{\rho} = H_0(D) \ket{\rho}$ 
\begin{align}
	\bra{1} \mathcal{C}_x \partial_t \ket{\rho} = \bra{1} [\mathcal{C}_x, H_0(D) ]\ket{\rho} = D\nabla_x^2  c(x) \,,
\end{align}	 
where we used the commutation relation from \cref{eq:commutdensopdiff} and that $\bra{1}\nabla_x^2(a^\dagger(x)) = \nabla_x^2(\bra{1}a^\dagger(x)) = 0$. This recovers the standard macroscopic diffusion equation.
The reaction generator is composed out of the loss and gain operators \cref{eq:lossdoigeneralonespecies,eq:gaindoigeneralonespecies} with $k=2$ and $l=0$:
\begin{align}
\begin{aligned}
    \mathcal{L}&=\frac{1}{2}\int_{\mathbb{X}^2} d\mb{z} \ a^\dagger(\mb{z})\lambda(\mb{z})a(\mb{z}), \\
    \mathcal{G}&=\frac{1}{2}\int_{\mathbb{X}^2} d\mb{z} \ \lambda(\mb{z})a(\mb{z}).
\end{aligned}
\end{align}
The reaction terms are treated similar to the diffusion term, using the property of the flat state $\bra{1}a^\dagger(x) = \bra{1}$
\begin{align}
	\begin{split}
	\bra{1} \mathcal{C}_x H(\lambda) \ket{\rho} 
    &=\bra{1} a(x) \mathcal{G}\ket{\rho}  - \bra{1} a(x)\mathcal{L} \ket{\rho}
    \end{split}
\end{align}
The first term is simply
\begin{align}
    \begin{split}
    \bra{1} a(x) \mathcal{G}\ket{\rho} 
    &= \frac{1}{2}\int_{\mathbb{X}^2} d\mb{z} \ \lambda(\mb{z}) \,  \bra{1} a(\mb{z}) a(x)\ket{\rho}.
    \end{split}
\end{align}
The second term is computed using the relations \cref{eq:a+a-CommutRels2} along with the properties of the flat state: 
\begin{align}
\begin{split} \label{eq:lossexample}
    &\bra{1} a(x) \mathcal{L}\ket{\rho} 
    = \int_{\mathbb{X}} dz \ \lambda(x,z) \bra{1} a(x)a(z)\ket{\rho}  \\
    & \qquad + \frac{1}{2}\int_{\mathbb{X}^2} d\mb{z} \ \lambda(\mb{z})  \bra{1}a(\mb{z}) a(x)\ket{\rho},
    \end{split}
\end{align}
When gathering the loss and gain terms, the last term in \cref{eq:lossexample} cancels and we are left with only the first integral. We recognize that the integrand $\bra{1} a(x)a(z)\ket{\rho}$ corresponds to 
the expectation of the second-order particle number/density operator $\mathrm{E}[ C(z)C(x) ]$, which is equivalent to the inclusive density from \cref{eq:inclusivedensDef} for $k=2$. In general $\mathrm{E}[ C(z)C(x)] = \mathrm{E}[ C(z)]\mathrm{E}[C(x)] + \mathrm{Cov}(C(z),C(x))$, thus 
\begin{align}
    \bra{1} \mathcal{C}_x H(\lambda) \ket{\rho} 
    &=  \int_{\mathbb{X}} \lambda(x,z) \  c(x) c(z) dz \\
    & \qquad  + \int_{\mathbb{X}} \ \lambda(x,z) \mathrm{Cov}( C(z),C(x)) dz. \nonumber
\end{align}
To obtain a closed equation, we need to choose the form of the reaction rate function $\lambda(x,z)$. We choose a simple indicator function $\lambda(x,z) = \kappa \mathbb{1}_{|x-z|\leq \sigma}$, which corresponds to particles reacting with constant rate $\kappa$ if their relative distance is less or equal to $\sigma$ \cite{doi1976stochastic}. Moreover, if the number of particles is large enough, the covariance is negligible in comparison with the product of the means due to the law of large numbers \cite{anderson2015stochastic,kostre2021coupling,del2024open}. Thus, assuming large copy numbers, $\mathrm{Cov}( C(z),C(x))\rightarrow 0$, and we are left with 
\begin{align}
	\begin{split}
		\bra{1} \mathcal{C}_x H_m(\lambda) \ket{\rho} 
		&=  \kappa \frac{4\pi}{3}\sigma^3 c(x)^2, 
	\end{split}
\end{align}
where we assumed three dimensional space and that the region sigma is small enough, so we can approximate $c(z)$ by its center value $c(x)$. Thus the emerging reaction-diffusion PDE for the mutual annihilation reaction is
\begin{align}
	\partial_t c(x) = D\nabla^2  c(x) - r c(x)^2 \quad \text{with} \quad r= \kappa \frac{4\pi}{3}\sigma^3,  
\end{align}
where $r$ is the macroscopic rate given in terms of the microscopic parameters $\sigma$ and $\kappa$. This result is consistent with previous results \cite{doi1976stochastic,smith2019spatial}. Moreover, the work \cite{kostre2021coupling} obtained the same result by writing the trajectory representation of a discretized convergent RDME \cite{isaacson2013convergent,isaacson2018unstructured} and then taking the expected value and large copy number limit. This result was also shown rigorously in \cite{isaacson2021reaction} by using mean-field limits of the underlying measure-valued stochastic process \cite{isaacson2022mean,popovic2023spatial} and also more recently in \cite{del2024open} using intuitive classical methods on the CDME. 

Reaction-diffusion PDEs have also been derived by applying mean-field limits to interacting particle systems with fixed total population but fluctuating subpopulations between species \cite{lim2020quantitative,oelschlager1989derivation,stevens2000derivation}, and more recently stochastic reaction-diffusion PDEs have been derived to include fluctuations
\cite{heldman2024fluctuation}. The field theory representation presented here is a straightforward method to perform many of these calculations. It can be easily employed to obtain the covariance and higher-order moments of any observable, or to construct alternative PDEs resulting from more complex reaction rate functions, yielding novel meso- and macroscopic equations with parameters that can be consistently translated to parameters of particle-based models.

\section{Concluding remarks and future scope}

We presented a unified and structured general framework for field theory representations of reaction-diffusion processes. Throughout this process, we reviewed a large body of previous work and placed this within the general framework (\cref{fig:maindiag}). In particular, from the basis-independent representation with continuous spatial resolution, we obtained various other field theory representations as special and/or limiting cases. We further exposed the relationships with well-known chemical physics models and showed how to obtain numerical discretizations, as well as meso- and/or macroscopic emerging models using field theory techniques. In addition, we overview the path-integral representations, which combined with the Galerkin discretization, yields an approach to obtain discretized meso/macroscopic models directly from the formulation at the particle level. To spark the interest of the reader, we delved briefly into some advanced topics of interest as large deviation theory, perturbation theory, renormalization group methods, and master equations for stochastic concentrations. 

The framework is presented in the context of reaction-diffusion, partly to keep the presentation simple and also due to the extensive applicability of reaction-diffusion processes. However, with a few modifications, one can extend the framework to a larger class of systems, while preserving the structure presented in \cref{fig:maindiag}. For instance, it is possible to take inertia into account by incorporating velocities into the dynamics. The diffusion terms would then model the Newtonian mechanics in the form of Liouville equations. If noise in the form of a thermostat is included, it transforms into Langevin dynamics whose probabilistic description is given by the Klein-Kramers equation \cite{ doi1976second,van1992stochastic}. On the other hand, the reaction part is not limited to model reactions; for instance, it could also model the addition or removal of particles due to interactions with a reservoir. We can thus extend the framework for these and other systems, as already partly explored for specific cases \cite{del2024open,doi1976stochastic,grassberger1980fock}. From a dynamical systems perspective, we can even apply it to any Markovian dynamical system---deterministic or stochastic---with variable particle numbers. Thus, this is a very general approach to handle classical dynamical systems with varying particle numbers and could be expanded to non-Markovian dynamics.

In the context of general dynamical systems, the application scope broadens considerably. The methods presented comprise a unified toolkit to operate on the probabilistic dynamics of particle or agent-based models with varying numbers: one can perform calculations, discretizations and derive emergent models at multiple scales without handling the combinatorics explicitly. Each of the different representations, such as the basis-independent, the Doi formulation or the Galerkin discretization, prove to be useful in different contexts and aid in establishing connections between the models at different scales. Using these methods and representations, it is possible to create mathematical bridges between particle or agent-based models of complex systems and their meso- and macroscopic descriptions. These bridges constitute the scaffolding to construct consistent numerical methods of complex systems across scales, as already shown for reaction-diffusion systems in specific settings \cite{erban2024multi,kang2019multiscale,kostre2021coupling,del2021multiscale,del2024open,smith2018spatially,winkelmann2020stochastic}. We foresee that the formulation of basis-independent models across different scales will be a key ingredient in the development of adaptive multiscale numerical methods, such as
adaptive mesh refinement \cite{bell1994three,hellander2020hierarchical} and adaptive moving mesh methods \cite{zegeling_2004}.

In particular, in the context of emerging models, we reviewed a consistent methodology to obtain meso- and macroscopic models using field theory techniques. The coarser descriptions emerge from specific limits or mean-field approximation of the finer levels. For instance, using the stochastic concentration description from \cref{sec:stochconcentrations}, we can take expectations by applying the flat state to obtain a macroscopic PDE. Moreover, it is possible to construct a stochastic PDE based on the stochastic concentration description itself. For instance, one could calculate the equations for the second and/or higher moments and characterize the features of the space-time noise term. For such tasks, the inclusive densities introduced in \cref{sec:inclusivedensity} might become essential to simplify calculations. Research on stochastic PDEs has been increasing considerably in the last decades \cite{hairer2009introduction,hairer2014theory,pavliotis2008stochastic}, specially in the context of multi-scale modeling \cite{abdulle2012numerical,helfmann2021interacting,baish2024fock,helfmann2023modelling,wehlitz2024approximating}, and it is a promising field of application for the techniques presented in this work. 

There are several relevant extensions of the techniques presented here in connection with other fields. Of particular interest is in the context of open driven (stochastic) quantum systems \cite{sieberer2016keldysh, van2015open, perfetto2023reaction, gerbino2024large} and their thermodynamics \cite{dann2021quantum,kosloff2019quantum}, which are often modeled with master equations of the Lindblad type \cite{manzano2020short}. There is on-going research on the correct form of this equation to model material exchange with reservoirs \cite{dann2021quantum,delle2024effective}, and it might be possible to reformulate the framework presented here in this setting. Another interesting extension is in the direction of stochastic thermodynamics \cite{pruessner2022field,van2013stochastic,zhang2023entropy}, which has not been fully established in the context of reaction-diffusion processes due to their inherent complexity. The formulations presented here provide an accessible framework for this development. In connection with mathematics, the groundbreaking theory of regularity structures \cite{hairer2014theory} provides an algebraic framework that gives a mathematically rigorous meaning to many stochastic PDEs. This theory was partly inspired by the mathematical analysis of field theories, so it is of interest to explore its connections with the work presented here. This could make the results of the theory accessible and applicable to a broader audience. Additional extensions consist of establishing deeper connections with Malliavin calculus \cite{lanconelli2023using},  incorporating the Kolmogorov backward equations in the general setting by following \cite{weber2017master}, as well as making the link between the generic field theory and large deviation theory more robust, see for instance \cite{smith2011large,assaf2017wkb,lazarescu2019large,bressloff2021coherent}.

Taking all this into consideration, we envision applications to multi-scale modeling and simulation of complex systems in a diverse range of domains. Especially interesting are those which cross the boundaries of existing disciplines, such as the use of stochastic predator-prey systems to model the onset of turbulence \cite{shih2016ecological}, or the renormalization group analysis of swarms of active particles \cite{cavagna2023natural}. We also foresee possible applications in enzyme kinetics and cascades \cite{lan2006variational}, cellular and molecular biology \cite{erban2009stochastic,bressloff2014stochastic}, tumor migration models \cite{deroulers2009modeling}, bacterial dynamics \cite{thompson2011lattice}, stochastic population dynamics \cite{assaf2017wkb}, gene expression models \cite{bressloff2017stochastic}, actin filament and microtubule growth \cite{pausch2019actin} and neurological applications from molecular transport and firing in neurons, to cognitive phenonema such as decision making \cite{chow2015path}. Ultimately, the methods described here could also be useful in multi-scale modeling of socio-economical phenomena such as segregation \cite{seara2025sociohydrodynamics}, polarization, opinion and voter dynamics \cite{helfmann2023modelling,redner2019reality} and economic inequality. With such a great potential for diverse applications, we hope that this work helps to connect communities from different disciplines working on mathematically similar systems, as it demonstrates how existing applications in diverse fields are connected in a single unifying formalism.

\section*{Acknowledgements}
The authors acknowledge support by the Dutch Institute of Emergent Phenomena (DIEP) cluster at the University of Amsterdam. MJR also acknowledges support from DFG Grant No. RA 3601/1-1.


%

\end{document}